\documentclass[aps,twocolumn]{revtex4-1}
\usepackage{amsmath,amssymb}
\usepackage{slashed}
\usepackage{graphicx}
\usepackage{bm}
\usepackage[T1]{fontenc}
\usepackage[utf8]{inputenc}
\usepackage{gauss} 
\usepackage[top=1.0in,bottom=1.0in,left=1.0in,right=1.0in]{geometry}
\usepackage[colorlinks,linkcolor=blue,citecolor=blue]{hyperref}
\usepackage{subfig}
\usepackage{dsfont}
\newcommand{\be}{\begin{equation}}
\newcommand{\ee}{\end{equation}}
\newcommand{\bea}{\begin{eqnarray}}
\newcommand{\eea}{\end{eqnarray}}
\newcommand{\ba}{\begin{eqnarray}}
\newcommand{\ea}{\end{eqnarray}}

\begin{document}

\title{Hadronic structure on the light-front  VII.\\
Pions and kaons  and their partonic distributions}

\author{Wei-Yang Liu}
\email{wei-yang.liu@stonybrook.edu}
\affiliation{Center for Nuclear Theory, Department of Physics and Astronomy, Stony Brook University, Stony Brook, New York 11794--3800, USA}

\author{Edward Shuryak}
\email{edward.shuryak@stonybrook.edu}
\affiliation{Center for Nuclear Theory, Department of Physics and Astronomy, Stony Brook University, Stony Brook, New York 11794--3800, USA}

\author{Ismail Zahed}
\email{ismail.zahed@stonybrook.edu}
\affiliation{Center for Nuclear Theory, Department of Physics and Astronomy, Stony Brook University, Stony Brook, New York 11794--3800, USA}

\begin{abstract}
This  work is a continuation in our  series of papers,  that addresses quark models of hadronic structure on the light front, motivated
by the QCD vacuum structure and lattice results. The spontaneous breaking of chiral symmetry on the light front, is shown to parallel
that in the rest frame,  where the non-local instanton induced $^\prime$t Hooft interaction plays a central role.  By rewriting this interaction solely
in terms of the good component of the fermionic field, a scalar chiral condensate emerges in  the mean-field approximation, which is identical
to the one obtained in the rest frame.  The pions and kaons emerge as deeply bound Goldstone modes in the chiral limit, with the scalar-isoscalar 
sigma meson mode as a threshold state with zero binding. We explicitly derive the light front distribution amplitudes (DAs) and partonic functions
(PDFs)  for these mesons. The DAs and PDFs  are in good agreement with those extracted from the QCD instanton vacuum in the rest frame, using the 
large momentum effective theory (LaMET). The QCD evolved DAs and PDFs compare well with available measurements, as well as recent lattice results.
 \end{abstract}
\maketitle

\section{Introduction}

Parton distribution functions (PDFs) are important for the description of hadrons at high energy.
They are currently central for the description of high energy cross sections at the Large Hadron
Collider (LHC). The PDFs describe  uni-dimensional distribution of the quark and gluons in the
infinite momentum frame, and are inherently non-perturbative. In the leading twist approximation,
they can be gleaned from experiments~\cite{Farrar:1979aw}, or more recently from first principle
lattice simulations~\cite{Zhang:2017bzy} using the LaMET  as  suggested  in~\cite{Ji:2013dva},
and some variants~\cite{Radyushkin:2017gjd,Nam:2017gzm}. The generalized PDFs offer a multi-dimensional description of the
quarks and gluons, but their extraction from current or future  experiments, as well as numerical lattice simulations, is more challenging.

A physical  understanding of the quark and gluon distributions in hadrons at low resolution, requires
an understanding of the QCD vacuum.  Cooled lattice simulations whereby the gauge configurations 
are iteratively pruned of quantum fluctuations~\cite{Chu:1994vi}, suggest a semi-classical landscape dominated by 
tunneling  instanton and anti-instanton  configurations, with large actions and finite topological charge.
Their inclusion in the determination of the PDFs for the light pseudoscalar mesons,
is one of the essential thrust of this work.

The QCD instanton liquid model (ILM) is a comprehensive model of QCD at low resolution, that is  currently supported 
in its details by current  lattice cooling simulations. It unequivocally captures the essentials of the spontaneous breaking of chiral symmetry, with
the emergence of the quark zero mode zone. It provides a semi-classical description of the QCD ground state
at low resolution, hence a well defined organizational principle that enforces 
 chiral and gauge Ward identities~\cite{Diakonov:1985eg,Shuryak:1988zx,Nowak:1989jd,Kacir:1996qn,Schafer:1996wv}.

However, the QCD instanton liquid model is inherently space-like, making the ensuing physics less transparent
time-like. This is particularly acute for the PDFs which capture the non-perturbative time-like structure the
partonic  constituents of hadrons, as probed by deep inelastic scattering. Recently, two of us in a series of
papers~\cite{Shuryak:2021fsu,Shuryak:2021hng,Shuryak:2021mlh,Shuryak:2022thi,Shuryak:2022wtk}, 
have put forth a program on how to export the successes of the QCD instanton liquid model, 
space-like. The program is based on the idea of deriving the essental of the central and spin forces on the light
front by analytically continuing in (Euclidean) rapidity pertinent correlators space-like. In a way, this construction
is similar in spirit to the one put forth by Ji~\cite{Ji:2013dva}.

This paper is a continuation of this series, whereby we show in details how chiral symmetry is spontaneously broken on the light front,
using solely the emergent $^\prime$t Hooft effective interactions from the QCD instanton liquid model. 
It fills for the schematic argument we put forth in~\cite{Shuryak:2021hng}. The light
front formulation of the QCD vacuum supports a scalar chiral condensate, which is identical to the one derived 
in the rest frame, as expected. The ensuing light front (LF) pion and kaon wavefunctions are characterized by
the same masses and decay constants as in the rest frame. In addition, they provide for a detailed description
of the partonic content of these mesons for the leading twist-2 operators. For the latter, a number of models 
have been also used~\cite{Chang:2013pq,Chen:2016sno,Ding:2019lwe,RuizArriola:2002bp,Dorokhov:2011ew,Broniowski:2017wbr,Broniowski:2017zqz,Praszalowicz:2002ct,Dumm:2013zoa,Petrov:1997ve,Petrov:1998kg,Dorokhov:1991nj,Dorokhov:1998up,Anikin:2000bn,Dorokhov:2000gu,Nam:2006au,Nam:2017gzm,Radyushkin:1994xv,Brodsky:2011yv,Brodsky:2014yha,Jia:2018ary,Lan:2019vui}.

The organization of the paper is as follows. In section~\ref{SEC_ILM} we briefly review the emergent 
$^\prime$t Hooft multi-flavor and non-local fermionic interactions in the QCD ILM vacuum.  In 
section~\ref{SEC_ILMLF} we address the role of these effective interactions on the light front. The 
central new element on the light front is the decomposition of the fermionic fields into good plus bad
components along the light cone directions. The bad component is a constraint, as it does not propagate
in the light front direction. The elimination of the constraint in the mean field approximation, yields a
 constituent quark mass much like in the rest frame, and  renormalized  $^\prime$Hooft multi-flavor
interactions, as initially proposed in~\cite{Bentz:1999gx,Itakura:2000te,Naito_2004} for local interactions 
of the Nambu-Jona-Lasinio (NJL) type. 
 In section~\ref{SEC_ILMH} we derive the corresponding light front Hamiltonian solely
in terms of the good fermionic component. In section~\ref{SEC_ILMGAP} we show that the the
emergent constituent quark mass and chiral condensate on the light front, are identical to the ones in the
rest frame. As expected, the spontaneous breaking of chiral symmetry is driven by the induced 
multi-flavor instanton and anti-instanton interactions in the ILM. It is universal and frame independent.
In section~\ref{SEC_ILMLFS} we  diagonalize the light front Hamiltonian in the 2-body sector for light
$u,d$ quarks, with  a strongly bound pion as a Goldstone mode. All other scalar and pseudoscalar 
modes areunbound on the light front, in the mean field approximation and away from the chiral limit.
The pion DA stemming from the diagonalization is also discussed.
In section~\ref{SEC_PART} we derive the pion PDF. The result is
QCD evolved with detailed comparison to existing measurements, lattice results and models. 
In section~\ref{SEC_GAPS} we extend our analysis to include strangeness, and derive the
pertinent gap equations and chiral condensates. In section~\ref{SEC_KAON} the light front
wavefunctions and masses for the kaons, using $U$- and $V$-spin are discussed. The
kaon DA and PDF are derived. The results are also evolved and compared with available 
empirical data,  recent lattice results and models. Our conclusions are in sec~\ref{SEC_CONC}. 
A number of Appendices are included to complement some of the derivations.

\section{Flavor interactions in ILM}
\label{SEC_ILM}
In the ILM, the QCD vacuum is composed of instanton and anti-instanton gauge fields,
tunneling and topologically active gauge configurations, surrounded by swaths of empty
space-time free of perturbative fields \cite{Leinweber:1999cw,Biddle:2019gke,Biddle:2020eec}. 
These configurations are self-dual and strong, with typical a size of $\frac 13$ fm, and a mean 
tunneling density of  $1/R^4\approx 1 {\rm fm}^{-4}$~\cite{Shuryak:1981fza}. As a result, the
starting and complex gauge dynamics can be reduced to the dynamics of a dilute ensemble
of pseudoparticles  on their pertinent moduli, and organized using the packing fraction
\bea
\label{KAPPA}
\kappa_{I+\bar I}=\frac{2\pi^2\rho^4}{R^4}\approx 0.1
\eea

The remarkable thingh about the self-dual instantons (anti-self-dual anti-instantons) is their
ability to trap quark states as zero modes with fixed helicity (left for instantons and right for 
anti-instantons). The collectivization of these zero modes, give rise a zero mode zone in
a narrow band  of virtualities
\bea
\label{BAND}
|\Delta\lambda |\sim {\rho^2\over R^3} \sim 20\, {\rm MeV}
\eea
with a mean  density $\varrho(0)$ of near-zero Dirac eigenvalues. As noted initially by
Banks and Casher~\cite{Banks:1979yr}, this mean density gives rise to a finite quark condensate
\bea
\langle \bar\psi\psi\rangle =-\pi\varrho(0)
\eea
(\ref{BAND}) is also characterized by universal fluctuations of the chiral condensate
in the microscopic limit, which are captured by chiral random matrix theory~\cite{Verbaarschot:1993pm}. 
In many ways, the spontaneous breaking of chiral symmetry in the ILM is tantamount
of the onset of conductivity in dirty metals.

The delocalization of the quark zero modes and the emergence of (\ref{BAND}), is a
direct prove of the importance of the gauge topology in the spontaneous breaking of
chiral symmetry. It provides for a  microscopic origin of the spontaneous breaking of chiral 
in QCD, as noted originally by   $^\prime$t Hooft ~\cite{tHooft:1976snw}. It generates 4-dimensional fermionic 
zero modes,  which lead to multi-fermion flavor mixing  interactions, 
$$ \bar u u \leftrightarrow \bar d d  \leftrightarrow \bar s s $$
which are quasi-local.

For  $N_f=3$ and in the instanton size zero size approximation, the interactions between
the $u,d,s$ quarks in the current mass limit is~\cite{Chernyshev:1995gj} 
 \begin{widetext}
 \bea
	\label{DETUDS}
	{\cal V}^{L+R}_{qqq}=\frac {G_{Hooft}}{N_c(N_c^2-1)}\bigg[
	\bigg(\frac{2N_c+1}{2(N_c+2)}\bigg)  {\rm det}(UDS) +\frac 1{2(N_c+1)}\,\bigg({\rm det}(U_{\mu\nu}D_{\mu\nu}S) + {\rm cyclic}\bigg)\bigg] +(L\leftrightarrow R)
\eea	
\end{widetext}
with $U=\overline u_R u_L$ and $U_{\mu\nu}=\overline u_R\sigma_{\mu\nu} u_L$, and similarly for $D,S$.
The strength of the 6-quark operators is related to the instanton plus anti-instanton density
\be
\label{DETUDSX}
 G_{\rm Hooft}={n_{I+\bar I} \over 2} \bigg(4\pi^2\rho^3 \bigg)^3 \bigg({1 \over m^*_u\rho}\bigg) \bigg({1 \over m^*_d\rho}\bigg) \bigg({1 \over m^*_s\rho}\bigg)
\ee
with the effective quark masses $m^*_q$.
For most of the analyses to follow in this work, we will specialize to 
$N_f=2$ with $u,d$ species with almost degenerate quark masses.
Strangeness will be addressed also in the same spirit, using $U$- and $V$-spin.
With this in mind, (\ref{DETUDS}) reduces to
\begin{widetext}
\bea
\label{TH2}
{\cal V}^{L+R}_{qq}=
{\kappa}_2\,\frac{(2N_c-1)}{2N_c(N_c^2-1)}\,
\bigg({\rm det}(UD)+\frac 1{4(2N_c-1)}\,{\rm det}(U_{\mu\nu}D_{\mu\nu})\bigg)
+(L\leftrightarrow R)
\eea
with $\kappa_2=3\,G_{\rm Hooft}\langle \overline s s\rangle<0$  and   attractive.
(\ref{TH2}) breaks explicity $U_A(1)$
axial symmetry, but otherwise preserves flavor left-right symmetry. This can be made explicit
through Fierzing, with the full Lagrangian now of the form
\begin{align}
\label{FTH}
   \mathcal{L}\propto
   \bar{\Psi}(i\slashed{\partial}-m)\Psi
   &+\frac{G}{8(N^2_c-1)}\left\{\frac{2N_c-1}{2N_c}\left[(\bar{\Psi}\Psi)^2-(\bar{\Psi}\tau^a\Psi)^2-(\bar{\Psi}i\gamma^5\Psi)^2+(\bar{\Psi}i\gamma^5\tau^a\Psi)^2\right]\right.\nonumber\\
   &\left.-\frac{1}{4N_c}\left[\left(\bar{\Psi}\sigma_{\mu\nu}\Psi\right)^2-\left(\bar{\Psi}\sigma_{\mu\nu}\tau^a\Psi\right)^2\right]\right\}
\end{align}
\end{widetext}
It is strongly attractive in the
sigma and pion channel, and repulsive in the $\eta^\prime$ channel thereby solving the $U_A(1)$ 
problem. (\ref{FTH}) in QCD is a replacement to the posited  Nambu-Jona-Lasinio model~\cite{Nambu:1961tp} of pre-QCD.

For the small size mesons such as pions and kaons (and also Upsilons), the quasi-local approximation is not
reliable, and the full non-locality of the instanton zero modes need to be retained. This amounts to the shift
\begin{equation}
\label{NLTH}
    \psi(x)\rightarrow\sqrt{\mathcal{F}(i\partial)}\psi(x)
\end{equation}
in (\ref{FTH}). The explicit form of $\mathcal{F}(k)$ in momentum space is
\begin{equation}
\label{FKK}
    \mathcal{F}(k)=\left[(zF'(z))^2\right]\bigg|_{z=\frac{k\rho}{2}}
\end{equation}
where $$F(z)=I_0(z)K_0(z)-I_1(z)K_1(z)$$ and $k=\sqrt{k^2}$ is the Euclidean $4$-momentum.

 \section{$^\prime$T Hooft effective Lagrangian on the light front}
 \label{SEC_ILMLF}
 On the light front, the old lore was that the QCD vacuum is trivial owing to the vanishing of the backward diagrams in perturbation theory~\cite{Lepage:1980fj}.
However, more carefull  analyses  reveal that the vacuum physics is encoded in the  longitudinal zero modes of the fields~\cite{Ji:2020baz}, 
which is rather manifest
 in two-dimensional QCD~\cite{Ji:2020bby} (and references therein). In effective models of QCD 
 such as the NJL model, the non-trivial aspects of the vacuum on the light front  are explicitly tied to the tadpole contributions,
 generated by the constrained part of the fermion field~\cite{Bentz:1999gx,Itakura:2000te,Naito_2004}.  Most of the analysis to follow in the ILM,
 will make use of this observation.

 More specifically, the projection of the fermion field along the light front,  splits the field into a good plus bad component, with the latter 
non-propagating or constraint. The elimination of the constaint, induces multi-fermion interactions in terms of the good component. In the mean-field 
approximation using $\frac 1{N_c}$ counting rules, these interactions account for the spontaneous breaking of chiral symmetry on the light front
 through tadpoles~\cite{Bentz:1999gx, Itakura:2000te,Naito_2004}.
  In this section, we will show that this approach can be applied to the emergent multi-flavor interactions in the ILM, with finite size form factors
 from the quark zero modes.

For simplicity and clarity of the analysis, we consider first the local form of (\ref{FTH}) in the large $N_c$ approximation.
The modifications for the finite instanton sizes will be quoted at the end. More specifically, we have
\begin{widetext}
\begin{equation}
\begin{aligned}
\label{THLOC}
    \mathcal{L}
    \propto\bar{\Psi}(i\slashed{\partial}-m)\Psi+\frac{G_S}{2}\left[(\bar{\Psi}\Psi)^2-(\bar{\Psi}\tau^a\Psi)^2-(\bar{\Psi}i\gamma^5\Psi)^2+(\bar{\Psi}i\gamma^5\tau^a\Psi)^2\right]\\
\end{aligned}
\end{equation}
with $G_S=\frac{G}{4N^2_c}$. In the mean-field or $\frac 1{N_c}$ approximation, it is customary to use the semi-bosonized form of (\ref{THLOC}),
through the use of the auxillary fields $\sigma$, $\sigma^a$, $\pi$, and $\pi^a$ 
\begin{equation}
\label{SEMI}
    \mathcal{L}\propto\bar{\Psi}(i\slashed{\partial}-m)\Psi+G_S\bar{\Psi}\left(\sigma-\sigma^a\tau^a-i\pi\gamma^5+i\pi^a\tau^a\gamma^5\right)\Psi-\frac{G_S}{2}\left[\sigma^2-(\sigma^a)^2-\pi^2+(\pi^a)^2\right]
\end{equation}
\end{widetext}
where $m=m_u=m_d$ represents the current mass of the $u,d$ quarks.  
Note that the combination
$$\sigma^2-(\sigma^a)^2-\pi^2+(\pi^a)^2$$
is flavor $U_A(1)$ violating. 
The extension to $s$ quarks with larger mass will be discussed below,
by  reduction to $U$- and $V$-spin.

To proceed, we now split the
fermionic field $\Psi$ into a good component $\Psi_+$ and a bad component $\Psi_-$
\bea
\label{GOODBAD}
\Psi=\Psi_++\Psi_-\equiv \frac{1}{2}\gamma^{-}\gamma^{+}\Psi+\frac{1}{2}\gamma^{+}\gamma^{-}\Psi
\eea
The bad component does not propagate along the light front $x^+$-direction, and therefore can be eliminated from (\ref{THLOC}) using the 
equation of motion. The latter is readily solved in terms of the good component  $\Psi_+$
\begin{equation}
\begin{aligned}
\label{BADSOLVED}
    \Psi_-=&\frac{\gamma^+}{2}\frac{-i}{\partial_-}\left(-i\gamma^i_\perp\partial_i+\hat{M}\right)\Psi_+\\
\end{aligned}
\end{equation}
where  $\hat{M}$ denotes
\begin{equation}
    \hat{M}=m-G_S\left(\sigma-\sigma^a\tau^a-i\pi\gamma^5+i\pi^a\tau^a\gamma^5\right)
\end{equation}
$\frac{-i}{\partial_-}$ is equal to the Green's function $G(x^-,y^-)$ for $i\partial_-$ where the Green's function can be defined as
\begin{equation}
    G(x^-,y^-)=\int_{-\infty}^\infty \frac{dk^+}{2\pi}\frac{1}{k^+}e^{-ik^+(x-y)^-}=\frac{-i}{2}\epsilon(x^--y^-)
\end{equation}
In terms of  (\ref{GOODBAD}), the semi-bosonized Lagrangian (\ref{SEMI})  can be solely rewritten in terms of the good component
\begin{widetext}
\begin{equation}
\begin{aligned}
\label{LGOOD}
     \mathcal{L}\propto &\bar{\Psi}_+i\gamma^+\partial_+\Psi_+-\frac{G_S}{2}\left[\sigma^2-(\sigma^a)^2-\pi^2+(\pi^a)^2\right]-\bar{\Psi}_+\left(i\gamma^i_\perp\partial_i-\hat{M}\right)\frac{\gamma^+}{2}\frac{-i}{\partial_-}\left(i\gamma^i_\perp\partial_i-\hat{M}\right)\Psi_+\\
\end{aligned}
\end{equation}
\end{widetext}

It is now clear, that the elimination of the Gaussian mesonic fields from (\ref{LGOOD}) generates strings of multi-fermion interactions, of increasing complexity on the light front.
Fortunately, they  are tractable in  $1/{N_c}$, with the leading order referred to as 
the  mean-field approximation. For that, we shift the scalar field by $\sigma=N_c\sigma_0+\delta\sigma$ in (\ref{LGOOD}) with a finite
vev $\sigma_0\sim N_c^0$, as all   other vevs are excluded by isospin symmetry and parity. We can now use the counting rules
$$\sigma^a, \pi, \pi^a, \delta\sigma\sim {\cal O}(\sqrt{N_c})$$
with $g_S=N_c G_S\sim N_c^0$, in the semi-bosonized Lagrangian (\ref{LGOOD}), to resum all the leading  tadpole diagrams, by setting the coefficient of $\delta\sigma$  to zero, 
\bea
\label{CONDZ}
\langle \bar{\psi}\psi\rangle-N_c\sigma_0=0
\eea
with now the effective fermionic field
\bea
\label{PSINEW}
\psi=\Psi_++\frac{\gamma^+}{2}\frac{-i}{\partial_-}(i\gamma^i_\perp\partial_i-M)\Psi_+
\eea
As a result, the good component of the quark field acquires a constituent mass of order $N_c^0$
\bea
\label{MCONST}
M=m-G_S\langle \bar\psi\psi\rangle
\eea
(\ref{CONDZ}) and (\ref{MCONST}) reflect on the spontaneous breaking of chiral symmetry. (\ref{MCONST}) is a gap equation as we detail below.

As  the remnant fluctuation of the bosonic fields $\sigma^a$, $\pi$, $\pi^a$, and $\delta\sigma=\sigma-N_c\sigma_0$ are of order $\mathcal{O}(\sqrt{N_c})$, 
these fields compensate the $\mathcal{O}(1/N_c)$ contribution from the 't-Hooft coupling $G_S=g_S/N_c$, leaving the semi-bosonized Lagrangian at the leading 
order of large $N_c$ of the form
\bea
\label{LFET}
    \mathcal{L}= &&
    \bar{\psi}(i\slashed{\partial}-M)\psi
    \nonumber\\
&&-\frac{1}{2}G_S\left[\delta\sigma\hat{D}_+\delta\sigma-\sigma^a\hat{D}_-\sigma^a-\pi\hat{D}_-\pi+\pi^a\hat{D}_+\pi^a\right]
    \nonumber\\
    &&+G_S\left[\bar{\psi}\psi\hat{\sigma}-\bar{\psi}\tau^a\psi\sigma^a-\bar{\psi}i\gamma^5\psi\pi+\bar{\psi}i\gamma^5\tau^a\psi\pi^a\right]\nonumber\\
\eea
Higher order contributions have not been retained. 
The factors $\hat{D}_\pm$ inside the quadratic bosonic potential are defined as
\begin{equation}
\label{FC_factor}
\begin{aligned}
    &\hat{D}_\pm=1\pm G_S\left\langle\bar{\psi}\gamma^+\frac{-i}{\overleftrightarrow{\partial_-}}\psi\right\rangle\\
\end{aligned}
\end{equation}
where ${-i}/{\overleftrightarrow{\partial_-}}$  in  (\ref{FC_factor}), is defined such that for any fields $\chi(x)$ and $\psi(x)$, 
\begin{equation}
    \chi(x)\frac{-i}{\overleftrightarrow{\partial_-}}\psi(x)=\frac{i}{\partial_-}\left[\chi(x)\right]\psi(x)+\chi(x)\frac{-i}{\partial_-}[\psi(x)]
\end{equation}
$\hat{D}_\pm$ follows from  the resummation of the tadpole diagrams in mean-field, or leading order in $1/N_c$. Indeed,  each virtual quark tadpole is of order
 ${\cal O}(N_c)$, thereby compensating  the $^\prime$t Hooft coupling $G_S\sim \mathcal{O}(1/N_c)$, with a net factor of $\mathcal{O}(N_c^0)$.
Note that the fermionic contributions   $\bar{\psi}\psi$, $\bar{\psi}\tau^a\psi$, $\bar{\psi}i\gamma^5\psi$, and $\bar{\psi}i\gamma^5\tau^a\psi$ 
in (\ref{LFET}), are all of the same order as the remnant quantum fluctuations or $\mathcal{O}(\sqrt{N_c})$.
With this in mind, we can now eliminate the auxillary bosonic fields, to obtain the light front Lagrangian 
\begin{widetext}
\begin{equation}
\label{LFEFT}
    \mathcal{L}=\bar{\psi}(i\slashed{\partial}-M)\psi+ \frac{G_S}{2}\left[\bar{\psi}\psi\hat{D}^{-1}_+\bar{\psi}\psi-\bar{\psi}\tau^a\psi\hat{D}^{-1}_-\bar{\psi}\tau^a\psi-\bar{\psi}i\gamma^5\psi\hat{D}^{-1}_+\bar{\psi}i\gamma^5\psi+\bar{\psi}i\gamma^5\tau^a\psi\hat{D}^{-1}_+\bar{\psi}i\gamma^5\tau^a\psi\right]
\end{equation}
\end{widetext}
In the mean-field or leading order in $1/N_c$ approximation, the light front Lagrangian can be solely written in terms of
the good fermionic component. The elimination of the bad component, generates a constituent mass, introduces an
effective quark field (\ref{PSINEW}) and renormalizes by $\hat{D}^{-1}_\pm$ (tadpole resummation) each of the original
multi-fermion interaction in the ILM in the zero size limit.

Most of the arguments presented above, carry for the non-local effective Lagrangian in the ILM. More specifically, 
the mean-field version of (\ref{LFET}) is now 
\begin{widetext}
\bea
\label{nonlocal_LFET}
   && \mathcal{L}=\bar{\psi}(i\slashed{\partial}-M)\psi-\frac{1}{2}G_S\left[\delta\sigma\hat{D}_+\delta\sigma-\sigma^a\hat{D}_-\sigma^a-\pi\hat{D}_-\pi+\pi^a\hat{D}_+\pi^a\right]
   \nonumber\\
    &&+G_S\left(\bar{\psi}\sqrt{\mathcal{F}(i\partial)}\delta\sigma\sqrt{\mathcal{F}(i\partial)}\psi-\bar{\psi}\sqrt{\mathcal{F}(i\partial)}\sigma^a\tau^a\sqrt{\mathcal{F}(i\partial)}\psi-\bar{\psi}\sqrt{\mathcal{F}(i\partial)}i\gamma^5\pi\sqrt{\mathcal{F}(i\partial)}\psi+\bar{\psi}\sqrt{\mathcal{F}(i\partial)}i\gamma^5\tau^a\pi^a\sqrt{\mathcal{F}(i\partial)}\psi\right)\nonumber\\
\eea
\end{widetext}
Here  $\sqrt{\mathcal{F}}(i\partial)$ is a derivative operator acting on all of the field on its right-hand side. In momentum space, it 
generates the pertinent form factors inherited from the underlying quark zero modes. Also,
\begin{equation}
\label{FC_factor}
\begin{aligned}
    &\hat{D}_\pm\rightarrow1\pm G_S\left\langle\bar{\psi}\gamma^+\mathcal{F}(i\partial)\frac{-i}{\overleftrightarrow{\partial_-}}[\mathcal{F}(i\partial)\psi]\right\rangle\\
\end{aligned}
\end{equation}
following from the mean-field resummation of the leading tadpoles. The auxillary bosonic fields can be eliminated by carrying explicitly
the Gaussian integration, as in the zero size limit. The astute reader may object that (\ref{nonlocal_LFET}) may suffer 
from abnormal characteristics and light propagation. However, since $\sqrt{\mathcal{F}(i\partial)}$ vanishes rapidly at large $|\rho\partial|$,
this is not the case.

\section{Light front Hamiltonian}
\label{SEC_ILMH}
The effective light front Hamiltonian associated to the mean-field Lagrangian (\ref{LFET}) in the zero size limit,
or (\ref{nonlocal_LFET}) in the finite size limit, can be deived using the canonical rules. The light front Hamiltonian allows
for the explicit derivation of the boost invariant meson spectra and their corresponding 
light-cone wavefunction (LCWFs). We now detail the canonical Legendre transformation from (\ref{LFET}) to the
Hamiltonian, and quote the results for (\ref{nonlocal_LFET}) at the end. 

Given a fermionic field theory in Lagrangian form,  the corresponding symmetric energy-momentum tensor is 
\begin{equation}
    T^{\mu\nu}=\frac{1}{2}\left[\bar{\psi}i\gamma^\mu\partial^\nu\psi+\bar{\psi}i\gamma^\nu\partial^\mu\psi\right]-g^{\mu\nu}\mathcal{L}
\end{equation}
The light front Hamiltonian  is then
\bea
        && P^-=\int dx^-d^2x_\perp~ T^{+-}=\nonumber\\
         &&\int dx^-d^2x_\perp \frac{1}{2}\left[\bar{\psi}i\gamma^+\partial_+\psi+\bar{\psi}i\gamma^-\partial_-\psi\right]-\mathcal{L}\nonumber\\
\eea
Applying this to (\ref{LFET}) gives
\begin{widetext}
\bea
P^-=&&\int dx^-d^2x_\perp \bar{\psi}\frac{\left(-\partial^2_\perp+M^2\right)}{2i\partial_-}\gamma^+\psi\nonumber\\
         &&-\frac{G_S}{2}\int dx^-d^2x_\perp\left[\bar{\psi}\psi\hat{D}^{-1}_+\bar{\psi}\psi-\bar{\psi}\tau^a\psi\hat{D}^{-1}_-\bar{\psi}\tau^a\psi-\bar{\psi}i\gamma^5\psi\hat{D}^{-1}_+\bar{\psi}i\gamma^5\psi+\bar{\psi}i\gamma^5\tau^a\psi\hat{D}^{-1}_+\bar{\psi}i\gamma^5\tau^a\psi\right]
\eea
or in momentum space
\begin{flalign}
&P^-=\int [d^3k]_+\int [d^3q]_+\frac{k^2_\perp+M^2}{2k^+}\bar{\psi}(k)\gamma^+\psi(q)(2\pi)^3 \delta^3_+(k-q)\\[5pt] \nonumber
&+\int [d^3k]_+\int [d^3q]_+\int [d^3p]_+\int [d^3l]_+(2\pi)^3
    \delta^3_+(p+k-q-l)V(k,q,p,l)
\end{flalign}
The first term in the light front Hamiltonian corresponds to the kinetic term. The second term is the two-body interaction relevant to two-particle bound states where the interaction kernel in the scalar and pseudoscakar channel can be defined as 
\begin{equation}
\begin{aligned}
    V(k,q,p,l)=-\frac{G_S }{2}\alpha_+(k^+-q^+)&\bar{\psi}_+(k)\left(\frac{\vec{\gamma}_\perp\cdot\vec{k}+M}{2k^+}\gamma^++ \gamma^+\frac{\vec{\gamma}_\perp\cdot\vec{q}+M}{2q^+}\right)\psi_+(q)\\
    &\times\bar{\psi}_+(p)\left(\frac{\vec{\gamma}_\perp\cdot\vec{p}+M}{2p^+}\gamma^++\gamma^+\frac{\vec{\gamma}_\perp\cdot\vec{l}+M}{2l^+}\right)\psi_+(l) \\
    +\frac{G_S }{2}\alpha_-(k^+-q^+)&\bar{\psi}_+(k)\left(\frac{\vec{\gamma}_\perp\cdot\vec{k}+M}{2k^+}i\gamma^+\gamma^5+i\gamma^5 \gamma^+\frac{\vec{\gamma}_\perp\cdot\vec{q}+M}{2q^+}\right)\psi_+(q)\\
    &\times\bar{\psi}_+(p)\left(\frac{\vec{\gamma}_\perp\cdot\vec{p}+M}{2p^+}i\gamma^+\gamma^5+i\gamma^5\gamma^+\frac{\vec{\gamma}_\perp\cdot\vec{l}+M}{2l^+}\right)\psi_+(l) \\
    +\frac{G_S }{2}\alpha_-(k^+-q^+)&\bar{\psi}_+(k)\left(\frac{\vec{\gamma}_\perp\cdot\vec{k}+M}{2k^+}\gamma^+ \tau^a+\tau^a \gamma^+\frac{\vec{\gamma}_\perp\cdot\vec{q}+M}{2q^+}\right)\psi_+(q)\\
    &\times\bar{\psi}_+(p)\left(\frac{\vec{\gamma}_\perp\cdot\vec{p}+M}{2p^+}\gamma^+\tau^a+\gamma^+\tau^a\frac{\vec{\gamma}_\perp\cdot\vec{l}+M}{2l^+}\right)\psi_+(l) \\
    -\frac{G_S }{2}\alpha_+(k^+-q^+)&\bar{\psi}_+(k)\left(\frac{\vec{\gamma}_\perp\cdot\vec{k}+M}{2k^+}i\gamma^+\gamma^5 \tau^a+\tau^a i\gamma^5\gamma^+\frac{\vec{\gamma}_\perp\cdot\vec{q}+M}{2q^+}\right)\psi_+(q)\\
    &\times\bar{\psi}_+(p)\left(\frac{\vec{\gamma}_\perp\cdot\vec{p}+M}{2p^+}i\gamma^+\gamma^5\tau^a+i\gamma^5\gamma^+\tau^a\frac{\vec{\gamma}_\perp\cdot\vec{l}+M}{2l^+}\right)\psi_+(l)
\end{aligned}
\end{equation}
\end{widetext}
Here $$\alpha_\pm(P^+)=\left[1\pm 2g_S\int\frac{dl^+d^2l_\perp}{(2\pi)^3}\frac{\epsilon(l^+)}{P^+-l^+}\right]^{-1}$$
The fermionic field in momentum space is defined as
\begin{equation}
    \psi(x^-,x_\perp)=\int [d^3k]_+\psi(k)e^{-ik^+x^-+ik_\perp\cdot x_\perp}
\end{equation}
It annihilates a particle ina  $u_s(k)$ mode,  or creates  an antiparticle in a $v_s(k)$ mode, i.e. 
\bea
\psi(k)=\sum_s u_s(k)b_s(k)\theta(k^+)+v_s(-k)c_s^\dagger(-k)\theta(-k^+)\nonumber\\
\eea
The measure in momentum space is
\bea
[d^3k]_+=\frac{dk^+d^2k_\perp}{(2\pi)^32k^+}\epsilon(k^+)
\eea
which sums over the  positive $k^+$ region for  particle modes,  and over the negative $k^+$ region for antiparticle modes.
For the   t'Hooft interaction in the zero size limit, the interaction is generically of the form
\bea
 &&V(k,q,p,l)=\nonumber\\
 &&\sum_{s_1,s_1^\prime,s_2,s_2^\prime}\mathcal{V}_{s_1,s_2,s_1^\prime,s_2^\prime}(k,q,p,l)b^\dagger_{s_1}(k) c^\dagger_{s_2}(q) c_{s_2'}(p) b_{s_1^\prime}(l)\nonumber\\
\eea
with
\begin{widetext}
\begin{equation}
\begin{aligned}
    \mathcal{V}_{s_1,s_2,s_1',s_2'}(k,q,p,l)=&-g_S \alpha_+(k^++q^+)\bar{u}_{s_1}(k)v_{s_2}(q)\bar{v}_{s_2'}(p)u_{s_1'}(l)\\
    &+g_S \alpha_-(k^++q^+)\bar{u}_{s_1}(k)i\gamma^5v_{s_2}(q)\bar{v}_{s_2'}(p)i\gamma^5u_{s_1'}(l)\\
    &+g_S \alpha_-(k^++q^+)\bar{u}_{s_1}(k)\tau^ai\gamma^5v_{s_2}(q)\bar{v}_{s_2'}(p)\tau^ai\gamma^5u_{s_1'}(l)\\
    &-g_S \alpha_+(k^++q^+)\bar{u}_{s_1}(k)\tau^ai\gamma^5v_{s_2}(q)\bar{v}_{s_2'}(p)\tau^ai\gamma^5u_{s_1'}(l)
\end{aligned}
\end{equation}
\end{widetext}
In the large $N_c$ limit, only the s-channel contribution of the $^\prime$t Hooft interaction dominates. Further details
regarding the interplay of the s- and t-channel exchanges, can be found in  Appendix~\ref{APP_HLF}.

These arguments carry to the mean field Lagrangian (\ref{nonlocal_LFET}) through the substitutions
\begin{equation}
    V(k,q,p,l)\rightarrow \sqrt{\mathcal{F}(k)\mathcal{F}(k)\mathcal{F}(k)\mathcal{F}(l)}V(k,q,p,l)
\end{equation}
and 
\begin{equation}
    \alpha_\pm(P^+)\rightarrow \left[1\pm 2g_S\int\frac{dl^+d^2l_\perp}{(2\pi)^3}\frac{\epsilon(l^+)}{P^+-l^+}\mathcal{F}\left(l\right)\mathcal{F}\left(P-l\right)\right]^{-1}
\end{equation}
so that
\begin{widetext}
\begin{flalign}
\label{LFHamiltonian}
&P^-=\int [d^3k]_+\int [d^3q]_+\frac{k^2_\perp+M^2}{2k^+}\bar{\psi}(k)\gamma^+\psi(q)(2\pi)^3 \delta^3_+(k-q)\\[5pt] \nonumber
&+\int [d^3k]_+\int [d^3q]_+\int [d^3p]_+\int [d^3l]_+(2\pi)^3
    \delta^3_+(p+k-q-l)\sqrt{\mathcal{F}(k)\mathcal{F}(q)\mathcal{F}(p)\mathcal{F}(l)}V(k,q,p,l)
\end{flalign}
The interaction kernel  is now
\begin{equation}
\begin{aligned}
        V(k,q,p,l)=-\frac{G_S}{2}&\bigg[\alpha_+(k^+-q^+)\bar{\psi}(k)\psi(q)\bar{\psi}(p)\psi(l)-\alpha^-(k^+-q^+)\bar{\psi}(k)i\gamma^5\psi(q)\bar{\psi}(p)i\gamma^5\psi(l)\\
        &-\alpha^-(k^+-q^+)\bar{\psi}(k)\tau^a\psi(q)\bar{\psi}(p)\tau^a\psi(l)+\alpha^+(k^+-q^+)\bar{\psi}(k)i\tau^a\gamma^5\psi(q)\bar{\psi}(p)i\tau^a\gamma^5\psi(l)\bigg] \\
\end{aligned}
\end{equation}
where  the tadpole resummed vertices
$$\alpha_\pm(P^+)=\left[1\pm 2g_S\int\frac{dl^+d^2l_\perp}{(2\pi)^3}\frac{\epsilon(l^+)}{P^+-l^+}\mathcal{F}\left(l\right)\mathcal{F}\left(P-l\right)\right]^{-1}$$ 
\end{widetext}

Again, the astute reader may have noticed that for the Lagrangian (\ref{nonlocal_LFET}) in non-local form, the symmetric form of the
energy momentum tensor may require further amendment in the presence of the non-local form factors. This is not the case, as we
now explain. Indeed, boost invariance and parity suggests the substitution
 the substitution
\bea
\label{FFSUB}
    \lim_{P^+\rightarrow\infty}\sqrt{\mathcal{F}(k)\mathcal{F}(P-k)}\rightarrow \nonumber\\
    \mathcal{F}\left(\frac{2P^+P^-}{\lambda^2_S}=\frac{k^2_\perp+M^2}{\lambda^2_Sx\bar{x}}\right)
\eea
The same substitution was developed in the analysis of the ILM using  the large momentum 
effective theory(LaMET)~\cite{Kock:2020frx,Kock:2021spt}.
The substitution (\ref{FFSUB}) garentees the consistency of the two approaches. 
$\lambda_S$  is a parameter of order 1. Remarkably, the same boost invariant substitution with $\lambda_S=1$
was argued long ago by Lepage and Brodsky in~\cite{Lepage:1980fj}, in their analysis of 2-body bound states
on the light front using Bethe-Salpeter vertices. Note that the substitution (\ref{FFSUB}) eliminates the metric component
$g_{-+}$ from the non-local form factors, with consequently no change in the symmetric energy-momentum tensor 
component $T^{+-}$.

Finally, we note that it is straightforward to generalize the Hamiltonian formalism to consider meson bound states in the  $U$-spin or $V$-spin sectors relevant for kaons. In the case of kaons, the coupling constant will be replaced by $g_K=N_cG_K$,  with a constituent mass   matrix $M=\mathrm{diag}(M_u,M_d)$ in the flavor basis, as the $s$ quark is significantly heavier than 
$u,d$ quarks. This construction will be detailed below.

\begin{figure*}
\subfloat[\label{fig_3pt1}]{%
  \includegraphics[height=5.5cm,width=.46\linewidth]{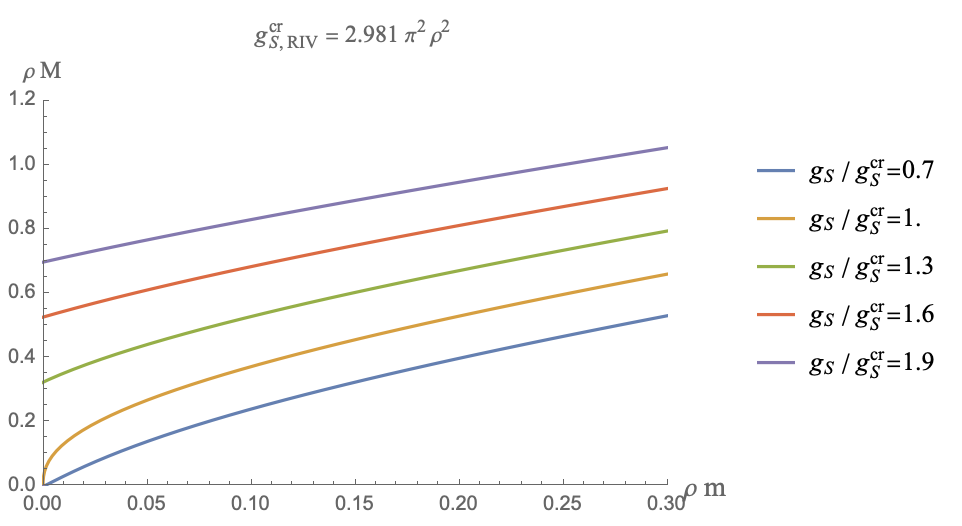}%
}\hfill
\subfloat[\label{fig_3pt6}]{%
  \includegraphics[height=5.5cm,width=.46\linewidth]{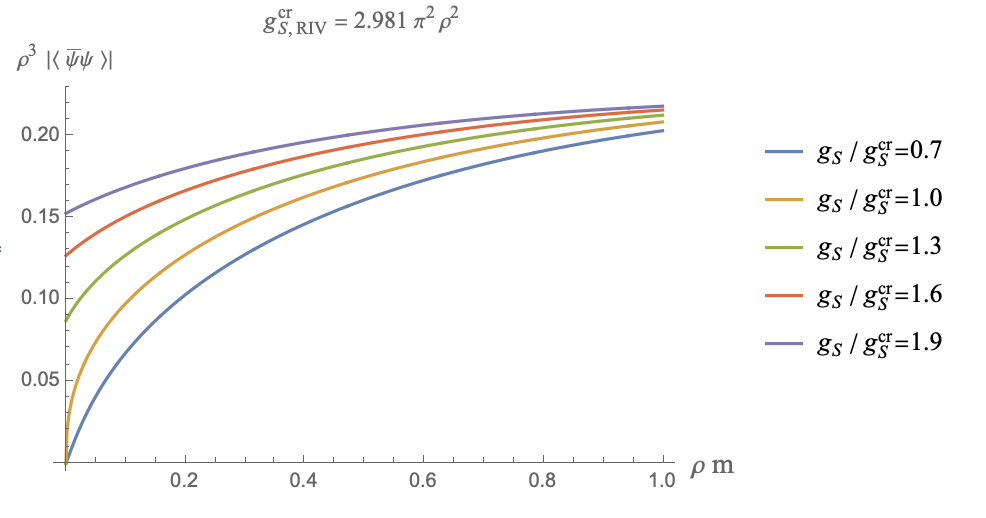}%
}
\caption{a: Quark constituent mass $M$ versus the current quark mass $m$
in the ILM,  for increasing strength of the $^\prime$t Hooft coupling $g_S$ (bottom-up),
with the critical coupling $g_{S,RIV}^{cr}=2.981\,\pi^2\rho^2$ and fixed instanton size $\rho$.
b: Quark condensate $\langle \bar \psi\psi\rangle$ versus the current quark mass in the ILM.}
\label{fig_Gap}
\end{figure*}

\section{Constituent mass and chiral condensate}
\label{SEC_ILMGAP}
The emergent  chiral condensate (\ref{CONDZ})  and constituent quark mass (\ref{MCONST}), are frame invariant scalars. 
Wether evaluated in the rest frame or on the light front, they should give the same results. We
now show that this is the case, for the resummed tadpoles, provided careful considerations are
given to the form factor arising from the zero mode in the ILM.

\subsection{Rest frame}
The emergent constituent quark mass 
in the rest frame, follows readily from (\ref{MCONST}) supplemented by (\ref{NLTH}),
 in the mean-field approximattion. In Euclidean signature, we have
\begin{equation}
\label{MKKX}
    M(k)=m+2g_S\mathcal{F}(k)\int\frac{d^4q}{(2\pi)^4}\frac{4M(q)}{q^2+M^2(q)}\mathcal{F}(q)
\end{equation}
where $g_S=G_S/N_c$ is the coupling strength for the $^\prime$t Hooft interaction.
At low momenta $k\rho\ll 1$, the  dynamical mass $M(k)$ is about constant $M=M(0)$. 
This is consistent with the 't-Hooft interaction in the zero instanton size limit.
At high momenta, the dynamical constituent mass asymptotes the  current mass $m$. 

The formal solution to (\ref{MKKX}) is
\begin{equation}
\label{MFK}
    M(k)=m\left[1-\mathcal{F}(k)\right]+M\mathcal{F}(k)\sim M\mathcal{F}(k)
\end{equation}
for which (\ref{MKKX}) turns to a gap equation for the constant part $M$
\begin{equation}
\label{MASSGAP}
    \frac{m}{M}=1-8g_S\int\frac{d^4k}{(2\pi)^4}\frac{\mathcal{F}^2(k)}{k^2+M^2}
\end{equation}
Solutions for $M$ only exists for a critical $g_S$, which in ILM is fixed by the
mean instanton-antiinstanton density. Following the arguments in~\cite{Kock:2021spt}, we have
dropped the k-dependence of the running mass in the denominator of (\ref{MASSGAP}), thanks to
the smallness of the packing fraction (\ref{KAPPA}).

The chiral condensate follows similarly
\bea
\label{PSIPSI}
    \langle\bar{\psi}\psi\rangle&=&-\int\frac{d^4k}{(2\pi)^4}\mathrm{Tr}S(k)\nonumber\\
    &=&-8N_cM\int\frac{d^4k}{(2\pi)^4} \frac{1}{k^2+M^2}\mathcal{F}(k)
\eea
It reduces the scalar mean field expectation $\langle \sigma\rangle=\langle\overline\psi\psi\rangle/N_c$ only in the zero instanton
size limit. In the latter, (\ref{MASSGAP}) and (\ref{PSIPSI}) are seen to diverge logarithmically, both in the IR and UV limits. So the quark zero mode 
induced form factor ${\mathcal F}(k)$ is key for a finite result.

In Fig.~\ref{fig_Gap}a we show the dependence of the constituent mass $M$ on the current mass, for different multi-fermion couplings $g_S$  in the
ILM of the QCD vacuum. In Fig.~\ref{fig_Gap}b we show the scalar quark condensate dependence on these parameters.

\subsection{Light front frame}
On the light front only the physical modes of  the good  fermionic component $\psi_+$ are present, after removal of the bad component
in the mean field approximation. The running quark mass on the light front is $M(k^-)\sim M{\mathcal F}(k^-)$ with the on-shell condition 
$2k^-k^+=k_\perp^2+M^2$, where  $M$ is fixed by the scalar gap equation  (\ref{MASSGAP}). In terms of the physical modes for $\psi_+$,
(\ref{MASSGAP}) can be re-written as
\begin{equation}
\label{gap_eq}
    \frac{m}{M}=1-2g_S\int \frac{dk^+d^2k_\perp}{(2\pi)^3}\frac{\epsilon(k^+)}{k^+}\mathcal{F}^2(k)\bigg|_{k^-=\frac{k_\perp^2+M^2}{2k^+}}
\end{equation}
assuming that ${\mathcal F}(k^2)$ is free of physical poles. This is the case of (\ref{FKK}) after analytical continuation, except for spurious
branch points, which contributions can be disregarded in leading order in the diluteness factor~(\ref{KAPPA}) as detailed in~\cite{Kock:2021spt}.

For the chiral condensate on the light front, the quark propagator for the effective light front effective field (\ref{PSINEW}), with the bad
component fully re-expressed in terms of the good component, reads
\begin{widetext}
\bea
\label{eq:prop}
    S(k)\rightarrow\left[\frac{i[\slashed{k}+M(k^2)]}{k^2-M(k^2)^2}-\frac{i\gamma^+}{2k^+}\right]
    \rightarrow \left[\frac{i[\slashed{k}+M(k^2)]}{k^2-M^2}-\frac{i\gamma^+}{2k^+}\right]
\eea
\end{widetext}
The quark condensate in the light front signature is then
\begin{equation}
\label{cond_eq}
    \langle\bar{\psi}\psi\rangle=-2N_cM\int\frac{dk^+d^2k_\perp}{(2\pi)^3}\frac{\epsilon(k^+)}{k^+}\mathcal{F}(k^-)
\end{equation}

More specifically, the gap equation with the ILM induced form factor is
\begin{widetext}
\bea
    \frac{m}{M}
    =1-\frac{4g_S}{\pi^2\rho^2}\int_0^{\infty} dz z\frac{z^3}{z^2+\frac{\rho^2M^2}{4}}|z(I_0(z)K_0(z)-I_1(z)K_1(z))'|^4
\eea
In the chiral limit, the constituent mass is nonzero only when the 't Hooft coupling is stronger than the critical coupling $g^{\mathrm{cr}}_{S,\mathrm{RIV}}$, which is defined as  
\bea
    g^{\mathrm{cr}}_{S,\mathrm{RIV}}=2\pi^2\rho^2\left[8\int_0^{\infty} dz z |z(I_0(z)K_0(z)-I_1(z)K_1(z))'|^4\right]^{-1}\approx2.981\pi^2\rho^2
\eea
 Similarly, the quark condensate is explicitly given by
\begin{equation}
    \rho^3\langle\bar{\psi}\psi\rangle=-\frac{4N_c}{\pi^2}\rho M\int_0^\infty dz \frac{z^3}{z^2+\frac{\rho^2M^2}{4}}|z(I_0(z)K_0(z)-I_1(z)K_1(z))'|^2
\end{equation}
\end{widetext}
In the small size expansion, apart from the quadratic dependence of $\rho$ in  leading order, the next-to-leading-order dependence in  $\rho M$ is dominated by polynomials.
In the ILM, a slightly stronger coupling $g_{S,\mathrm{RIV}}^\mathrm{cr}=2.981\pi^2\rho^2$ (instanton density) 
 is required to produce a quark condensate that breaks chiral symmetry spontaneously. 

For comparison, we will also quote the results in the zero instanton size limit for which ${\cal F}\rightarrow 1$.
This case necessitates the introduction of a sharp cuoff.  For vacuum loops in the 0-body sector, we will always
enforce the on-shell condition $k^2=2k^+k^--k_\perp^2$, and use the the boost invariant and parity even cutoffs 
$$|\sqrt{2}k^\pm|\leq \Lambda\sim 1/\rho$$
More discussions on this point can be found in Appendix~\ref{APP_SHARP0}. In the 2-body sector and higher,
boost invariance requires a different cutoff as we detail in Appendix~\ref{APP_SHARP2}.

 \section{Light front spectrum}
 \label{SEC_ILMLFS}
Now, we can use the light front Hamiltonian (\ref{LFHamiltonian}) to solve the eigenvalue equation for the meson wave functions,
\begin{equation}
\label{PMINUS}
    P^-|X,P\rangle=\frac{m_X^2}{2P^+}|X,P\rangle
\end{equation}
In the valence approximation,which is compatible with the mean-field analysis in leading order in $1/N_c$,
the mesons  state is dominated by valence constituent quark dynamics, 
\begin{widetext}
\begin{equation}
\label{Meson_bound_state}
    |\mathrm{Meson} ~X,P\rangle=\frac{1}{\sqrt{N_c}}\int_0^1 \frac{dx}{\sqrt{2x\bar{x}}}\int\frac{d^2k_\perp}{(2\pi)^3}\sum_{s_1,s_2}\Phi_X(x,k_\perp,s_1,s_2)b^\dagger_{s_1}(k) c^\dagger_{s_2}(P-k)|0\rangle
\end{equation}
The wavefunction is normalized by $\langle P|P'\rangle=(2\pi)^32P^+\delta^{3}(P-P')$
 
\begin{equation}
\label{normal}
    \int_0^1dx\int\frac{d^2k_\perp}{(2\pi)^3}\sum_{s_1,s_2}\left|\Phi_X(x,k_\perp,s_1,s_2)\right|^2=1
\end{equation}

\subsection{Boost invariant bound state equations}
for two light flavors  $(u,d)$ with equal current masses, we expect light scalar and pseudoscalar mesons
 $\sigma$, $\sigma_5$, $\pi^{\pm,0}$ and $\pi_5^{\pm,0}$.  A heavy scalar $\sigma$ meson, commensurate
 with the order parameter of the spontaneous breaking of $SU(2)_V\times SU(2)_A$ to $SU(2)_V$. 
 The broken $SU(2)_A$ symmetry generates near massless Goldstone modes $\pi^{\pm,0}$, with
 the valence quark assignments
\bea
    \sigma=\frac{1}{\sqrt{2}}(u\bar{u}+d\bar{d})\qquad 
    \pi^{\pm,0}=u\bar{d},\ d\bar{u},\ \frac{1}{\sqrt{2}}(u\bar{u}-d\bar{d})
\eea
Both of these channels are strongly attractive in the ILM. In contrast, the ILM interaction in 
the  pseudoscalar meson $\sigma_5$ ($\eta^\prime$), and flavor non-singlet scalar partners $\pi^{\pm,0}_5$ 
is strongly repulsive. They are unbound both in the rest frame and on the light front. Irrespective of binding, the generic light front
vertices $\Phi_X$ in (\ref{Meson_bound_state}) for all these mesonic channels are
\bea
\label{PHIX}
    \Phi_\sigma(x,k_\perp,s_1,s_2)&=&\phi_\sigma(x,k_\perp)\bar{u}_{s_1}(k)\frac{1}{\sqrt{2}}v_{s_2}(P-k)\nonumber\\
    \Phi_{\sigma_5}(x,k_\perp,s_1,s_2)&=&\phi_{\sigma_5}(x,k_\perp)\bar{u}_{s_1}(k)\frac{1}{\sqrt{2}}i\gamma^5v_{s_2}(P-k)\nonumber\\
    \Phi_\pi^a(x,k_\perp,s_1,s_2)&=&\phi_\pi(x,k_\perp)\bar{u}_{s_1}(k)\frac{\tau^a}{\sqrt{2}}i\gamma^5v_{s_2}(P-k)\nonumber\\
    \Phi^a_{\pi_5}(x,k_\perp,s_1,s_2)&=&\phi_{\pi_5}(x,k_\perp)\bar{u}_{s_1}(k)\frac{\tau^a}{\sqrt{2}}v_{s_2}(P-k)
\eea
For convenience the spin-flavor part is made explicit using the light front spinors $u_s(k)$ for a particle spinor,
and $v_s(k)$ for an anti-particle spinor, with the remaining $\phi_X$ a scalar-isoscalar wavefunction. The spinors are explicity given in Appendix \ref{Appx:LFspinor}.

Inserting (\ref{PHIX}) in (\ref{Meson_bound_state}) and then in (\ref{PMINUS}), and unwinding the various contractions from the
4-Fermi interaction terms, yield the boost invariant eigenvalue equation
\begin{equation}
\label{eq:bound_state}
\begin{aligned}
        &m_X^2\Phi_X(x,k_\perp,s_1,s_2)=\frac{k_\perp^2+M^2}{x\bar{x}}\Phi_X(x,k_\perp,s_1,s_2)\\
        &+\frac{1}{\sqrt{2x\bar{x}}}\sqrt{\mathcal{F}(k)\mathcal{F}(P-k)}\int_0^1 \frac{dy}{\sqrt{2y\bar{y}}}\int\frac{d^2q_\perp}{(2\pi)^3}\sum_{s,s'}\mathcal{V}_{s_1,s_2,s,s'}(k,P-k,q,P-q)\Phi_X(y,q_\perp,s,s')\sqrt{\mathcal{F}(q)\mathcal{F}(P-q)}
\end{aligned}
\end{equation}
with the ILM interaction kernel $\mathcal{V}_{s_1,s_2,s_1',s_2'}$ 
\begin{equation}
\begin{aligned}
    \mathcal{V}_{s,s',s_1,s_2}(q,q',k,k')=&    -G_S\bigg[\alpha_+(P^+)\bar{u}_{s_1}(k)v_{s_2}(k')\bar{v}_{s'}(q')u_{s}(q)-\alpha_-(P^+)\bar{u}_{s_1}(k)i\gamma^5v_{s_2}(k')\bar{v}_{s'}(q')i\gamma^5u_{s}(q)\\
    &+\alpha_+(P^+)\bar{u}_{s_1}(k)\tau^ai\gamma^5v_{s_2}(k')\bar{v}_{s'}(q')\tau^ai\gamma^5u_{s}(q)-\alpha_-(P^+)\bar{u}_{s_1}(k)\tau^av_{s_2}(k')\bar{v}_{s'}(q')\tau^au_{s}(q)\bigg]\\
\end{aligned}
\end{equation} 
Channel by channel, the explicit form of the kernel is
\begin{equation}
\begin{aligned}
        &\sum_{s,s'}\mathcal{V}_{s,s',s_1,s_2}(q,q',k,k')\Phi_{\sigma}(y,q_\perp,s,s')\\
        =&-g_S\alpha_+(P^+)\mathrm{Tr}\left[(\slashed{q}+M)(\slashed{q'}-M)\right]\phi_{\sigma}(y,q_\perp)\bar{u}_{s_1}(k)\mathbf{1}\mathrm{tr}(\mathbf{1})v_{s_2}(k')\\
        =&-4g_S\alpha_+(P^+)\left(\frac{q_\perp^2+(y-\bar{y})^2M^2}{y\bar{y}}\right)\phi_{\sigma}(y,q_\perp)\bar{u}_{s_1}(k)v_{s_2}(k')
\end{aligned}
\end{equation}
\begin{equation}
\begin{aligned}
    &\sum_{s,s'}\mathcal{V}_{s,s',s_1,s_2}(q,q',k,k')\Phi_{\sigma_5}(y,q_\perp,s,s')\\
    =&g_S\alpha_-(P^+)\mathrm{Tr}\left[(\slashed{q}+M)(\slashed{q'}+M)\right]\phi_{\sigma_5}(y,q_\perp)\bar{u}_{s_1}(k)i\gamma^5\mathbf{1}\mathrm{tr}(\mathbf{1})v_{s_2}(k')\\
    =&4g_S\alpha_-(P^+)\left(\frac{q_\perp^2+M^2}{y\bar{y}}\right)\phi_{\sigma_5}(y,q_\perp)\bar{u}_{s_1}(k)i\gamma^5v_{s_2}(k')
\end{aligned}
\end{equation}
\begin{equation}
\begin{aligned}
        &\sum_{s,s'}\mathcal{V}_{s,s',s_1,s_2}(q,q',k,k')\Phi^a_{\pi_5}(y,q_\perp,s,s')\\
        =&g_S\alpha_-(P^+)\mathrm{Tr}\left[(\slashed{q}+M)(\slashed{q'}-M)\right]\phi_{\pi_5}(y,q_\perp)\bar{u}_{s_1}(k)\tau^b\mathrm{tr}(\tau^a\tau^b)v_{s_2}(k')\\
        =&4g_S\alpha_-(P^+)\left(\frac{q_\perp^2+(y-\bar{y})^2M^2}{y\bar{y}}\right)\phi_{\pi_5}(y,q_\perp)\bar{u}_{s_1}(k)\tau^av_{s_2}(k')
\end{aligned}
\end{equation}
\begin{equation}
\begin{aligned}
    &\sum_{s,s'}\mathcal{V}_{s,s',s_1,s_2}(q,q',k,k')\Phi^a_{\pi}(y,q_\perp,s,s')\\
    =&-g_S\alpha_+(P^+)\mathrm{Tr}\left[(\slashed{q}+M)(\slashed{q'}+M)\right]\phi_{\pi}(y,q_\perp)\bar{u}_{s_1}(k)i\gamma^5\tau^b\mathrm{tr}(\tau^a\tau^b)v_{s_2}(k')\\
    =&-4g_S\alpha_+(P^+)\left(\frac{q_\perp^2+M^2}{y\bar{y}}\right)\phi_{\pi}(y,q_\perp)\bar{u}_{s_1}(k)i\gamma^5\tau^av_{s_2}(k')
\end{aligned}
\end{equation}
The corresponding eigenvalue equations for the scalar-isoscalar wavefunctions $\phi_X$ are
\begin{equation}
\label{1X}
\begin{aligned}
        m_\sigma^2\phi_{\sigma}(x,k_\perp)=&\frac{k^2_\perp+M^2}{x\bar{x}}\phi_{\sigma}(x,k_\perp)\\
        &-\frac{2g_S\alpha_+(P^+)}{\sqrt{x\bar{x}}}\sqrt{\mathcal{F}(k)\mathcal{F}(P-k)}\int \frac{dy}{\sqrt{y\bar{y}}} \int\frac{d^2q_\perp}{(2\pi)^3}\left(\frac{q_\perp^2+(y-\bar{y})^2M^2}{y\bar{y}}\right)\phi_\sigma(y,q_\perp)\sqrt{\mathcal{F}(q)\mathcal{F}(P-q)}
\end{aligned}
\end{equation}
\begin{equation}
\label{2X}
\begin{aligned}
        m_{\sigma_5}^2\phi_{\sigma_5}(x,k_\perp)=&\frac{k^2_\perp+M^2}{x\bar{x}}\phi_{\sigma_5}(x,k_\perp)\\
        &+\frac{2g_S\alpha_-(P^+)}{\sqrt{x\bar{x}}}\sqrt{\mathcal{F}(k)\mathcal{F}(P-k)}\int \frac{dy}{\sqrt{y\bar{y}}} \int\frac{d^2q_\perp}{(2\pi)^3} \left(\frac{q_\perp^2+M^2}{y\bar{y}}\right)\phi_{\sigma_5}(y,q_\perp)\sqrt{\mathcal{F}(q)\mathcal{F}(P-q)}
\end{aligned}
\end{equation}
\begin{equation}
\label{3X}
\begin{aligned}
        m_{\pi_5}^2\phi_{\pi_5}(x,k_\perp)=&\frac{k^2_\perp+M^2}{x\bar{x}}\phi_{\pi_5}(x,k_\perp)\\
    &+\frac{2g_S\alpha_-(P^+)}{\sqrt{x\bar{x}}}\sqrt{\mathcal{F}(k)\mathcal{F}(P-k)}\int \frac{dy}{\sqrt{y\bar{y}}} \int\frac{d^2q_\perp}{(2\pi)^3}\left(\frac{q_\perp^2+(y-\bar{y})^2M^2}{y\bar{y}}\right)\phi_{\pi_5}(y,q_\perp)\sqrt{\mathcal{F}(q)\mathcal{F}(P-q)}
\end{aligned}
\end{equation}
\begin{equation}
\label{4X}
\begin{aligned}
        m_{\pi}^2\phi_{\pi}(x,k_\perp)=&\frac{k^2_\perp+M^2}{x\bar{x}}\phi_{\pi}(x,k_\perp)\\
        &-\frac{2g_S\alpha_+(P^+)}{\sqrt{x\bar{x}}}\sqrt{\mathcal{F}(k)\mathcal{F}(P-k)}\int \frac{dy}{\sqrt{y\bar{y}}} \int\frac{d^2q_\perp}{(2\pi)^3} \left(\frac{q_\perp^2+M^2}{y\bar{y}}\right)\phi_{\pi}(y,q_\perp)\sqrt{\mathcal{F}(q)\mathcal{F}(P-q)}
\end{aligned}
\end{equation}
\end{widetext}

\subsection{Meson masses}
The generic solution to the boost invariant equations (\ref{1X}-\ref{4X}) is
\begin{equation}
    1=\int_0^1dy\int d^2q_\perp\frac{V_X(y,q_\perp)}{y\bar{y}m_X^2-(q_\perp^2+M^2)}\mathcal{F}(q)\mathcal{F}(P-q)
\end{equation}
where for each of the mesonic channel we have for $V_X$ in the numerator
\begin{widetext}
\bea
    V_X=\begin{cases}
    \mp \frac{2g_S}{(2\pi)^3}\alpha_\pm(P^+)\left(\frac{q_\perp^2+(y-\bar{y})^2M^2}{y\bar{y}}\right)\ ,~\ \text{scalars}~\ \sigma, \pi_5\\
    \mp \frac{2g_S}{(2\pi)^3}\alpha_\pm(P^+)\left(\frac{q_\perp^2+M^2}{y\bar{y}}\right)\ ,~\ \text{pseudo scalars}~\ \pi, \sigma_5
    \end{cases}
\eea
\end{widetext}
For the scalar $\sigma$ and pseudoscalar $\pi$ channel, the potential is negative and a bound solution exists, while for $\sigma_5$ and $\pi_5$, the potential is positive
and a solution is ruled out in the mean-field approximation.

\begin{figure}
    \centering
     \includegraphics[scale=0.6]{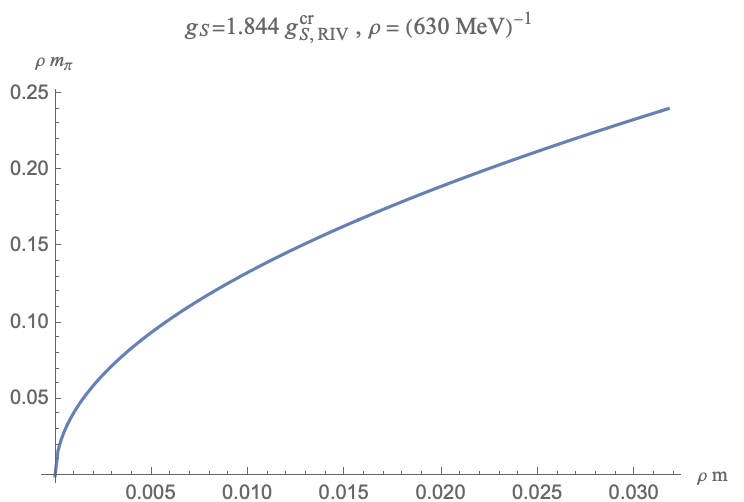}
     \caption{ Change of the pion mass with the current quark mass $m$, for a fixed fermionic coupling $g_S=1.844 g_{S,RIV}^{cr}$
     and instanton size $\rho=(630\,{\rm MeV})^{-1}$.}
     \label{Pion_mass}
\end{figure}

To deal with the coupling renormalization $\alpha(P^+)$ induced by the bad component through the constraint equation, we separate the  $k^+$-integral into an integral in the physical range $0<k^+<P^+$ where the quark momentum can be associated with the momentum fraction $x$ in the bound state,  plus an integral outside the physical range $k^+<0$ or $k^+>P^+$. The latter
contribution can be identified with a similar contribution in the mass gap equation
\begin{widetext}
\begin{equation}
\label{5X}
\begin{aligned}
        \alpha_\pm(P^+)^{-1}=&1\pm2g_S\int\frac{dk^+d^2k_\perp}{(2\pi)^3}\frac{\epsilon(k^+)}{P^+-k^+}\mathcal{F}(k)\mathcal{F}(P-k)\\
        =&1\pm\frac{2g_S}{(2\pi)^3}\int d^2k_\perp\left[\int_0^\infty dk^+\frac{1}{P^+-k^+}\mathcal{F}(k)\mathcal{F}(P-k)-\int_{-\infty}^0 dk^+\frac{1}{P^+-k^+}\mathcal{F}(k)\mathcal{F}(P-k)\right]\\
        =&1\pm\frac{2g_S}{(2\pi)^3}\int d^2k_\perp\left(\int_0^{P^+} dk^+\frac{2}{k^+}\mathcal{F}(k)\mathcal{F}(P-k)-\int_{-\infty}^\infty dk^+\frac{\epsilon(k^+)}{k^+}\mathcal{F}^2(k)\right)\\
\end{aligned}
\end{equation}
\end{widetext}
When the integration runs outside of the physical range of the 2-body bound state, the virtual quark momentum will overtake the  two-body bound light front mmentum $P^+$ logarithmically. For fixed $P^+$ but large,  the difference between $\mathcal{F}(k)$ and $\mathcal{F}(P-k)$ is then negligible. Hence, in the second part of the integral in (\ref{5X}) where $k^+$ runs outside $0<k^+<P^+$, we have replaced $\mathcal{F}(k)\mathcal{F}(P-k)$ by $\mathcal{F}^2(k)$, which gives a contribution identical to that in the mass gap equation (\ref{gap_eq}). With this in mind, (\ref{5X}) simplifies 
\begin{widetext}
\begin{equation}
\begin{aligned}
    \alpha_\pm(P^+)^{-1}=&\begin{cases}
    \frac{m}{M}+\frac{4g_S}{(2\pi)^3}\int d^2k_\perp\int_0^1 dx\frac{1}{x}\mathcal{F}(k)\mathcal{F}(P-k) \\
    2-\frac{m}{M}-\frac{4g_S}{(2\pi)^3}\int d^2k_\perp\int_0^1 dx\frac{1}{x}\mathcal{F}(k)\mathcal{F}(P-k)
    \end{cases}
\end{aligned}
\end{equation}
where $x=k^+/P^+$ is the momentum fraction of the quark inside the bound state.
\\
\\
{\bf Scalars:}
\\
For the scalar type particle $\sigma$ and $\pi^a_5$, the mass eigenvalue equations are
\label{SCALARX1}
\begin{equation}
\begin{aligned}
           0=&\alpha_\pm(P^+)^{-1}\pm\frac{2g_S}{(2\pi)^3}\int dy\int d^2q_\perp\frac{\left(q_\perp^2+(y-\bar{y})^2M^2\right)/y\bar{y}}{y\bar{y}m_X^2-(q_\perp^2+M^2)}\mathcal{F}(q)\mathcal{F}(P-q)\\
           =&\begin{cases}
    \frac{m}{M}+\frac{2g_S}{(2\pi)^3}\int_0^1dy\int d^2q_\perp\left[\frac{2}{y}-\frac{1}{y\bar{y}}+\frac{m_\sigma^2-(1-(y-\bar{y})^2)M^2/y\bar{y}}{y\bar{y}m_X^2-(q_\perp^2+M^2)}\right]\mathcal{F}(q)\mathcal{F}(P-q)\\[7pt]
    2-\frac{m}{M}-\frac{2g_S}{(2\pi)^3}\int_0^1dy\int d^2q_\perp\left[\frac{2}{y}-\frac{1}{y\bar{y}}+\frac{m_{\pi_5}^2-(1-(y-\bar{y})^2)M^2/y\bar{y}}{y\bar{y}m_X^2-(q_\perp^2+M^2)}\right]\mathcal{F}(q)\mathcal{F}(P-q)
    \end{cases}\\
    =&\begin{cases}
    \frac{m}{M}+\frac{2g_S}{(2\pi)^3}\int_0^1dy\int d^2q_\perp\left[\frac{m_\sigma^2-4M^2}{y\bar{y}m_X^2-(q_\perp^2+M^2)}\right]\mathcal{F}(q)\mathcal{F}(P-q)\ , & \sigma~\mathrm{meson}\\[7pt]
    2-\frac{m}{M}-\frac{2g_S}{(2\pi)^3}\int_0^1dy\int d^2q_\perp\left[\frac{m_{\pi_5}^2-4M^2}{y\bar{y}m_X^2-(q_\perp^2+M^2)}\right]\mathcal{F}(q)\mathcal{F}(P-q)\ , & \pi_5~\mathrm{meson}
    \end{cases}
\end{aligned}
\end{equation}
The boost-invariant property of the form factor guarantees the cancellation between the integral $\int_0^1dy\frac{2}{y}$ and $\int_0^1dy \frac{1}{y\bar{y}}$.
\\
\\
{\bf Pseudoscalars:}
\\
For the pseudo scalar type particle $\sigma_5$ and $\pi^a$, the mass eigenvalue equations are

\begin{equation}
\label{PSEUDOSCALARX2}
\begin{aligned}
           0=&\alpha_\pm(P^+)^{-1}\pm\frac{2g_S}{(2\pi)^3}\int_0^1dy\int d^2q_\perp\frac{\left(q_\perp^2+M^2\right)/y\bar{y}}{y\bar{y}m_X^2-(q_\perp^2+M^2)}\mathcal{F}(q)\mathcal{F}(P-q)\\
           =&\begin{cases}
2-\frac{m}{M}-\frac{2g_S}{(2\pi)^3}\int_0^1dy\int d^2q_\perp\left[\frac{2}{y}-\frac{1}{y\bar{y}}+\frac{m_{\sigma_5}^2}{y\bar{y}m_{\sigma_5}^2-(q_\perp^2+M^2)}\right]\mathcal{F}(q)\mathcal{F}(P-q)\\[7pt]
\frac{m}{M}+\frac{2g_S}{(2\pi)^3}\int_0^1dy\int d^2q_\perp\left[\frac{2}{y}-\frac{1}{y\bar{y}}+\frac{m_\pi^2}{y\bar{y}m_\pi^2-(q_\perp^2+M^2)}\right]\mathcal{F}(q)\mathcal{F}(P-q)
\end{cases}\\
=&\begin{cases}
2-\frac{m}{M}-\frac{2g_S}{(2\pi)^3}\int_0^1dy\int d^2q_\perp\left[\frac{m_{\sigma_5}^2}{y\bar{y}m_{\sigma_5}^2-(q_\perp^2+M^2)}\right]\mathcal{F}(q)\mathcal{F}(P-q)\ , & \sigma_5~\mathrm{meson}\\[7pt]
\frac{m}{M}+\frac{2g_S}{(2\pi)^3}\int_0^1dy\int d^2q_\perp\left[\frac{m_\pi^2}{y\bar{y}m_\pi^2-(q_\perp^2+M^2)}\right]\mathcal{F}(q)\mathcal{F}(P-q)\ , & \pi~\mathrm{meson}
\end{cases}
\end{aligned}
\end{equation}
\\
\\
{\bf sigma meson:}
\\
With the transverse cut-off regularization from the ILM, the  mass eigenvalue of the $\sigma$ meson is
\begin{equation}
\begin{aligned}
\label{RIV_eig_eq}
    \frac{m}{M}=&-\frac{g_S}{4\pi^2}\int_{0}^{1}dy\int_0^{\infty} dq^2_\perp\left[\frac{m_\sigma^2-4M^2}{y\bar{y}m_\sigma^2-(q_\perp^2+M^2)}\right]\left[(zF'(z))^4\right]\bigg|_{z=\frac{\rho \sqrt{q_\perp^2+M^2}}{2\lambda_S\sqrt{y\bar{y}}}}\\
    =&-\frac{g_S}{2\pi^2}(4M^2-m_\sigma^2)\int_0^1dy\int_{\frac{\rho M}{2\lambda_S\sqrt{y\bar{y}}}}^\infty dz\frac{z}{z^2-\frac{\rho^2m_\sigma^2}{4\lambda_S^2}}\left[zF'\left(z\right)\right]^4~,\ m_\sigma^2<4M^2\\
\end{aligned}
\end{equation}
In the chiral limit $m=0$, fthe scalar-isoscalar $\sigma$ meson is a treshold state, with mass $m_\sigma=2M$. The same treshold mass has been observed for
the standard ILM in the rest frame, as it should. For $m\neq0$, since the integration in the second line of (\ref{RIV_eig_eq}) is always positive, the sigma meson unbinds in the ILM.
In the mean-field approximation and away from the chiral limit,  the $^\prime$t Hooft interaction in the ILM is not strong enough to bind
$\sigma=\bar{u}u+\bar{d}d$.
\\
\\
{\bf $\sigma_5$ and $\pi_5$ mesons:}
\\
The same observation applies to the 
scalar-isovector  $\pi_5$ and the pseudoscalar-isoscalar $\sigma_5$ channels. The eigenvalue equation for $\pi_5$ is
\begin{equation}
\begin{aligned}
    2-\frac{m}{M}=&\frac{g_S}{4\pi^2}\int_{0}^{1}dy\int_0^{\infty} dq^2_\perp\left[\frac{m_{\pi_5}^2-4M^2}{y\bar{y}m_{\pi_5}^2-(q_\perp^2+M^2)}\right]\left[(zF'(z))^4\right]\bigg|_{z=\frac{\rho \sqrt{q_\perp^2+M^2}}{2\lambda_S\sqrt{y\bar{y}}}}\\
    =&\frac{g_S}{2\pi^2}(4M^2-m_{\pi_5}^2)\int_0^1dy\int_{\frac{\rho M}{2\lambda_S\sqrt{y\bar{y}}}}^\infty dz\frac{z}{z^2-\frac{\rho^2m_{\pi_5}^2}{4\lambda_S^2}}\left[zF'\left(z\right)\right]^4~,\ m_{\pi_5}^2<4M^2\\
\end{aligned}
\end{equation}
and the eigenvalue equation for $\pi_5$ is
\begin{equation}
\begin{aligned}
    2-\frac{m}{M}=&\frac{g_S}{4\pi^2}\int_0^1dy\int_0^{\infty} dq^2_\perp\left[\frac{m_{\sigma_5}^2}{y\bar{y}m_{\sigma_5}^2-(q_\perp^2+M^2)}\right]\left[(zF'(z))^4\right]\bigg|_{z=\frac{\rho \sqrt{q_\perp^2+M^2}}{2\lambda_S\sqrt{y\bar{y}}}}\\
    =&-\frac{g_S}{2\pi^2}m_{\sigma_5}^2\int_0^1dy\int_{\frac{\rho M}{2\lambda_S\sqrt{y\bar{y}}}}^\infty dz\frac{z}{z^2-\frac{\rho^2m_{\sigma_5}^2}{4\lambda_S^2}}\left[zF'\left(z\right)\right]^4
\end{aligned}
\end{equation}
Regardless of the chiral limit, $\pi_5$ and $\sigma_5$ cannot be bound in the mean field approximation of the ILM on the light front.
\\
\\
{\bf pi meson:}
\\
The interaction induced by the ILM in the pion channel is very strong and attractive, leading to a triplet of massless Nambu-Goldstone
modes $\pi^{\pm, 0}$. The pion decay constant in the chiral limit is soley given by the zero mode form factor on the light front
\bea
\label{eq:fpi}
    f_\pi&=&\frac{\sqrt{N_c}M}{\sqrt{2}\pi}
  \times\left[\int_0^1dx\int_0^\infty dk_\perp^2\left(\frac{1}{k_\perp^2+M^2}\right)\mathcal{F}(k)\mathcal{F}(P-k)\right]^{1/2}\nonumber\\
  &=&\frac{\sqrt{N_c}M}{\pi}\left[\int_0^1 dx\int_\frac{\rho M}{2\lambda_S\sqrt{x\bar{x}}}^\infty dz z^3(F'(z))^4\right]^{1/2}\approx\frac{\sqrt{N_c}M}{\sqrt{2}\pi}\sqrt{\ln\frac{C}{\rho^2 M^2}}+\mathcal{O}(\rho)
\eea
In the small $\rho$ expansion, it can be defined as $$f^2_\pi=\frac{1}{2\pi^2}M^2N_c\ln\left(\frac{C}{\rho^2M^2}\right)$$ where $C\approx0.361$ is the constant determined numerically with the input parameter $\lambda_S=3.285$, in total agreement with the result obtained also in the ILM using the large momenetum effective theory~\cite{Kock:2020frx}.
The pion mass is a solution to the eigenvalue equation
\begin{equation}
\label{pion_mass_RIV}
\begin{aligned}
    \frac{m}{M}=&-\frac{g_S}{4\pi^2}\int_{0}^{1}dy\int_0^{\infty} dq^2_\perp\left[\frac{m_\pi^2}{y\bar{y}m_\pi^2-(q_\perp^2+M^2)}\right]\left[(zF'(z))^4\right]\bigg|_{z=\frac{\rho}{2\lambda_S}\sqrt{\frac{ q_\perp^2+M^2}{y\bar{y}}}}\\
    =&\frac{g_S}{2\pi^2}m_\pi^2\int_0^1dy\int_{\frac{\rho M}{2\lambda_S\sqrt{y\bar{y}}}}^\infty dz\frac{z}{z^2-\frac{\rho^2m_{\pi}^2}{4\lambda_S^2}}\left[zF'\left(z\right)\right]^4~,\ m_\pi^2<4M^2\\
\end{aligned}
\end{equation}
\end{widetext}
with clearly a massless pion in the chiral limit.
Away from the chiral limit and with a non-vanishing constituent quark mass $M$, the pion
mass following from (\ref{pion_mass_RIV}) can be assessed in perturbation theory, with the result
\begin{equation}
\label{eq:GOR}
    m^2_\pi=\frac{2m}{f^2_\pi}|\langle \bar{\psi}\psi\rangle| +\mathcal{O}(m^2)
\end{equation}
where the quark condensate $$|\langle \bar{\psi}\psi\rangle|=|\langle \bar{u}u\rangle+\langle \bar{d}d\rangle|=\frac{N_c}{g_S}(M-m)$$ is given  by the gap equation.
(\ref{eq:GOR}) is the expected Gell-Mann-Oakes-Renner relation for Nambu-Goldstone modes.
In Fig.~\ref{Pion_mass} we show the pion mass solution to (\ref{pion_mass_RIV}) in the ILM on the light front, as a function of the current quark mass $m$, for 
the fermionic coupling $g_S=1.844 g_{S,RIV}^{cr}$, and a fixed instanton size $\rho=(630\,{\rm MeV})^{-1}$.

The parameters $\rho, m, g_S, \lambda_S$ in the mean-field approximation of the ILM are fixed as follows:
the instanton size iis set at its mean canonical value $\rho=0.313$ fm; the light current quark mass is set to $m=16.0$ MeV; 
the multi-fermion coupling is set to $g_S=1.844~g_{S,\mathrm{RIV}}^{cr}$ to give a constituent quark mass
of $M=444.8$ MeV  and a chiral condensate $|\langle \bar{\psi}\psi\rangle|^{1/3}=337.9$ MeV; 
the cutoff parameter is set to $\lambda_S=3.285$ to give a pion decay constant
 $f^{\mathrm{RIV}}_\pi=130.3$ MeV and pion mass $m^{\mathrm{RIV}}_\pi=135$ MeV, very close
 to the empirical results. In Fig.~\ref{Pion_mass} we show the change of the pion mass with the current quark mass.
 We note the rapid vanishing of the mass in the chiral limit, for a non-vanishing constituent quark mass $M$ and
 a chiral condensate, as expected for a Goldstone mode.

\begin{figure*}
\subfloat[\label{fig_3pt1}]{%
  \includegraphics[height=5.5cm,width=.46\linewidth]{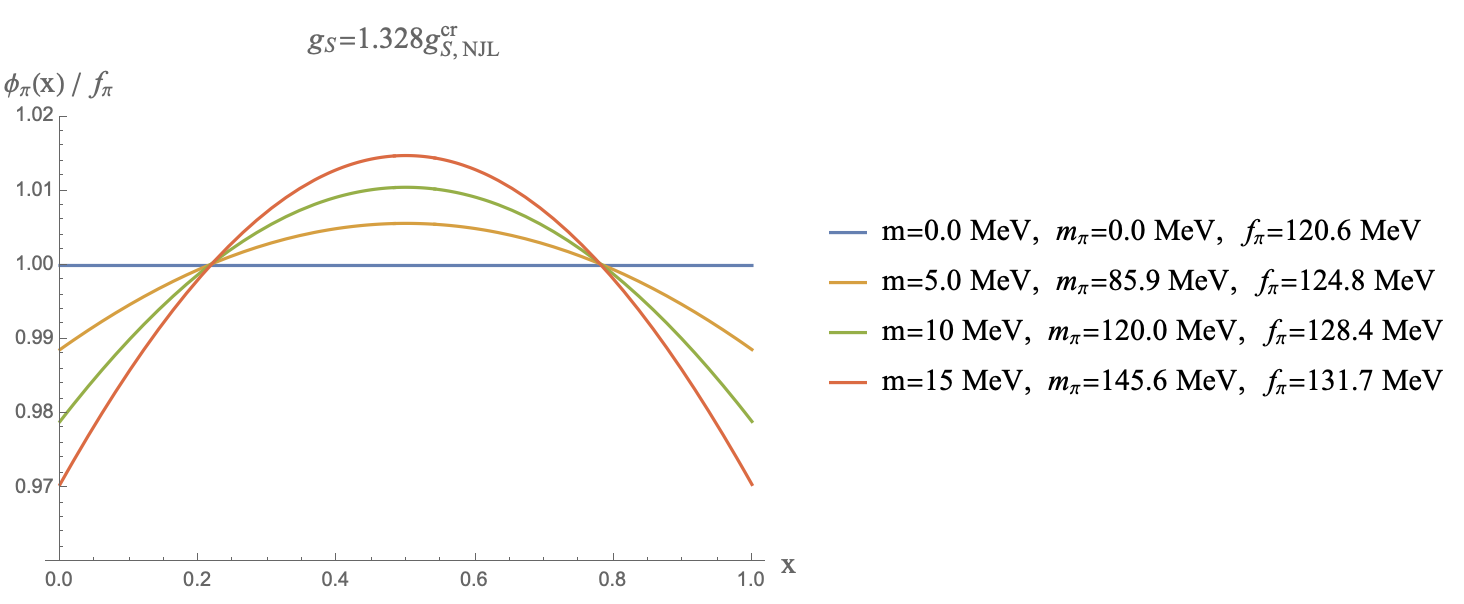}%
}\hfill
\subfloat[\label{fig_3pt6}]{%
  \includegraphics[height=5.5cm,width=.46\linewidth]{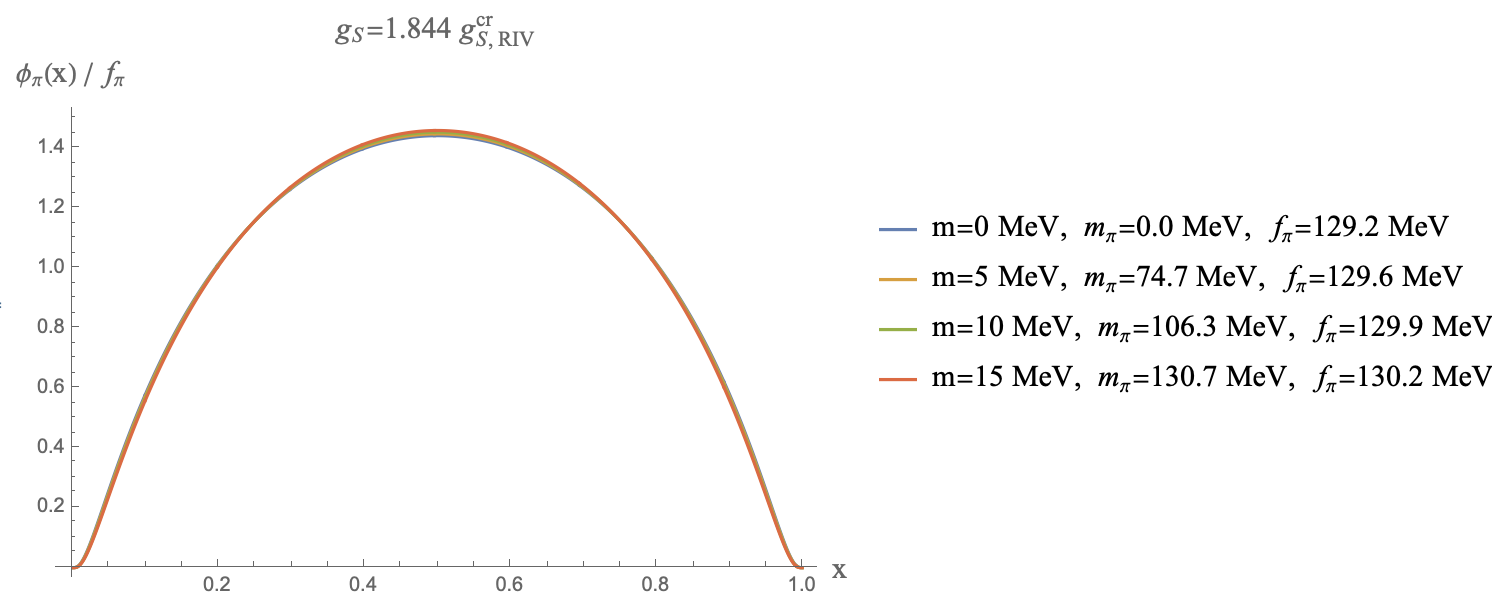}%
}\hfill
\subfloat[\label{fig_3pt6}]{%
  \includegraphics[height=5.5cm,width=.46\linewidth]{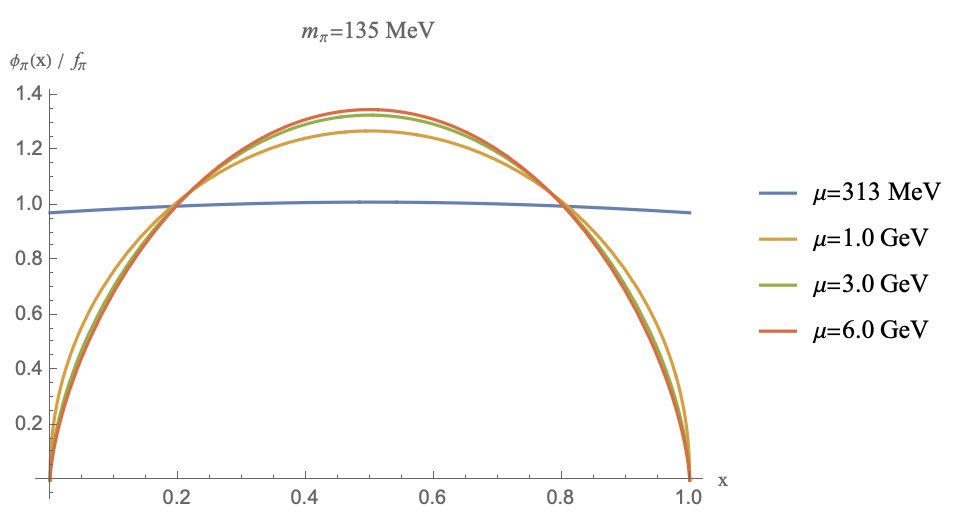}%
}\hfill
\subfloat[\label{fig_3pt6}]{%
  \includegraphics[height=5.5cm,width=.46\linewidth]{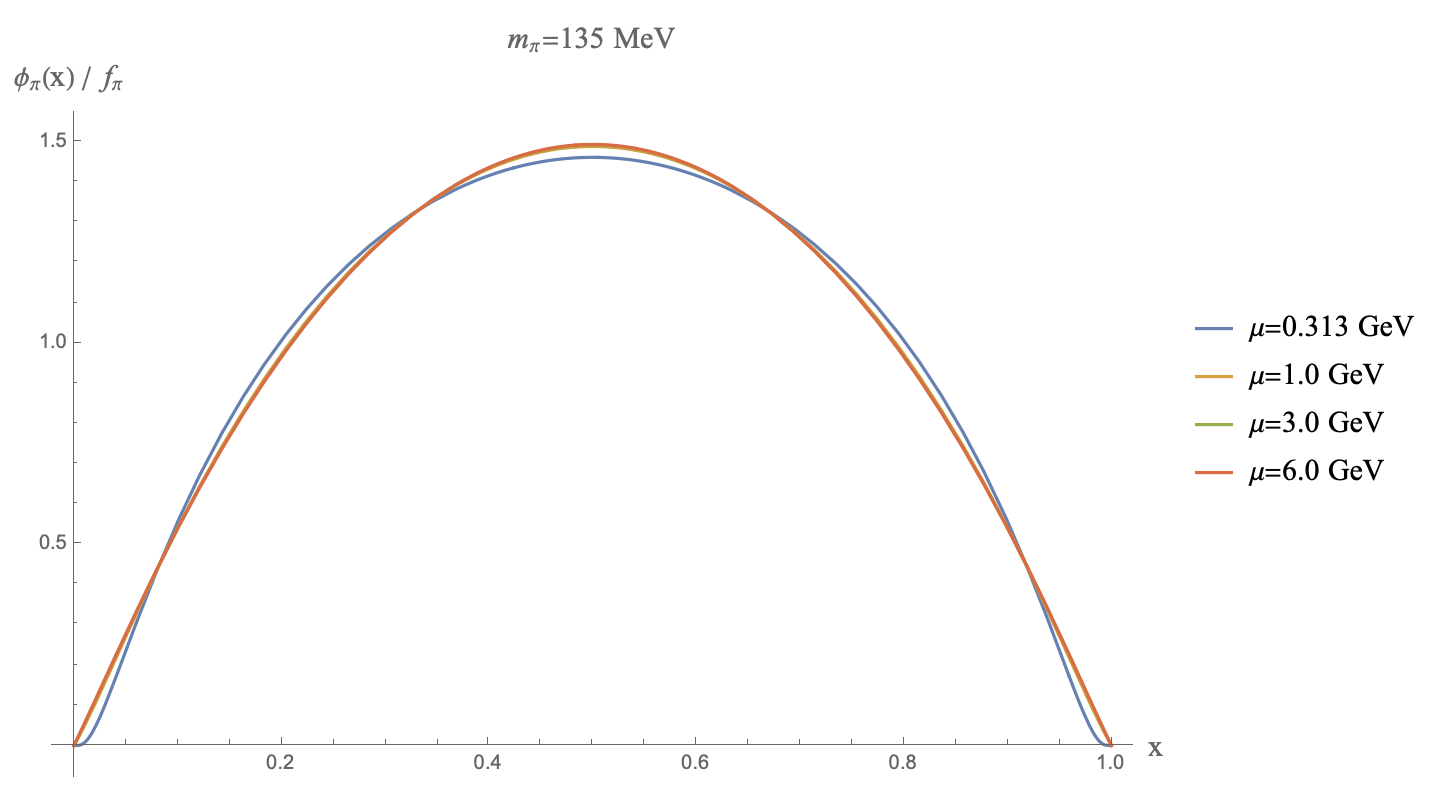}%
}
\caption{
a: The pion DA with different pion mass $m_\pi$ in the zero instanton size limit;
b: The pion DA with different pion mass $m_\pi$ in the ILM with finite instanton size $\rho=(630\mathrm{MeV})^{-1}$;
c: The ERBL evolution of pion DA in zero instanton size limit;
d: The ERBL evolution of pion DA in RIV with finite instanton size $\rho=(630\mathrm{MeV})^{-1}$.
}
\label{fig_pionERBL}
\end{figure*}

\subsection{Pion DA from LFWF}
In the mean-field approximation to the ILM, the generic solution for the valence light  front wavefunctions (LFWFs),
is generically of the form
\begin{widetext}
\begin{equation}
    \phi_{X}(x,k_\perp)=\frac{1}{\sqrt{2x\bar{x}}}\frac{C_{X}}{m^2_{X}-\frac{k_\perp^2+M^2}{x\bar{x}}}\sqrt{\mathcal{F}\left(k\right)\mathcal{F}\left(P-k\right)}
\end{equation}
with the constant $C_X$ fixed by the normalization (\ref{normal}). Alternatively, the LFWFs  can also be  deduced from the quark-meson interaction
amplitude (by integration over $k^-$) in which case  $C_X$ corresponds to the effective quark-meson coupling $g_X$ in the interaction, i.e.
$ C_X=-\sqrt{N_c}g_X$. The minus sign is chosen to insure a  positive-definite LFWF.

Since of all the light scalars and pseudoscalars made out of $u,d$ light quarks, only the pion is strongly bound in the mean field approximation
(the sigma meson is a treshold state), we now construct the pion DA using the pion LFWF. More specifically, using the pion DA as  defined through
 the forward matrix element of the twist-2 operator on the light cone, 
\begin{equation}
\label{PITWIST2}
\begin{aligned}
        \phi_{\pi}(x)=&-i\int\frac{d\xi^-}{2\pi}e^{ixP^+\xi^-}\langle0|\bar{\psi}(0)\gamma^+\gamma^5\frac{\tau^a}{\sqrt{2}}W(0,\xi^-)\psi(\xi^-)|\pi^a(P)\rangle
\end{aligned}
\end{equation}
with the normalization  fixed by the pion weak decay constant
\begin{equation}
\label{PIWEAK}
    \langle0|\bar{\psi}\gamma^+\gamma^5\frac{\tau^a}{\sqrt{2}}\psi|\pi^a(P)\rangle=if_\pi P^+
\end{equation}
we obtain
\begin{equation}
\begin{aligned}
        \phi_{\pi}(x)&=\frac{\sqrt{N_c}M}{2\sqrt{2}\pi^2}\int_0^{\infty} dk^2_\perp \frac{C_{\pi}}{x\bar{x}m^2_{\pi}-(k_\perp^2+M^2)}\mathcal{F}\left(k\right)\mathcal{F}\left(P-k\right)\\
\end{aligned}
\end{equation}
The on-shell normalization (\ref{PIWEAK}) fixes the dependence of the pion weak decay constant $f_\pi(m_\pi)$ on the pion mass $m_\pi$
\begin{equation}
    f_\pi(m_\pi)=f_\pi\frac{\int_0^1dx\int_0^\infty dk_\perp^2\left(\frac{1}{k_\perp^2+M^2-x\bar{x}m_\pi^2}\right)\mathcal{F}\left(k\right)\mathcal{F}\left(P-k\right)}{\left[\int_0^1dx\int_0^\infty dk_\perp^2\frac{k_\perp^2+M^2}{(k_\perp^2+M^2-x\bar{x}m_\pi^2)^2}\mathcal{F}\left(k\right)\mathcal{F}\left(P-k\right)\right]^{1/2}\left[\int_0^1dx\int_0^\infty dk_\perp^2\left(\frac{1}{k_\perp^2+M^2}\right)\mathcal{F}\left(k\right)\mathcal{F}\left(P-k\right)\right]^{1/2}}
\end{equation}
with $f_\pi$
given in (\ref{eq:fpi}).

In the zero instanton size approximation with a fixed transverse momentum cutoff $|k_\perp|<\Lambda$, the pion DA amplitude is a step function
\bea
        \phi_{\pi}(x)
        =-\frac{\sqrt{N_c}MC_{\pi}}{2\sqrt{2}\pi^2}\theta(x\bar{x})\ln\left(1+\frac{\Lambda^2}{M^2-x\bar{x}m_\pi^2}\right)
        \xrightarrow{m_\pi\rightarrow0}\frac{\sqrt{N_c}M}{\sqrt{2}\pi}\left[\ln\left(1+\frac{\Lambda^2}{M^2}\right)\right]^{1/2}\theta(x\bar{x})
\eea
in agreement with a result established first in the local NJL model~\cite{Broniowski:2017wbr}. The pion  weak decay constant is
\begin{align}
    f_\pi=-\frac{\sqrt{N_c}MC_{\pi}}{\sqrt{2}\pi^2}
    &\left[\frac{1}{2}\ln\left(1+\frac{\Lambda^2}{M^2}\right)-\sqrt{\frac{4M^2-m_\pi^2}{m_\pi^2}}\tan^{-1}\frac{1}{\sqrt{\frac{4M^2-m_\pi^2}{m_\pi^2}}}\right. \\
    &\left.+\sqrt{\frac{4(M^2+\Lambda^2)-m_\pi^2}{m_\pi^2}}\tan^{-1}\frac{1}{\sqrt{\frac{4(M^2+\Lambda^2)-m_\pi^2}{m_\pi^2}}}\right]
    \xrightarrow{m_\pi\rightarrow0}\frac{\sqrt{N_c}M}{\sqrt{2}\pi}\left[\ln\left(1+\frac{\Lambda^2}{M^2}\right)\right]^{1/2}   
\end{align}
\end{widetext}

\begin{figure*}
\subfloat[\label{fig_3pt1}]{%
  \includegraphics[height=5.5cm,width=.46\linewidth]{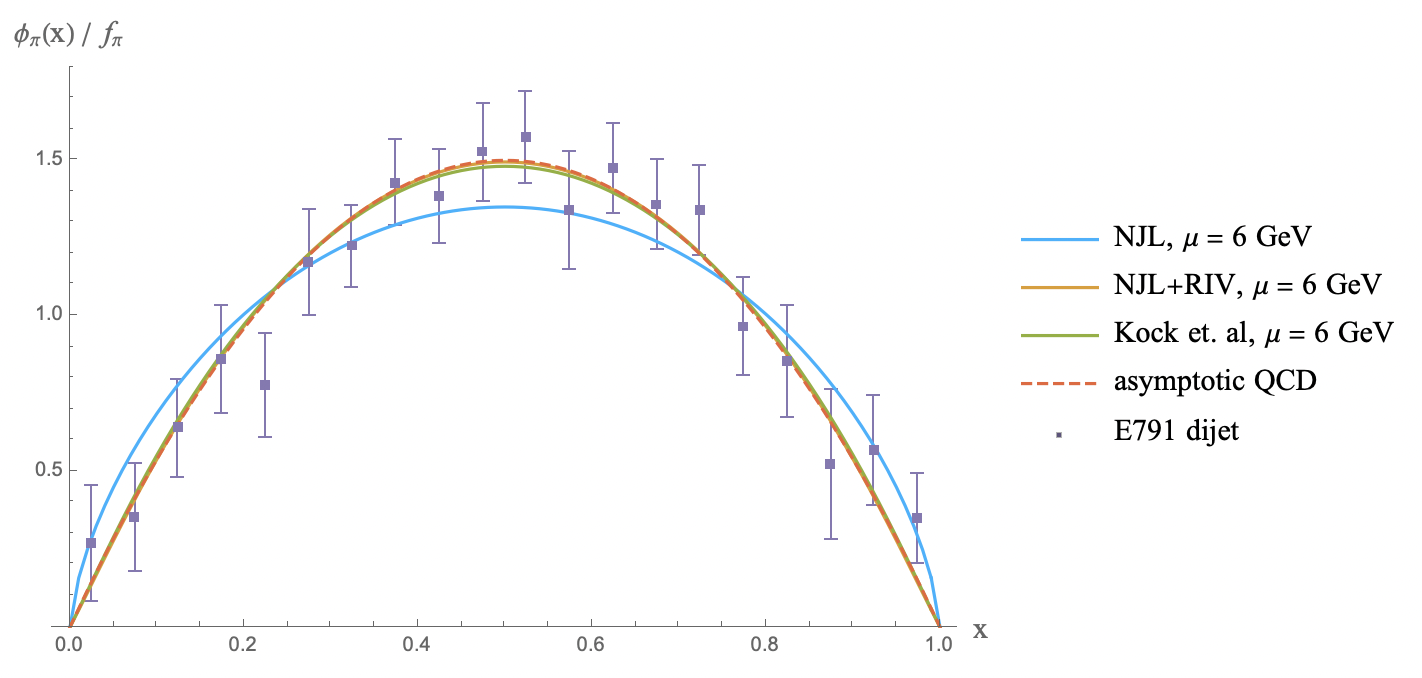}%
}\hfill
\subfloat[\label{fig_3pt6}]{%
  \includegraphics[height=5.5cm,width=.46\linewidth]{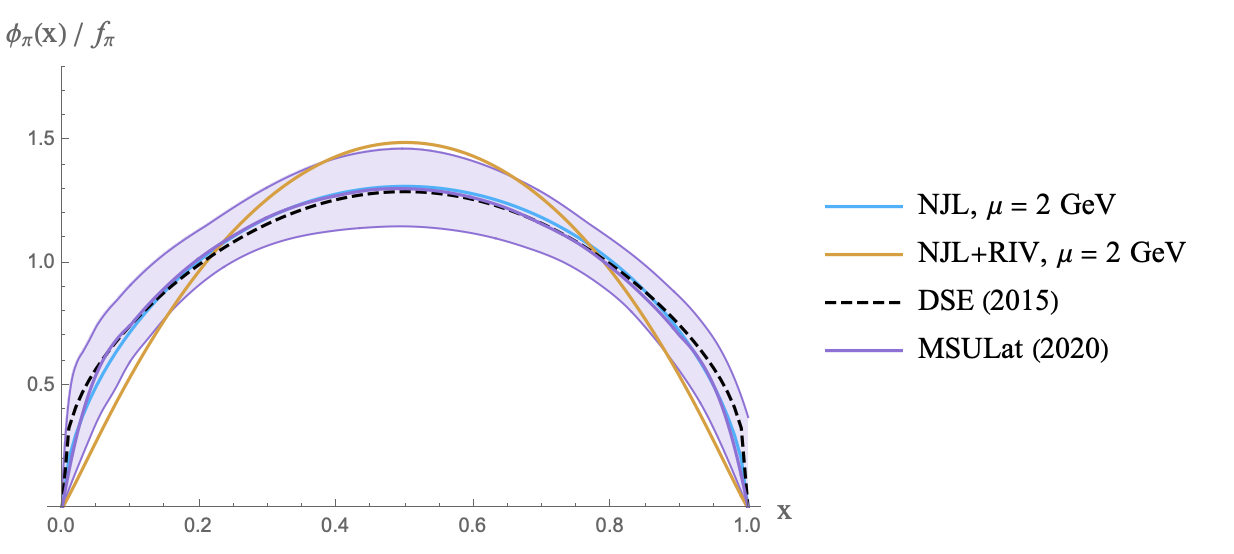}%
}
\caption{a: Pion  DA in the ILM with zero instanton size (NJL-like) as given in~(\ref{pionPDF}) (solid-blue), in the ILM with finite instanton size
as given in~(\ref{pionPDF_RIV}) (solid-orange), after ERBL evolution from $\mu_0=0.313$ GeV to $\mu=6$ GeV. The pion mass is $m_\pi=135$ MeV. 
The results are compared to the DAs obtained in~\cite{Kock:2020frx,Kock:2021spt} (green-solid) by applying the LaMET in  the ILM, also evolved 
to $\mu=6$ GeV. We also show the asymptotic pQCD result of $6x\bar x$ from~\cite{EFREMOV1980245} (dashed-red). 
The measured DA (purple) is  from $\pi^-$ into di-jets via diffractive dissociation, with invariant dijet mass of $6$ GeV \cite{E791:2000xcx},
as extracted and normalized in~\cite{Broniowski2008}. \\
b: The same evolved pion DAs   from~(\ref{pionPDF}) (solid-blue) and from~(\ref{pionPDF_RIV}) (solid-orange) now evolved to only
$\mu=2$ GeV, are compared to recent lattice results using LaMET (purple-band) from~\cite{Zhang2020}. 
 The DSE curve (dashed-black) follows from  Dyson-Schwinger equations with Bethe-Salpeter wavefunctions \cite{Shi2015}.}
\label{fig_piLATDATX}
\end{figure*}

In the ILM with a finite instanton size, the pion DA  is
\begin{equation}
\label{PIXX}
\begin{aligned}
        \phi_{\pi}(x)&=\frac{\sqrt{N_c}M}{\sqrt{2}\pi^2}C_{\pi}\int_{\frac{\rho M}{2\lambda\sqrt{x}\bar{x}}}^{\infty} dz  \frac{z^5}{\frac{\rho^2m^2_\pi}{4\lambda^2}-z^2} (F'(z))^4 \\
\end{aligned}
\end{equation}
which simplifies in the chiral limit
\begin{equation}
\label{PIXXX}
\begin{aligned}
        \phi_{\pi}(x)&=-\frac{\sqrt{N_c}M}{\sqrt{2}\pi^2}C_{\pi}\int_{\frac{\rho M}{2\sqrt{x}\bar{x}}}^{\infty} dz  z^3 (F'(z))^4 \\
\end{aligned}
\end{equation}
In general, we note that the induced form factors $\sqrt{{\mathcal F}(i\partial)}$ give rise to  extra contributions to the 
Noether axial vector current, and possibly the axial source current in (\ref{PITWIST2}). Indeed, the semi-bosonized 
Lagrangian $\mathcal{L}$ in (\ref{nonlocal_LFET}) with the minimal substitution  
$$i\partial_\mu\rightarrow i\partial_\mu+\gamma^5\tau^aA_\mu^a$$  where $A_\mu^a$ is a local external flavor
gauge field, yields the conserved $SU(2)_A$ current in the chiral limit,
\begin{widetext}
\label{PIFFF}
\begin{equation}
    \frac{\partial \mathcal{L}}{\partial A^a}\bigg|_{\substack{A,\ \pi=0\\ \sigma=\bar{\sigma}}}=\bar{\psi}\gamma^\mu\tau^a\psi+\bar{\sigma}\bar{\psi}\sqrt{\mathcal{F}(i\partial)}'\tau^a\sqrt{\mathcal{F}(i\partial)}\psi+\bar{\sigma}\bar{\psi}\sqrt{\mathcal{F}(i\partial)}\tau^a\sqrt{\mathcal{F}(i\partial)}'\psi
\end{equation}
\end{widetext}
with  $\sqrt{\mathcal{F}(x)}'=\frac{d}{dx}\sqrt{\mathcal{F}(x)}$. From (\ref{FFSUB}) it follows that the extra contributions in (\ref{PIFFF}) are 
absent for the leading twist-2 operator on the light front. Therefore, the GOR relation and the normalization of the pion DA are unchanged.
The results (\ref{PIXX}-\ref{PIXXX}) are in agreement with those derived from the ILM using the large momentum effective theory,
in the dilute approximation~\cite{Kock:2020frx}.

\subsection{ERBL evolution of the pion DA}
The evolution of the pion DA is governed by ERBL equation
\begin{equation}
\label{eq:ERBL}
    \phi(x,Q)=6x\bar{x}\sum_{n=0}^\infty a_n(Q_0)\left(\frac{\alpha_s(Q^2)}{\alpha_s(Q^2_0)}\right)^{\gamma_n/\beta_0}C_n^{3/2}(x-\bar{x})
\end{equation}
which is an expansion in Gegenbauer polynomials $C_n^{m}(z)$ of increasing powers,
with anomalous dimension 
\begin{equation}
    \gamma_n=C_F\left[-3+4\sum_{j=1}^{n+1}\frac{1}{j}-\frac{2}{(n+1)(n+2)}\right]
\end{equation}
where $C_F=\frac{N_c^2-1}{2N_c}$, $$\alpha_s(Q)=\frac{4\pi}{\beta_0\ln\left(\frac{Q^2}{\Lambda_{QCD}^2}\right)}$$
and $\beta_0=\frac{11}{3}N_c-\frac{2}{3}n_f$, and $\Lambda_{QCD}=226$ MeV.
 Due to the orthogonality of the Gegenbauer polynomials, the initial coefficients can be evaluated by 
\begin{equation}
    a_n(Q_0)=\frac{2(2n+3)}{3(n+1)(n+2)}\int_0^1dyC^{3/2}_n(y-\bar{y})\,\phi(y,Q_0)
\end{equation}

In Fig.~\ref{fig_pionERBL}a we show the pion DA versus $x$ in the zero instanton size limit, for 
fixed fermionic coupling $g_S=1.32\, g_{S,NJL}^{cr}$ and different current quark masses. 
In Fig.~\ref{fig_pionERBL}b we show the  same pion DA versus $x$ in the ILM for a similar 
fermionic coupling $g_S=1.844 \,g_{S,RIV}^{cr}$  and different current quark masses. There
is a dramatic change at the end-points following the emergence of the instanton induced form factors
from the quark zero modes. In Fig.~\ref{fig_pionERBL}c we show the ERBL evolved pion DA in the 
zero instanton size limit, and in Fig.~\ref{fig_pionERBL}d the same evolution in the ILM. 
The initial scale used is $\mu_0=0.313$ GeV to $\mu=6$ GeV. The pion mass is $m_\pi=135$ MeV

In Fig.~\ref{fig_piLATDATX}a we compare the ERBL evolved pion DAs to $\mu=6$ GeV,  with the QCD asymptotic result of 
$6 x\bar x$\cite{EFREMOV1980245} (dashed-red curve), 
and the measured pion DA by the E791 dijet collaboration~\cite{E791:2000xcx} (purple points). 
The data are from $\pi^-$ into di-jets via diffractive dissociation with invariant dijet mass $6$ GeV~\cite{E791:2000xcx}, as compiled  
in~\cite{Broniowski2008}. The evolved DAs are for zero size insantons (blue-solid), the current light front analysis of the
ILM (orange solid) and the LaMET analysis of the ILM in~\cite{Kock:2020frx,Kock:2021spt}.
In Fig.~\ref{fig_piLATDATX}b we compare our pion DA to the Dyson-Schwinger result~\cite{Shi2015} (dashed-black), and
the more recent MSULAT lattice results using the LaMET procedure~\cite{Zhang2020} (filled-purple-band).
The evolution is now from $\mu_0=0.313$ GeV to $\mu=2$ GeV. Our  evolved results are for zero size instantons (solid-blue),
and the ILM (solid-orange).

.

\section{Parton distribution functions}
\label{SEC_PART}
In general, the partonic structures in a hadron, are related to pertinent leading twist matrix elements,
following factorization of  various observables. The distribution functions or PDFs are  Fourier 
transforms of fermionic (gluonic) correlators at fixed separation on the light cone. They follow from
the LFWFs. More specifically, the  quark twist-2 PDF in a bound meson state, is given by
\begin{widetext}
\begin{equation}
    q_{X}(x)=\int_{-\infty}^\infty\frac{d\xi^-}{4\pi}e^{ix P^+\xi^-}\langle P|\bar{\psi}(0)\gamma^+ W(0,\xi^-)\psi(\xi^-)|P\rangle=\int\frac{d^2k_\perp}{(2\pi)^3}\left|\Phi_X(x,k_,s,s')\right|^2
\end{equation}
while the anti-quark PDF is given by
\begin{equation}
    \bar{q}_X(x)=\int_{-\infty}^\infty\frac{d\xi^-}{4\pi}e^{-ix P^+\xi^-}\langle P|\bar{\psi}(0)\gamma^+ W(0,\xi^-)\psi(\xi^-)|P\rangle=\int\frac{d^2k_\perp}{(2\pi)^3}\left|\Phi_X(\bar{x},k_,s,s')\right|^2
\end{equation}
\end{widetext}
Here  $$W(\xi^-,0)=\mathrm{exp}\left[-ig\int_0^{\xi^-}d\eta^- A^+(\eta^-)\right]\rightarrow 1$$ is a gauge link,
which will be set to 1 throughout.  For the meson PDFs, the 
quark and antiquark distributions are related by charge symmetry 
$$\bar{q}_X(x)=q_{\bar{X}}(x)=q_X(1-x)=\bar{q}_{\bar{X}}(1-x)$$
 Also, the PDFs are normalized to $1$ by  charge conservation
\begin{equation}
    \langle X(P)|\bar{\psi}\frac{1}{2}\gamma^+\psi|X(P)\rangle=1
\end{equation}

\begin{figure*}
\subfloat[\label{fig_3pt1}]{%
  \includegraphics[height=5.5cm,width=.46\linewidth]{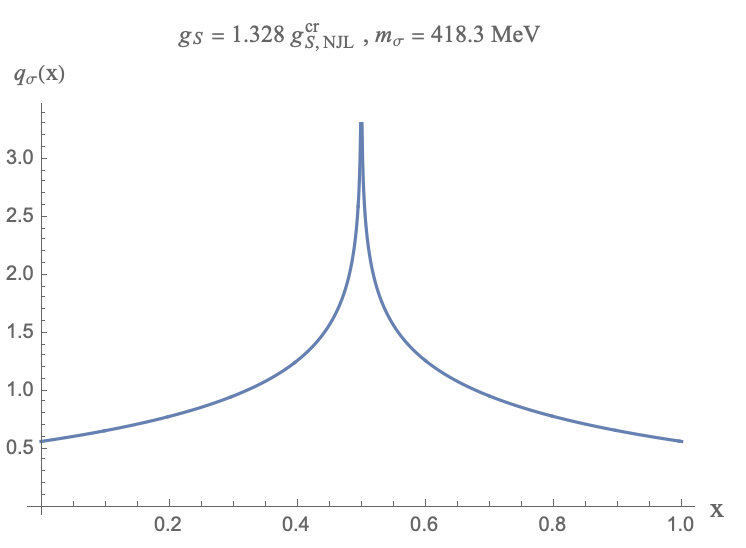}%
}\hfill
\subfloat[\label{fig_3pt6}]{%
  \includegraphics[height=5.5cm,width=.46\linewidth]{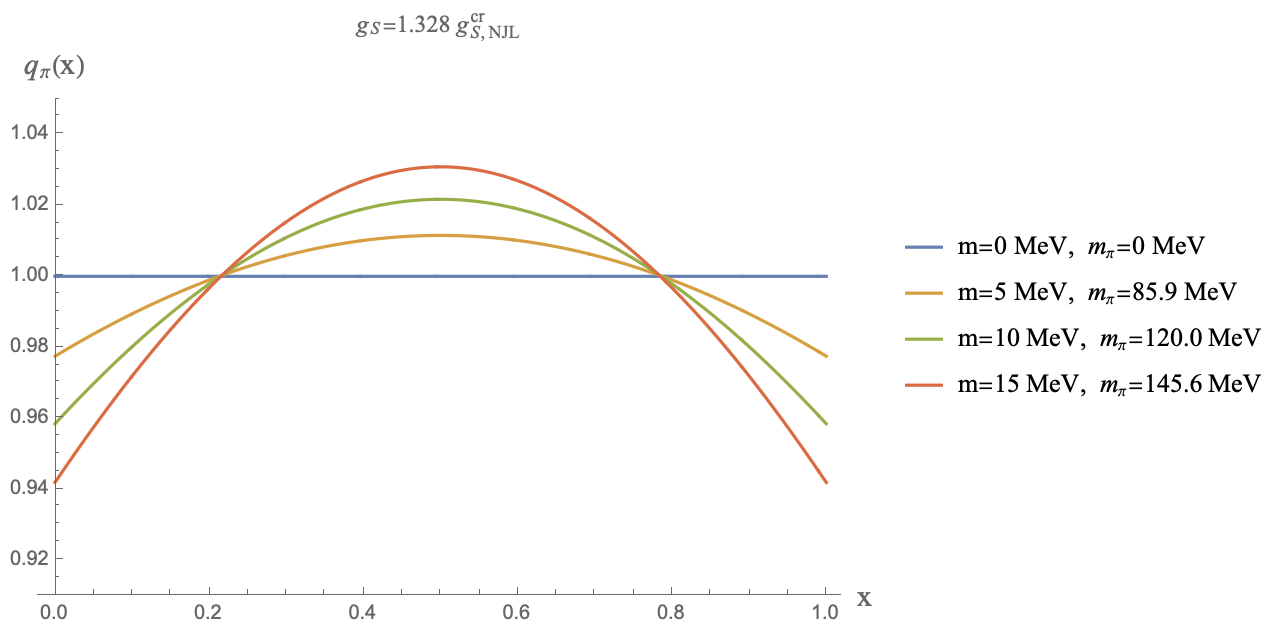}%
}\hfill
\subfloat[\label{fig_3pt6}]{%
  \includegraphics[height=5.5cm,width=.46\linewidth]{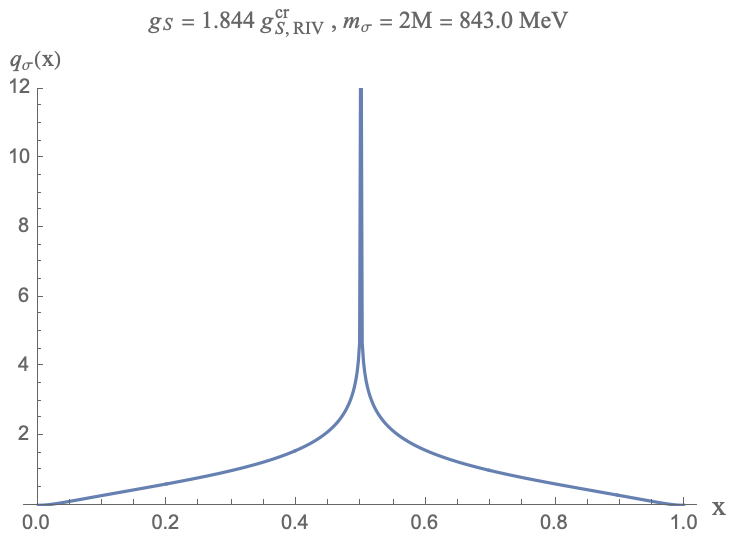}%
}\hfill
\subfloat[\label{fig_3pt6}]{%
  \includegraphics[height=5.5cm,width=.46\linewidth]{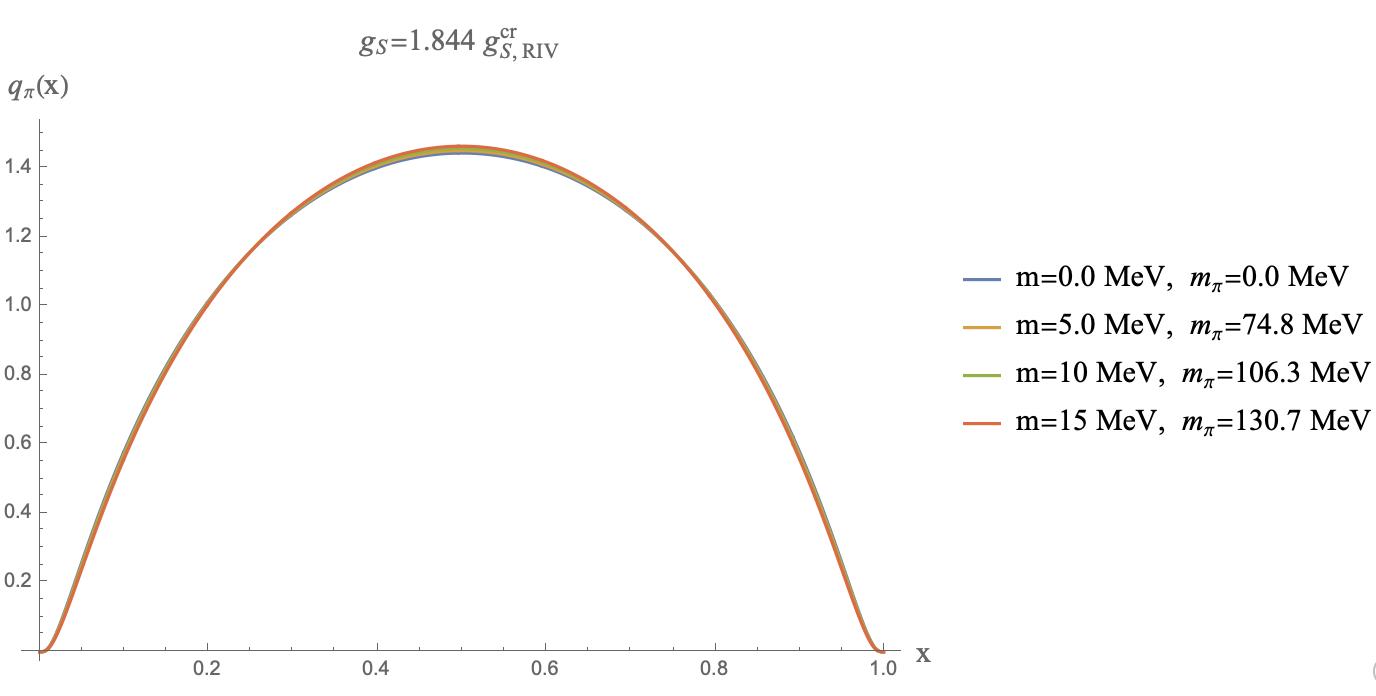}%
}
\caption{a: The sigma PDF in zero instanton size and in the chiral limit, with a  treshold sigma  mass $2M=418.9$ MeV;
b: The pion PDF with zero instanton size, but with different pion masses varying with quark current masses;
c: The sigma PDF with the finite instanton size $1/\rho=630$ MeV, with a  treshold sigma  mass $2M=418.9$ MeV in the chiral limit;
d: The pion PDF with the finite instanton size  $\rho=0.313$ fm, with different pion masses.}
\label{fig_sigpi}
\end{figure*}

\begin{widetext}
\subsection{sigma and pion PDFs}
The $\sigma$ PDF is
\begin{equation}
\begin{aligned}
    q_\sigma(x)=&\frac{1}{4\pi^2}\int_0^{\infty} dk^2_\perp\frac{1}{x\bar{x}}\left|\frac{C_\sigma}{m^2_\sigma-\frac{k_\perp^2+M^2}{x\bar{x}}}\right|^2\frac{k_\perp^2+(x-\bar{x})^2M^2}{x\bar{x}}\mathcal{F}\left(k\right)\mathcal{F}\left(P-k\right)\\
\end{aligned}
\end{equation}
Also, the pion PDF is
\begin{equation}
\begin{aligned}
    q_\pi(x)=&\frac{1}{4\pi^2}\int_0^{\infty} dk^2_\perp\frac{1}{x\bar{x}}\left|\frac{C_\pi}{m^2_\pi-\frac{k_\perp^2+M^2}{x\bar{x}}}\right|^2\frac{k_\perp^2+M^2}{x\bar{x}}\mathcal{F}\left(k\right)\mathcal{F}\left(P-k\right)\\
\end{aligned}
\end{equation}
\end{widetext}
In the chiral limit, the normalization constant can be determined by the cut-off function induced by the non-local quark form factor as well. Thus, the normalization constant can be related to the pion decay constant $f_\pi$ and the constituent mass $M$,
\bea
    C_\pi&=&-2\pi\int_0^{\infty} dk^2_\perp\frac{1}{k_\perp^2+M^2}\mathcal{F}\left(k\right)\mathcal{F}\left(P-k\right)\nonumber\\
    &=&-\frac{\sqrt{2N_c}M}{f_\pi}
\eea
Again, the minus sign is chosen to make the wave function positive-definite for convenience.

In the zero instanton
size limit, we can use the same hard cut-off in described in Appendix~\ref{APP_SHARP0}, with the result for the   sigma PDF as
\begin{equation}
\begin{aligned}
    q_\sigma(x)
    =&\frac{C_\sigma^2}{4\pi^2}\theta(x\bar{x})\ln\left(1+\frac{\Lambda^2}{(x-\bar{x})^2M^2}\right)
\end{aligned}
\end{equation}
where $$-C_\sigma=2\pi\left[\ln\left(1+\frac{\Lambda^2}{M^2}\right)+2\frac{\Lambda}{M}\tan^{-1}\frac{M}{\Lambda}\right]^{-1/2}$$
Recall that the sigma meson  is only bound in the chiral limit with $m_\sigma=2M$. For comparison, we note tha in the same approximation
of zero instanton size, the pion  PDF is
\begin{widetext}
\begin{equation}
\label{pionPDF}
\begin{aligned}
    q_\pi(x)
    =&\frac{C_\pi^2}{4\pi^2}\theta(x\bar{x})\left[\frac{x\bar{x}m^2_\pi\Lambda^2}{(M^2-x\bar{x}m^2_\pi)(M^2+\Lambda^2-x\bar{x}m^2_\pi)}-\ln\left(\frac{M^2-x\bar{x}m^2_\pi}{M^2+\Lambda^2-x\bar{x}m^2_\pi}\right)\right]\\
    \xrightarrow{m_\pi\rightarrow0}&\frac{C_\pi^2}{4\pi^2}\theta(x\bar{x})\ln\left(1+\frac{\Lambda^2}{M^2}\right)
\end{aligned}
\end{equation}
\end{widetext}
where in the chiral limit, $$C_\pi=-2\pi\left[\ln\left(1+\frac{\Lambda^2}{M^2}\right)\right]^{-1/2}$$
This result agrees with the result in the zero instanton size limit $\rho\rightarrow0$ in~\cite{Kock:2020frx,Kock:2021spt},
using the large momentum effective theory (LaMET). It was first established in the NJL model in~\cite{RuizArriola:2002bp}.
However, it is at variance with the end-point expectation
 $f_{q/\pi}(x\rightarrow1,Q^2)\sim(1-x)$ at  $Q^2\rightarrow\infty$, from the Drell-Yan-West result~\cite{Drell:1969km,West:1970av}. 
The discrepancy is expected to wane out with QCD evolution, which depletes the large-x part of the PDF as we show below.

For a finite instanton size, all integrations are tamed by the induced zero mode form factor $\sqrt{{\mathcal F}(i\partial)}$
discussed earlier. The ensuing non-local interactions are expected to contribute to the currents, hence to the twist-2 part
of the PDFs in general~\cite{BOWLER1995655,Hell_2009}, as we noted earlier for the axial-vector current in~(\ref{PIFFF}). For
the vector current, the extra contributions follow from minimal substitution
\begin{equation}
    i\partial_\mu\rightarrow i\partial_\mu+V_\mu
\end{equation}
where $V_\mu$ is the external $U(1)$ flavor gauge field. The vector current for the non-local semi-bosonized Lagrangian (\ref{nonlocal_LFET}), follows
from the Noether construction in the form
\begin{widetext}
\begin{equation}
\label{VDD}
    \frac{\partial \mathcal{L}}{\partial V}\bigg|_{\substack{V,\ \pi=0\\ \sigma=\bar{\sigma}}}=\bar{\psi}\gamma^\mu\psi+\bar{\sigma}\bar{\psi}\sqrt{\mathcal{F}(i\partial)}'\sqrt{\mathcal{F}(i\partial)}\psi+\bar{\sigma}\bar{\psi}\sqrt{\mathcal{F}(i\partial)}\sqrt{\mathcal{F}(i\partial)}'\psi
\end{equation}
\end{widetext}
where again, the derivative on the functional form of the form factor is defined as $\sqrt{\mathcal{F}(x)}'=\frac{d}{dx}\sqrt{\mathcal{F}(x)}$. 
Now we recall that in the 2-body channel, the boost invariant substitution (\ref{FFSUB}) is only dependent on the relative transverse momentum of the
pair. From (\ref{VDD}), it follows that the leading twist-2 light front current is not modified, leaving the chiral relation and normalization of the twist-$2$ meson PDF and DA unchanged.

\begin{widetext}
With this in mind, the sigma PDF in the ILM in the chiral limit is
\begin{equation}
\begin{aligned}
    q_{\sigma}(x)=\frac{1}{4\pi^2}\int_0^{\infty} dk^2_\perp C_\sigma^2\frac{1}{k_\perp^2-(4x\bar{x}-1)M^2}\mathcal{F}\left(k\right)\mathcal{F}\left(P-k\right)
    =\frac{C_\sigma^2}{2\pi^2}\int_{\frac{\rho M}{2\lambda_S\sqrt{x\bar{x}}}}^{\infty} dz\frac{z^5}{z^2-\rho^2M^2/\lambda_S^2}(F^\prime(z))^4
\end{aligned}
\end{equation}
while the pion PDF,  in general is
\begin{equation}
\label{pionPDF_RIV}
\begin{aligned}
    q_{\pi}(x)=&\frac{1}{4\pi^2}\int_0^{\infty} dk^2_\perp\frac{C_\pi^2(k_\perp^2+M^2)}{(x\bar{x}m^2_\pi-k_\perp^2-M^2)^2}\mathcal{F}\left(k\right)\mathcal{F}\left(P-k\right)
=\frac{C_\pi^2}{2\pi^2}\int_\frac{\rho M}{2\lambda_S\sqrt{x\bar{x}}}^\infty dz \frac{z^3}{\left(\frac{\rho^2 m_\pi^2}{4\lambda_S^2}-z^2\right)^2}[zF^\prime(z)]^4
\nonumber\\
   \end{aligned}
\end{equation}
\end{widetext}
In the chiral limit,
\bea
q_{\pi}(x) \xrightarrow{m_\pi\rightarrow0}&\frac{C_\pi^2}{2\pi^2}\int_\frac{\rho M}{2\lambda_S\sqrt{x\bar{x}}}^\infty dz z^3(F'(z))^4
\eea
with the normalization constant
\bea
C_\pi&=&-\sqrt{2}\pi\left[\int_0^1 dx\int_\frac{\rho M}{2\lambda_S\sqrt{x\bar{x}}}^\infty dz z^3(F'(z))^4\right]^{-1/2}\nonumber\\
&=&-\sqrt{2N_c}M/f_\pi\approx-7.993
\eea
with  $M=421.5$ MeV, $\rho=0.313$ fm and $f_\pi=130.3$ MeV.
At the end point $x,\bar{x}=0$, the asymptotic form of $F(z)\sim1/4z^3$ dominates the integral. 
The endpoint behavior of the pion PDF in the ILM at a resolution of $Q^2\sim 1/\rho^2$,
is softer than the one  expected from the Drell-Yan-West  relation
at much larger resolution~\cite{Drell:1969km,West:1970av}, with
\begin{equation}
    q_{\pi}(x)\sim \frac{C_\pi^2}{2\pi^2}\frac{108\lambda_S^{12}}{\rho^{12}M^{12}}(x\bar{x})^6
\end{equation}

In Fig.~\ref{fig_sigpi}a we show the sigma PDF in the chiral limit, with a  treshold sigma  mass $2M=418.3$ MeV, in the instanton zero size approximation.
The PDF is sharply picked at $x=\frac 12$, and does not vanish at the end points $x=0,1$. In Fig.~\ref{fig_sigpi}b we show the pion PDF for different current quark masses, also in the zero instanton size approximation. In Fig.~\ref{fig_sigpi}c we show the sigma PDF in the ILM, with a much sharper distribution at
$x=\frac 12$ that reflects on the treshold state, in the zero instanton size limit. In Fig.~\ref{fig_sigpi}d the pion PDF in the ILM is shown for different current quark masses.  The instanton induced form factors cause it to vanish at the end-points, with little sensitivity to the current quark masses.

\begin{figure*}
\subfloat[\label{fig_3pt1}]{%
  \includegraphics[height=5.5cm,width=.46\linewidth]{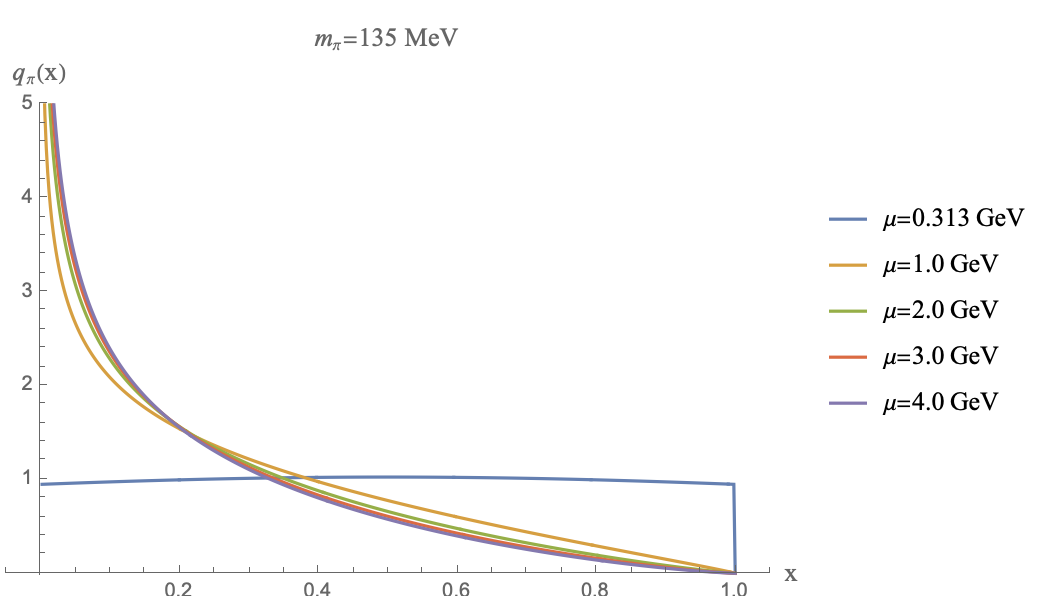}%
}\hfill
\subfloat[\label{fig_3pt6}]{%
  \includegraphics[height=5.5cm,width=.46\linewidth]{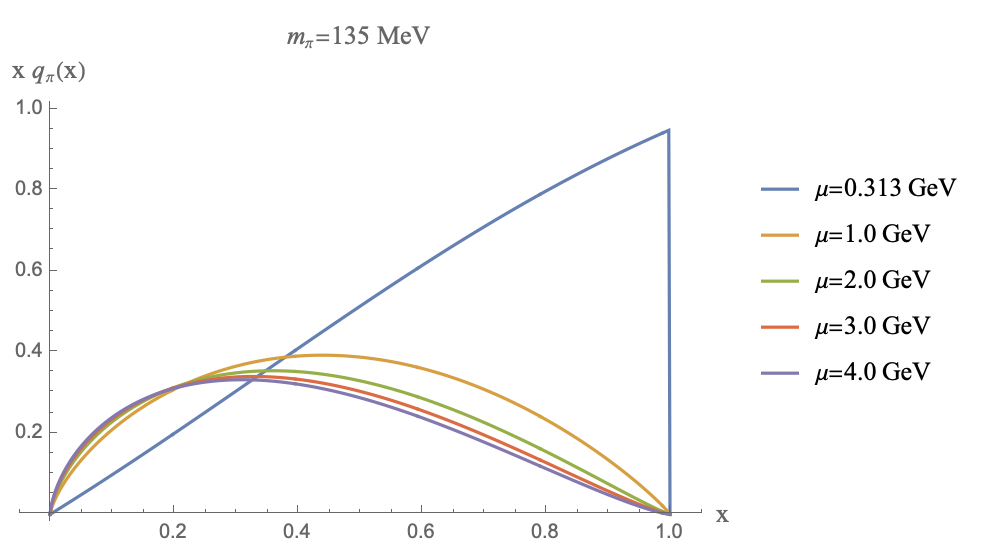}%
}\hfill
\subfloat[\label{fig_3pt6}]{%
  \includegraphics[height=5.5cm,width=.46\linewidth]{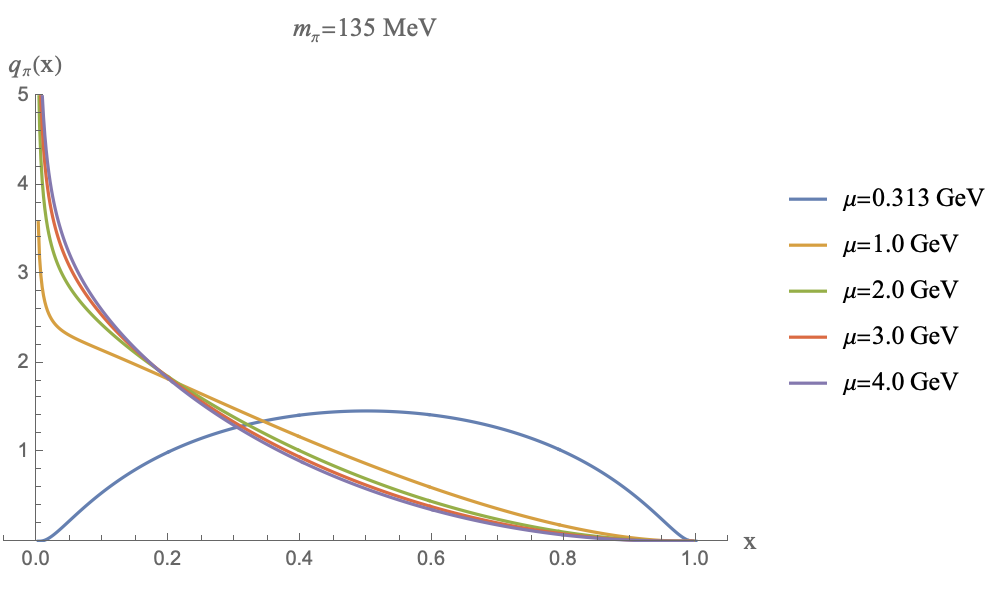}%
}\hfill
\subfloat[\label{fig_3pt6}]{%
  \includegraphics[height=5.5cm,width=.46\linewidth]{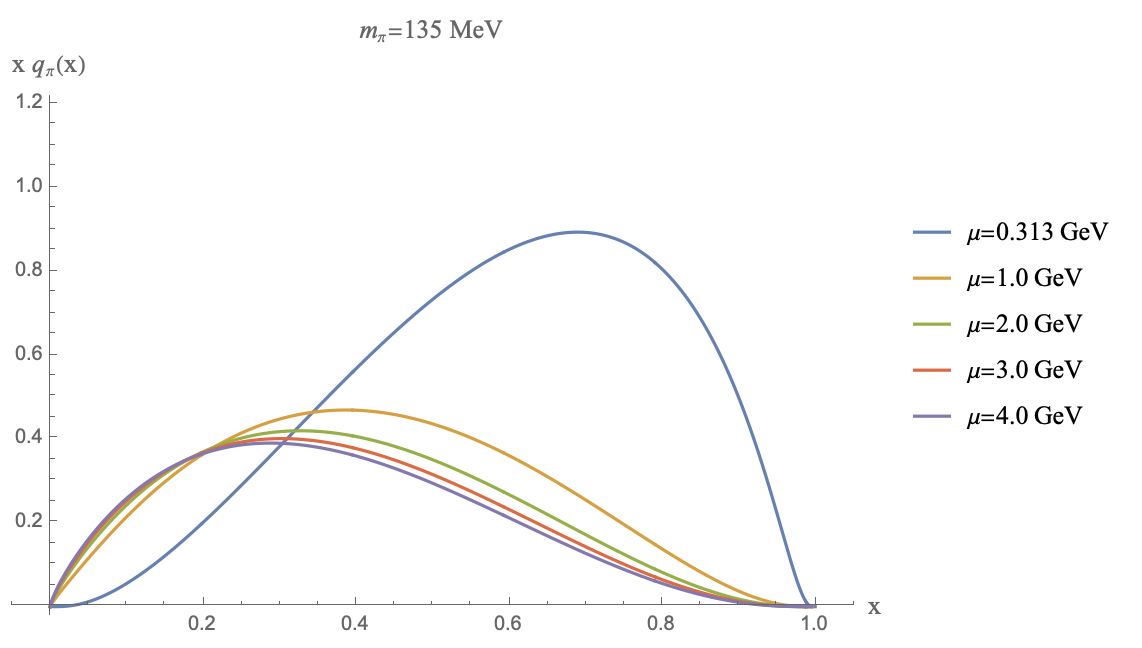}%
}
\caption{a: DGLAP evolution of pion PDF for zero instanton size  and in the  chiral limit;
b: DGLAP evolution of te pion valence quark momentum distribution  for zero instanton size  and in the chiral limit;
c: DGLAP evolution for the pion PDF with a finite instanton size  $\rho=0.313$ fm;
d: DGLAP evolution of tof the pion valence quark momentum distribution with a  finite instanton size $\rho=0.313$ fm,
in the chiral limit. All evolutions start from the initial scale  $\mu_0=0.313$ GeV.}
\label{fig_piondglap}
\end{figure*}

\begin{figure*}
\subfloat[\label{fig_3pt1}]{%
  \includegraphics[height=5.5cm,width=.46\linewidth]{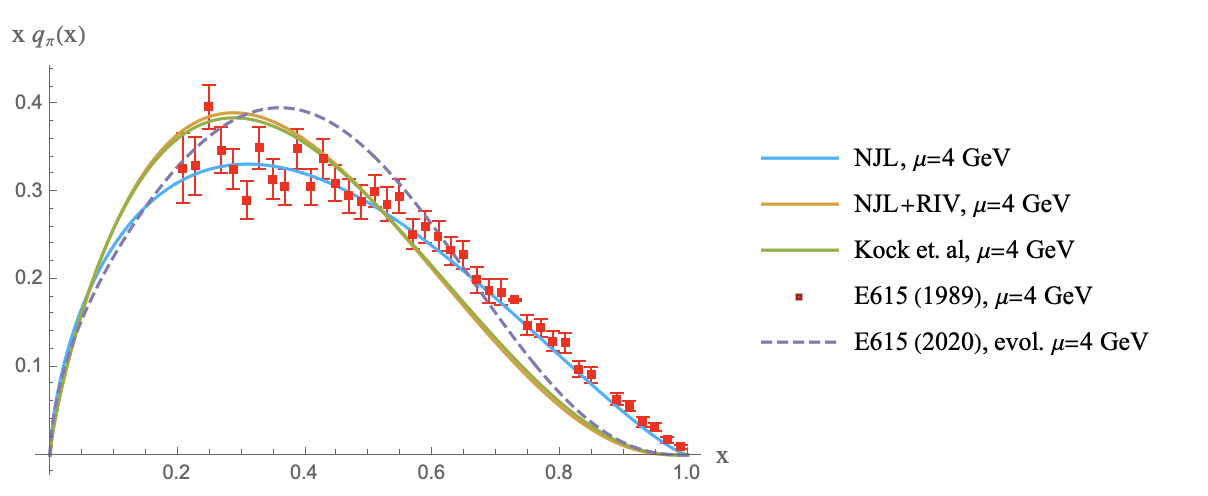}%
}\hfill
\subfloat[\label{fig_3pt6}]{%
  \includegraphics[height=5.5cm,width=.46\linewidth]{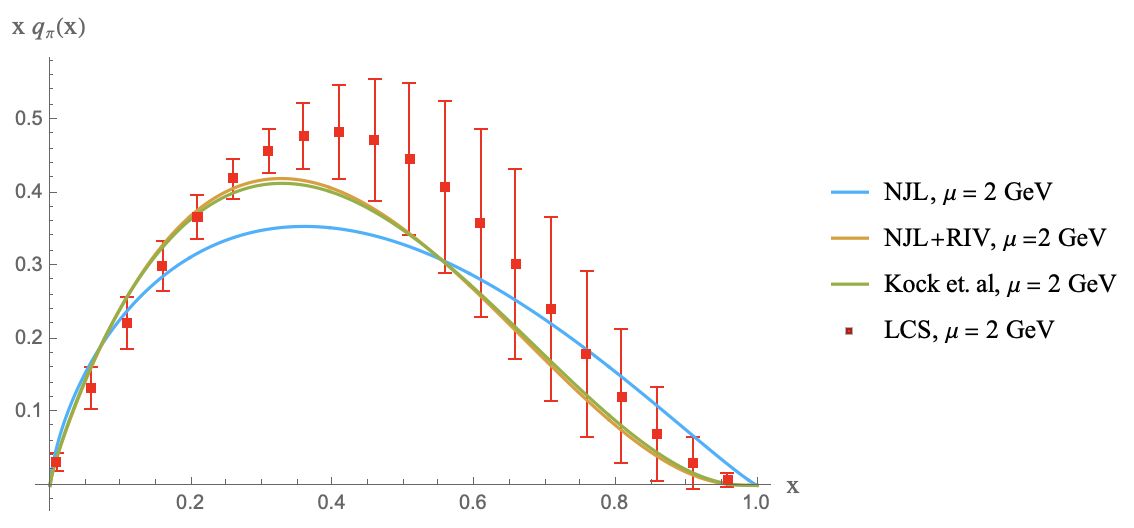}%
}
\caption{a: Pion parton momentum distribution function for zero instanton size (solid-blue) and finite instanton size 
of $\rho=0.317$ fm (solid-orange), both of which are evolved to $\mu=4$ GeV with a pion mass $m_\pi=135$ MeV.
The results are compared to the those extracted from the ILM using the LaMET (solid-green)~\cite{Kock:2020frx,Kock:2021spt},
also evolved to $\mu=4$ GeV.  The E615 data from 1989 (red) are from \cite{Conway:1989fs},  corresponding to a  fixed invariant muon pair mass $m_{\mu^+\mu^-}\geq 4.05$ GeV. The improved E615 data from 2020 (dashed-purple) are from  \cite{Aicher:2010cb}, using the original E615 experimental data from~\cite{Conway:1989fs}.\\
b: The same pion parton distribution functions as in a, but now evolved to $\mu=2$ GeV, for comparison with 
the lattice data using  cross sections (LCS) (red) from~\cite{Sufian:2020vzb}.
}
\label{fig_piLATDAT}
\end{figure*}

\subsection{DGLAP evolution}
In the ILM the partonic distributions are defined at a factorization scale $\mu_0$ which is smaller than the inverse size $1/\rho\sim 630$ MeV
of an instanton. To compare our result with the available experiments in~\cite{Conway:1989fs,Lan:2019vui,Aicher:2010cb}, we evolve the PDFs in (\ref{pionPDF}) and (\ref{pionPDF_RIV}), starting from say $\mu_0=313$ MeV, a non-perturbative scale that is not that large enough to  resolve the instantons
and anti-instantons in the ILM. The  PDFs are evolved to  $2$ GeV for comparison to also lattice data \cite{Sufian:2020vzb}. Since only the valence quark dynamics is retained in our analysis, we will only keep the quark splitting  in the DGLAP evolution. The reader may be concerned that at such a low energy scale, an altogether nonperturbative evolution of the type discussed recently in~\cite{Shuryak:2022wtk}, maybe required. This point will be addressed in a sequel. Here, we will only implement the perurbative DGLAP evolution for a qualitative comparison.

In Figs.~\ref{fig_piondglap}a, b we show the DGLAP evolved pion PDF in the zero instanton size approximation, starting from the
initial scala $\mu_0=313$ MeV up to 4 GeV. In Figs.~\ref{fig_piondglap}c,d we show the same DGLAP evolution for the results in the ILM,
with a finite instanton size $\rho=0.313$ fm.
In both cases, the pion mass  is fixed at its physical value $m_\pi=135$ MeV.

In Fig.~\ref{fig_piLATDAT}a we compare our DGLAP evolved results with the original E615(1989) data~\cite{Conway:1989fs} (red), and the
subsequent E165(2020) improved analysis~\cite{Aicher:2010cb} (dashed-purple)  at $\mu=4$ GeV. The data are compiled from measurements 
of an invariant muon pair mass $m_{\mu^+\mu^-}\geq 4.05$ GeV.  Our evolved results  are for the zero instanton size (solid-blue), finite
instanton size in the ILM model (solid-orange). They are also compared to the extracted pion PDF from the ILM model using LaMET ~\cite{Kock:2020frx,Kock:2021spt} (solid-green). Our results are evolved from  an initial scale $\mu_0=313$ MeV below the instanton resolution scale $1/\rho=631$ MeV, to a final scale of 4 GeV. In Fig.~\ref{fig_piLATDAT}b our evolved results are compared to the LCS lattice results~\cite{Sufian:2020vzb} (red) at $\mu=2$ GeV. All the
theoretical results are also evolved to the same scale.

\section{$N_f=2$ Instanton-induced Interaction with Unequal Masses}
\label{SEC_GAPS}
To construct the kaon partonic distributions, we need to address the three flavor case with $u,d,s$ quarks.
For simplicity, we will consider the reduced mass case with light $m_u=m_d$ and heavier $m_s$. In this
case, the 2-flavor kaon sectors of $SU(3)$ with $U$-spin and $V$-spin are emanable to the same light front
effective Lagrangian (\ref{LFEFT}). The only difference is the large difference in the assigned current quark masses.
More specifically, we have
\begin{widetext}
\begin{equation}
\label{LFEFT_K}
    \mathcal{L}=\bar{\psi}(i\slashed{\partial}-M)\psi+ \frac{G_K}{2}\left[\bar{\psi}\psi\hat{D}^{-1}_{K+}\bar{\psi}\psi-\bar{\psi}\tau^a\psi\hat{D}^{-1}_{K-}\bar{\psi}\tau^a\psi-\bar{\psi}i\gamma^5\psi\hat{D}^{-1}_{K+}\bar{\psi}i\gamma^5\psi+\bar{\psi}i\gamma^5\tau^a\psi\hat{D}^{-1}_{K-}\bar{\psi}i\gamma^5\tau^a\psi\right]
\end{equation}
\end{widetext}
with the tadpole resummed vertices
\begin{equation}
\label{D}
    \hat{D}^{-1}_{K\pm}=\frac{1}{1\pm \frac{g_K}{N_c}\left\langle\bar{\psi}\gamma^+\frac{-i}{\overleftrightarrow{\partial_-}}\psi\right\rangle}
\end{equation}
$G_K$ is the corresponding 't Hooft coupling strength inthe  Kaon channel.
The quark constituent mass is now matrix valued $M=\mathrm{diag}(M_u,M_s)$,
 where $u$ denotes the lighter quark $u$ or $d$, and $s$ denotes the heavier strange quark.

\begin{figure*}
\subfloat[\label{fig_3pt1}]{%
  \includegraphics[height=4cm,width=.46\linewidth]{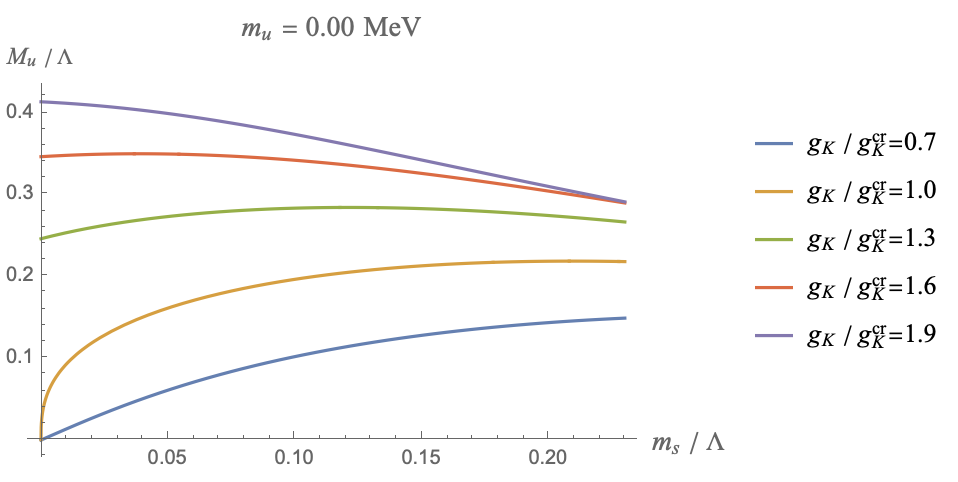}%
}\hfill
\subfloat[\label{fig_3pt6}]{%
  \includegraphics[height=4cm,width=.46\linewidth]{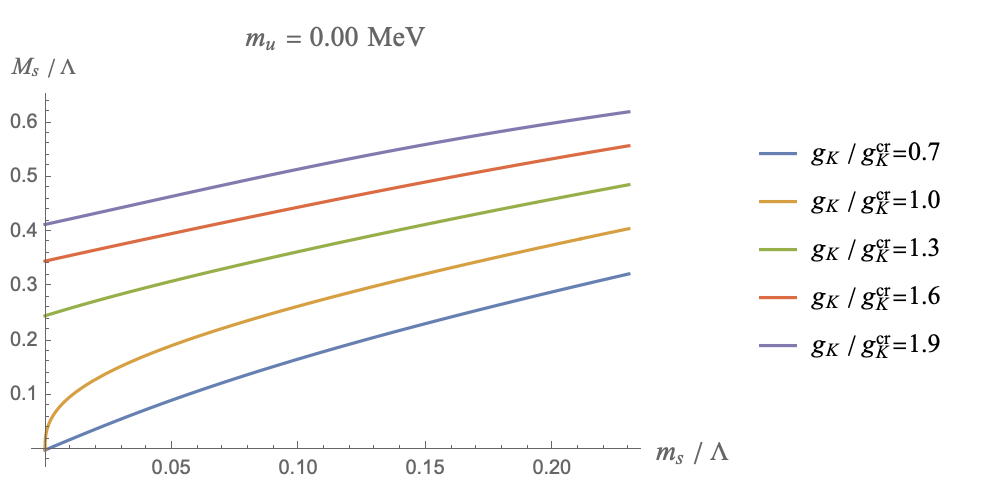}%
}\hfill
\subfloat[\label{fig_3pt6}]{%
  \includegraphics[height=4cm,width=.46\linewidth]{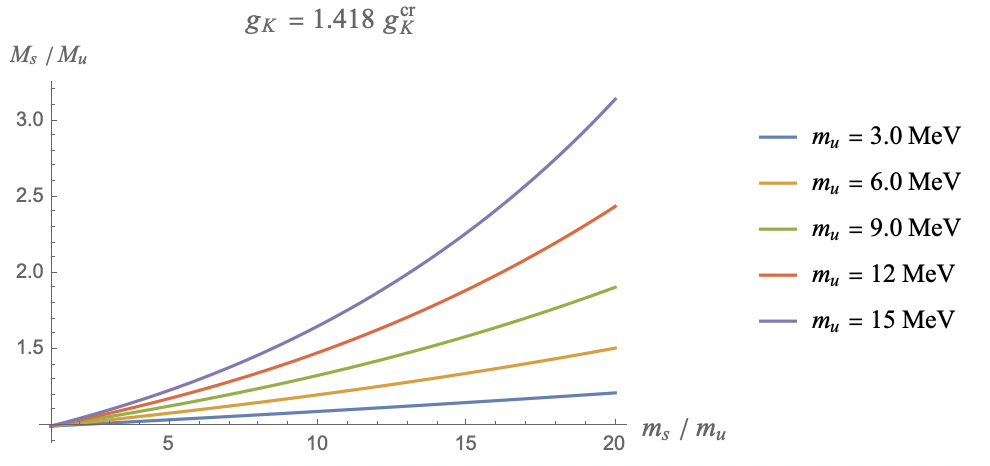}%
}\hfill
\subfloat[\label{fig_3pt6}]{%
  \includegraphics[height=4cm,width=.46\linewidth]{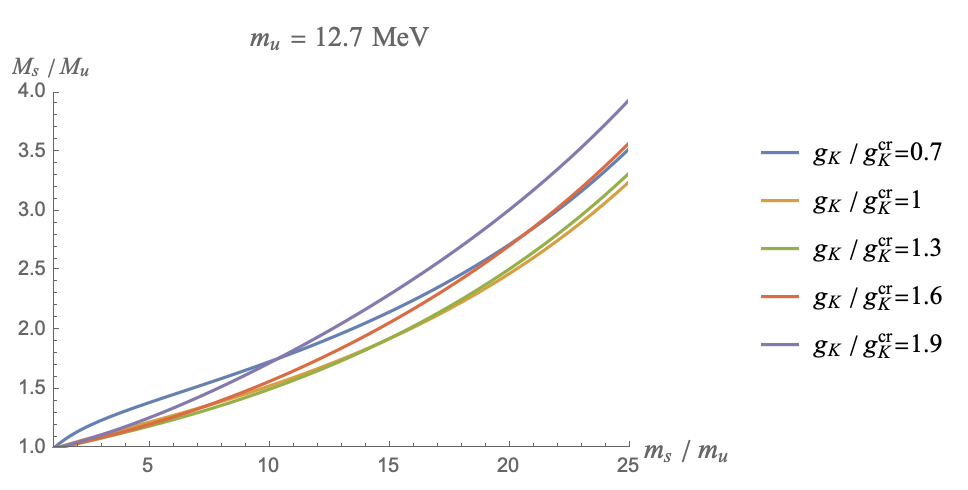}%
}\hfill
\subfloat[\label{fig_3pt1}]{%
  \includegraphics[height=4cm,width=.46\linewidth]{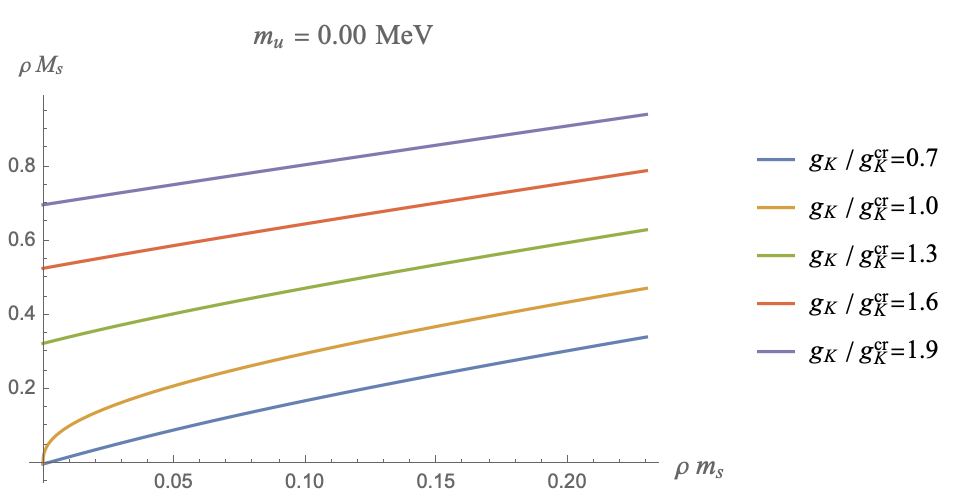}%
}\hfill
\subfloat[\label{fig_3pt6}]{%
  \includegraphics[height=4cm,width=.46\linewidth]{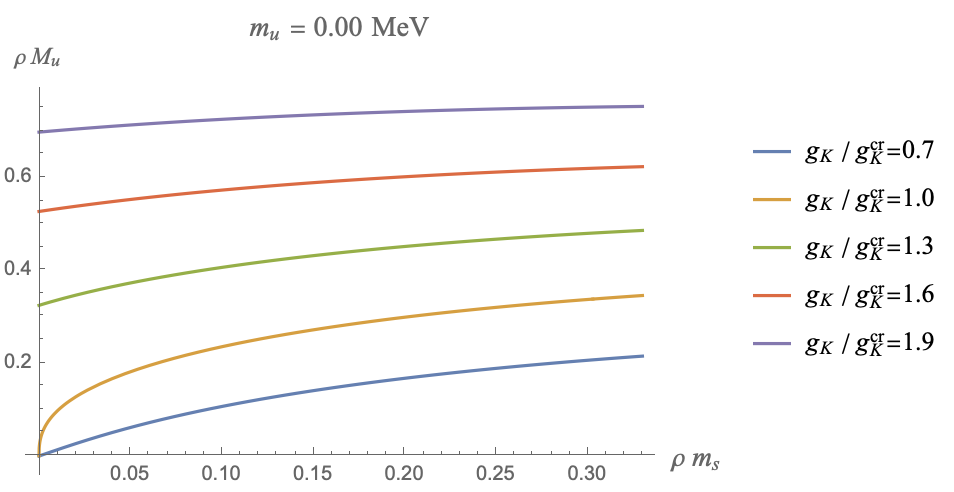}%
}\hfill
\subfloat[\label{fig_3pt6}]{%
  \includegraphics[height=4cm,width=.46\linewidth]{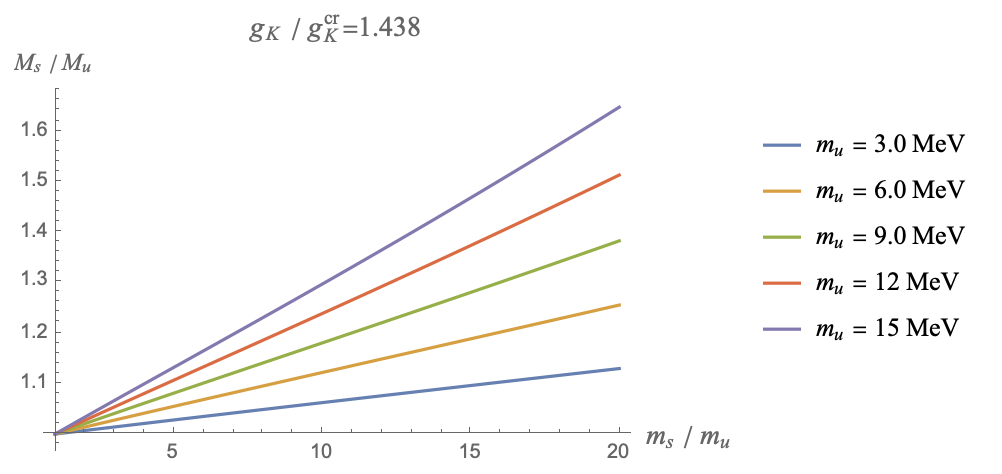}%
}\hfill
\subfloat[\label{fig_3pt6}]{%
  \includegraphics[height=4cm,width=.46\linewidth]{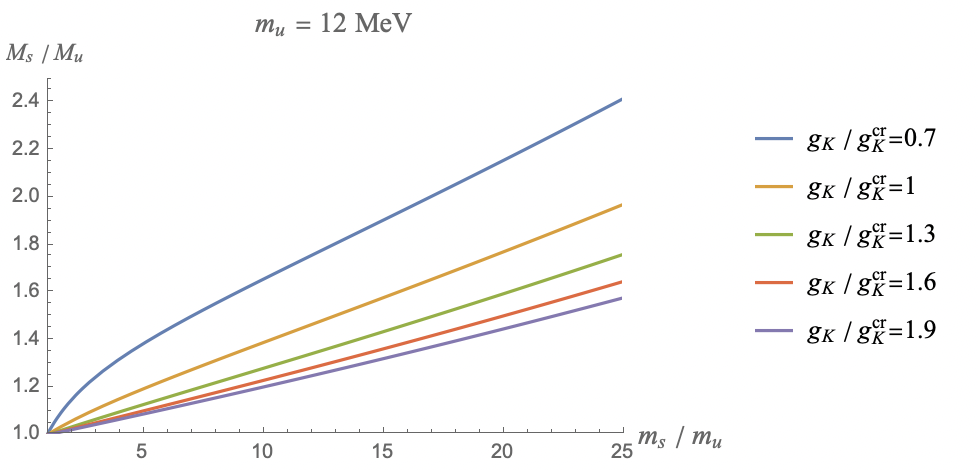}%
}
\caption{
a,b: The constituent masses $M_{u,s}$ versus the strange quark mass $m_s$  for different couplings $g_K/g_{K}^{cr}$,  fixed $m_u=0$,
in the zero instanton size limit; 
c,d: The constituent mass ratio $M_s/M_u$ versus $m_s/m_u$,
 for fixed coupling $g_K/g_{K}^{cr}=1.418$ with varying $m_u$,  and fixed $m_u=12.7$ MeV with varying $g_K/g_{K}^{cr}$, in the zero instanton size limit;
 e,f,g,h: The same constituent masses as in a,b,c,d, in the ILM with a finite instanton size.}
\label{fig_Mus}
\end{figure*}

\begin{figure*}
\subfloat[\label{fig_3pt1}]{%
  \includegraphics[height=5.5cm,width=.46\linewidth]{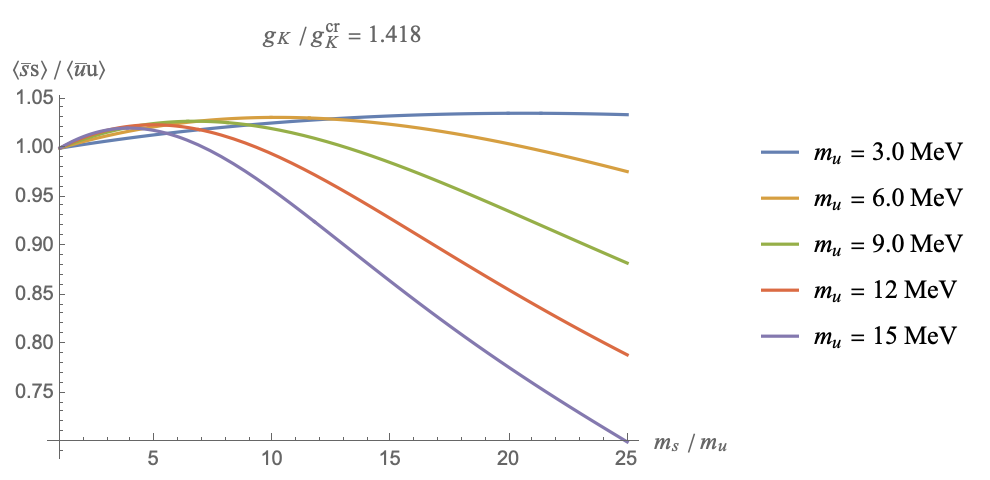}%
}\hfill
\subfloat[\label{fig_3pt6}]{%
  \includegraphics[height=5.5cm,width=.46\linewidth]{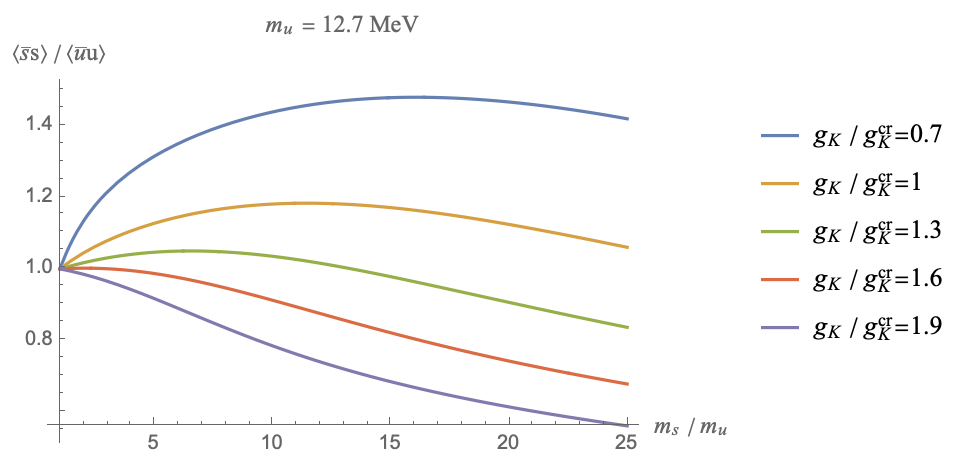}%
}
\caption{
a: The ratio of the quark condensates $\langle\bar ss\rangle/\langle\bar u u\rangle$
versus $m_s/m_u$, for fixed $g_K/g_{K}^{cr}=1.418$ and different $m_u$;
b: The ratio of the quark condensates $\langle\bar ss\rangle/\langle\bar u u\rangle$
versus $m_s/m_u$, for fixed $m_u=12.7$ MeV and different  $g_K/g_{K}^{cr}$.
}
\label{fig_uscondRIV}
\end{figure*}

\subsection{Gap equation}
In the $U$-spin ($V$-spin) sector, the effective momentum dependent form factors  $\sqrt{\mathcal{F}_{u/s}(k)}$ for each flavor will also be introduced to the interacting quark fields. Effectively, the quark fields in the 't Hooft interaction terms will be dressed by a form factor,

\begin{equation}
    \psi(x)\rightarrow\sqrt{\mathcal{F}(i\partial)}\psi(x)
\end{equation} 
The assignment suggested on mass shell in (\ref{FFSUB}) that enforces boost invariance in the 2-body state, implies that 
\begin{equation}
    \sqrt{\mathcal{F}(i\partial)}\rightarrow \begin{pmatrix}
    \sqrt{\mathcal{F}_{u}(i\partial)} & 0\\
    0& \sqrt{\mathcal{F}_{s}(i\partial)}
    \end{pmatrix}
\end{equation}
The non-local 't-Hooft interaction given by the momentum dependent form factors, yield  momentum-dependent gap equations, with running constituent masses
 $M_{u/s}(k^2)$ for each flavor,
\begin{equation}
\label{M_cut_off_s}
    M_{u/s}(k^2)=M_{u/s}\mathcal{F}_{u/s}(k^2)=M_{u/s}\left[(zF'(z))^2\right]\bigg|_{z=\frac{k\rho}{2}}
\end{equation}
In the low momentum regime ($k\rho\ll1$), the dynamical constituent masses are constant, and solution to the gap  equations in the mean field approximation
\bea
\label{gap_eq_u}
    \frac{M_s-m_s}{M_u}&=&2g_K\int \frac{dk^+d^2k_\perp}{(2\pi)^3}\frac{\epsilon(k^+)}{k^+}\mathcal{F}^2_u(k^2)\nonumber\\
    \frac{M_u-m_u}{M_s}&=&2g_K\int \frac{dk^+d^2k_\perp}{(2\pi)^3}\frac{\epsilon(k^+)}{k^+}\mathcal{F}^2_s(k^2)\nonumber\\
\eea
Since the two-flavor quark doublet carries  different masses, both  $\langle\bar{\psi}\psi\rangle$  and $\langle\bar{\psi}\tau^3\psi\rangle$ 
receive contributions from the explicit flavor symmetry breaking. In the mean field approximation, they amount to different scalar condensates
\bea
\label{condensate_u}
    \langle\bar{u}u\rangle&=&-N_cM_u\int\frac{dk^+d^2k_\perp}{(2\pi)^3}\frac{\epsilon(k^+)}{k^+}\mathcal{F}_u(k^2)\nonumber\\
    \langle\bar{s}s\rangle&=&-N_cM_s\int\frac{dk^+d^2k_\perp}{(2\pi)^3}\frac{\epsilon(k^+)}{k^+}\mathcal{F}_s(k^2)
\eea
with 
\bea
\langle\bar{s}s\rangle&=&\frac{1}{2}\left(\langle\bar{\psi}\psi\rangle-\langle\bar{\psi}\tau^3\psi\rangle\right)\nonumber\\
\langle\bar{u}u\rangle&=&\frac{1}{2}\left(\langle\bar{\psi}\psi\rangle+\langle\bar{\psi}\tau^3\psi\rangle\right)
\eea
To accommodate the effective dynamical mass with the light front formalism, we have to analytically continue the Euclidean momentum dependence $k^2$ to the Minkowski space. The argument of the form factor will become $2k^+k^-$.

For simplicity, we first analyze the gap equations and quark condensates
in the zero instanton size limit.  Using the   boost-invariant cut-off $\Lambda$ scheme discussed in  Appendix~\ref{APP_SHARP0},
we have for the gap equations
\begin{widetext}
\bea
    \frac{M_s-m_s}{M_u}&=&
    2g_K\int \frac{dk^+d^2k_\perp}{(2\pi)^3}\frac{\epsilon(k^+)}{k^+}\theta(\Lambda/\sqrt{2}-k^+)\theta(\Lambda/\sqrt{2}-k^-)\bigg|_{k^-=\frac{k_\perp^2+M_u^2}{2k^+}}\nonumber\\
    \frac{M_u-m_u}{M_s}&=&
    2g_K\int \frac{dk^+d^2k_\perp}{(2\pi)^3}\frac{\epsilon(k^+)}{k^+}\theta(\Lambda/\sqrt{2}-k^+)\theta(\Lambda/\sqrt{2}-k^-)\bigg|_{k^-=\frac{k_\perp^2+M_s^2}{2k^+}}
\eea
\end{widetext}
Using the result  in~(\ref{mass_integral}), the solution of the gap equations in zero size limit are
\bea
    \frac{M_s-m_s}{M_u}&=&\frac{g_K\Lambda^2}{2\pi^2}\left[1-\frac{M_u^2}{\Lambda^2}+\frac{M_u^2}{\Lambda^2}\ln\frac{M_u^2}{\Lambda^2}\right]\nonumber\\
    \frac{M_u-m_u}{M_s}&=&\frac{g_K\Lambda^2}{2\pi^2}\left[1-\frac{M_s^2}{\Lambda^2}+\frac{M_s^2}{\Lambda^2}\ln\frac{M_s^2}{\Lambda^2}\right]\nonumber\\
\eea
In the chiral limit $m_u=m_s=0$, the constituent masses are equal. The non-zero solution in the chiral limit exists only for sufficiently strong couplings, larger than 
$g_{K,\mathrm{NJL}}^\mathrm{cr}=\frac{2\pi^2}{\Lambda^2}$. When $g_K$ is less than $g_{K,\mathrm{NJL}}^\mathrm{cr}=\frac{2\pi^2}{\Lambda^2}$, no quark condensate will be formed in the chiral limit. However, as long as the coupling $g_K$ is strong enough,  chiral symmetry is dynamically broken, with a non-vanishing scalar quark condensate, in the chiral limit.
The constituent mass ratio of two quarks $M_s/M_u$ will be controlled by the coupling $g_K$ and the two current quark mass $m_u$, $m_s$.
The quark condensates for two flavors are
\bea
    \frac{\langle\bar{u}u\rangle}{\Lambda^3}&=&-\frac{N_c}{4\pi^2} \left(\frac{M_u}{\Lambda}\right)\left[1-\frac{M_u^2}{\Lambda^2}+\frac{M_u^2}{\Lambda^2}\ln\frac{M_u^2}{\Lambda^2}\right]\nonumber\\
    \frac{\langle\bar{s}s\rangle}{\Lambda^3}&=&-\frac{N_c}{4\pi^2} \left(\frac{M_s}{\Lambda}\right)\left[1-\frac{M_s^2}{\Lambda^2}+\frac{M_s^2}{\Lambda^2}\ln\frac{M_s^2}{\Lambda^2}\right]\nonumber\\
\eea

For a finite instanton size, the integrals in the emergent constituent masses and chiral condensates,
are naturally regulated by the boost invariant cutoffs  given in (\ref{M_cut_off_s}). More specifically, the mass gaps are given by
    \bea
   && \frac{M_s-m_s}{M_u}=\nonumber\\
    &&\frac{4g_K}{\pi^2\rho^2}\int_0^{\infty} dz z\frac{z^3}{z^2+\frac{\rho^2M_u^2}{4}}|z(I_0(z)K_0(z)-I_1(z)K_1(z))'|^4  \nonumber\\
    &&\frac{M_u-m_u}{M_s}=\nonumber\\
    &&\frac{4g_K}{\pi^2\rho^2}\int_0^{\infty} dz z\frac{z^3}{z^2+\frac{\rho^2M_s^2}{4}}|z(I_0(z)K_0(z)-I_1(z)K_1(z))'|^4\nonumber \\   
     \eea
and the scalar condensates are given by
\bea
  && \rho^3\langle\bar{u}u\rangle= \frac{2N_c}{\pi^2}\rho M_u\nonumber\\
  &&\times \int_0^{\infty} dz z\frac{z^3}{z^2+\frac{\rho^2M_u^2}{4}}|z(I_0(z)K_0(z)-I_1(z)K_1(z))'|^2\nonumber\\
   &&\rho^3\langle\bar{s}s\rangle=\frac{2N_c}{\pi^2}\rho M_s\nonumber\\
   &&\times\int_0^{\infty} dz z\frac{z^3}{z^2+\frac{\rho^2M_s^2}{4}}|z(I_0(z)K_0(z)-I_1(z)K_1(z))'|^2\nonumber\\
\eea

In Figs.~\ref{fig_Mus}a and b we show the change of the constituent $u$ quark mass $M_u$, and
the constituent $s$ quark mass $M_s$  with the strange quark mass $m_s$, for different
multi-fermion couplings $g_K$, respectively. The current $u$ quark mass is set to zero. 
In Fig.~\ref{fig_Mus}c we show the ratio $M_s/M_u$ for fixed $g_K/g_K^{cr}$ but
varying $m_u$. In Fig.~\ref{fig_Mus}d we show the ratio $M_s/M_u$ for fixed $m_u=12.7$ MeV and varying $g_K$.
All Figs.~\ref{fig_Mus}a-d are for the zero instanton size limit, but a finite transverse cutoff $\Lambda$. 
Figs.~\ref{fig_Mus}e-h display the same results, for the ILM with a finite instanton size $\rho$.

In Fig.~\ref{fig_uscondRIV}a  we show the change in the ratio of the strange  to non-strange condensates
$\langle\bar ss\rangle/\langle\bar uu\rangle$, with
the ratio of the strange to non-strange current masses $m_s/m_u$, for different $u$ quark masses but fixed coupling
$g_K/g_K^{cr}=1.418$. In Fig.~\ref{fig_uscondRIV}b we show the same but now with fixed $u$ quark mass $m_u=12.7$ MeV and increasing coupling $g_K/g_K^{cr}$.

\section{Kaons from $U$- and $V$-spin: \\ LFWFs and masses}
\label{SEC_KAON}
The general light fron state (\ref{Meson_bound_state}) can be readily adapted to kaons
with $K^{\pm}=u\bar{s}$, $s\bar{u}$ in the $U$-spin sector. In the $V$-spin sector, we have the pseudoscalars $K^0,\bar{K}^0= d\bar{s} ,s\bar{d}$,
 with their scalar counterparts $K_5^0$, $\bar{K}_5^0$. Without loss of generality, we will assume that isospin symmetry still holds. Thus, the result in $U$-spin and $V$-spin will be identical,  $\Phi_{K^\pm}=\Phi_{K^0}=\Phi_{K}$. With this in mind, the LFWFs for the pseudoscalar kaons $K^\pm$,  and their scalar counterparts $K_5$, are given by

    \bea
    \Phi_{K_5}(x,k_\perp,s_1,s_2)&=&\phi_{K_5}(x,k_\perp)\bar{u}_{s_1}(k)\tau^{\pm}v_{s_2}(P-k)\nonumber\\
    \Phi_{K}(x,k_\perp,s_1,s_2)&=&\phi_{K}(x,k^\pm_\perp)\bar{u}_{s_1}(k)i\gamma^5\tau^{\pm}v_{s_2}(P-k)\nonumber\\
    \eea
    where $\tau^{\pm}=\frac{1}{2}\left(\tau^1\pm i\tau^2\right)$  are  the Pauli matrices  for $U$- or $V$-spin.
Again, we have separated the spin independent part of the wave functions, and express the spin part in terms of light front Dirac spinors.

\subsection{Bound state equations for kaons}
In the kaon channels, there are still two quark species, but their masses are different, compared with the isospin sector. The constituent mass of the $s$ quark is heavier than the $u$ and $d$ quarks. However, we can still assume that  isospin symmetry holds for simplicity, i.e.  the induced instanton coupling $G_K$ is the same for $U$-spin and $V$-spin sector. With this in mind, we can generalize the light front Hamiltonian in(\ref{LFHamiltonian}) to the  $U$-spin and $V$-spin sectors. Since the current mass of the $s$ quark is significantly different from that of the $u$ and $d$ quarks, we will treat $M$ as a diagonal matrix $M=\mathrm{diag}(M_u,M_s)$. This will slightly break the $U$-spin in the $K^\pm$ channels, or the $V$-spin symmetry fthe $K^{0}$ and $\bar{K}^0$ channels.

To be consistent with the boost-invariant light front symmetry requirement in the two-body process, the momentum dependent cut-off function $\mathcal{F}(k)$ induced from the finite size effect of the instanton ensemble has to be a function of boost-invariant variables $\frac{k_\perp^2+M_u^2}{x}$ and $\frac{k_\perp^2+M_s^2}{\bar{x}}$ related to the quark kinetic energies. In light of
the substitution (\ref{}), we use 
\begin{widetext}
\bea
    \lim_{P^+\rightarrow\infty}\sqrt{\mathcal{F}_u(k)\mathcal{F}_s(P-k)}\rightarrow \mathcal{F}\left(2P^+P^-=\frac{k^2_\perp+M_u^2}{x}+\frac{k^2_\perp+M_s^2}{\bar{x}}\right)
\eea
As a result,  the bound-state equation for the kaon$^\prime$s ($K^\pm$) is
\begin{equation}
\begin{aligned}
        m_K^2\Phi_K&(x,k_\perp,s_1,s_2)=\left[\frac{k_\perp^2+M_u^2}{x}+\frac{k_\perp^2+M_{s}^2}{\bar{x}}\right]\Phi_K(x,k_\perp,s_1,s_2)\\
        &+\frac{1}{\sqrt{2x\bar{x}}}\sqrt{\mathcal{F}_u(k)\mathcal{F}_s(P-k)}\int_0^1 \frac{dy}{\sqrt{2y\bar{y}}}\int\frac{d^2q_\perp}{(2\pi)^3}\sum_{s,s'}\mathcal{V}_{s,s',s_1,s_2}(q,q',k,k')\Phi_K(y,q_\perp,s,s')\sqrt{\mathcal{F}_u(q)\mathcal{F}_s(P-q)}
\end{aligned}
\end{equation}
The interaction kernel includes the $K$ pseudoscalar meson channel and the $K_5$ scalar meson channel.
\begin{equation}
\begin{aligned}
    \mathcal{V}_{s,s',s_1,s_2}(q,q',k,k')=&    -g_K\bigg[\alpha_{K+}(P^+)\bar{u}_{s_1}(k)\tau^+i\gamma^5v_{s_2}(k')\bar{v}_{s'}(q')\tau^-i\gamma^5u_{s}(q)-\alpha_{K-}(P^+)\bar{u}_{s_1}(k)\tau^+v_{s_2}(k')\bar{v}_{s'}(q')\tau^-u_{s}(q)\\
    &+\alpha_{K+}(P^+)\bar{u}_{s_1}(k)\tau^-i\gamma^5v_{s_2}(k')\bar{v}_{s'}(q')\tau^+i\gamma^5u_{s}(q)-\alpha_{K-}(P^+)\bar{u}_{s_1}(k)\tau^-v_{s_2}(k')\bar{v}_{s'}(q')\tau^+u_{s}(q)\bigg]\\
\end{aligned}
\end{equation} 
where
\begin{equation}
    \alpha_{K\pm}=\left\{1\pm g_K\int\frac{dk^+d^2k_\perp}{(2\pi)^3}\frac{\epsilon(k^+)}{P^+-k^+}\left[\mathcal{F}_u(k)\mathcal{F}_s(P-k)+\mathcal{F}_s(k)\mathcal{F}_u(P-k)\right] \right\}^{-1}
\end{equation}
The interaction kernels for $K^+$ are
\begin{equation}
\begin{aligned}
        &\sum_{s,s'}\mathcal{V}_{s,s',s_1,s_2}(q,q',k,k')\Phi_{K_5}(y,q_\perp,s,s')\\
        =&2\alpha_{K-}(P^+)\mathrm{Tr}\left[(\slashed{q}+M_u)(\slashed{q'}-M_s)\right]\phi_{K_5}(y,q_\perp)\bar{u}_{s_1}(k)\lambda_U^\pm v_{s_2}(k')\\
        =&4\alpha_{K-}(P^+)\left(\frac{q_\perp^2+y^2M^2_s-2y\bar{y}M_uM_s+\bar{y}^2M_u^2}{y\bar{y}}\right)\phi_{K_5}(y,q_\perp)\bar{u}_{s_1}(k)\lambda_U^+ v_{s_2}(k')
\end{aligned}
\end{equation}

\begin{equation}
\begin{aligned}
    &\sum_{s,s'}\mathcal{V}_{s,s',s_1,s_2}(q,q',k,k')\Phi_{K}(y,q_\perp,s,s')\\
    =&-2\alpha_{K+}(P^+)\mathrm{Tr}\left[(\slashed{q}+M_u)(\slashed{q'}+M_s)\right]\phi_{K}(y,q_\perp)\bar{u}_{s_1}(k)i\gamma^5\lambda_U^\pm v_{s_2}(k')\\
    =&-4\alpha_{K+}(P^+)\left(\frac{q_\perp^2+y^2M^2_s+2y\bar{y}M_uM_s+\bar{y}^2M_u^2}{y\bar{y}}\right)\phi_{K}(y,q_\perp)\bar{u}_{s_1}(k)i\gamma^5\lambda_U^+ v_{s_2}(k')
\end{aligned}
\end{equation}
The kernel for $K^-$ follows similarly from the exchange $u\leftrightarrow s$ and $\tau^+\leftrightarrow\tau^-$. Hence,  the bound state equations
\bea
       m_{K_5^+}^2\phi_{K_5^+}(x,k_\perp)=&&\left[\frac{k^2_\perp+M_{u}^2}{x}+\frac{k^2_\perp+M_{s}^2}{\bar{x}}\right]\phi_{K_5^+}(x,k_\perp)\nonumber\\
       &&+\frac{2g_K\alpha_{K-}(P^+)}{\sqrt{x\bar{x}}}\sqrt{\mathcal{F}_u(k)\mathcal{F}_s(P-k)}\int_0^1 \frac{dy}{\sqrt{y\bar{y}}}\int\frac{d^2q_\perp}{(2\pi)^3}\nonumber\\
       &&\times\left(\frac{q_\perp^2+y^2M_{s}^2-2y\bar{y}M_{u}M_{s}+\bar{y}^2M_{u}^2}{y\bar{y}}\right)\phi_{K_5^+}(y,q_\perp)\sqrt{\mathcal{F}_u(q)\mathcal{F}_s(P-q)}
\nonumber\\
       m_{K^+}^2\phi_{K^+}(x,k_\perp)=&&\left[\frac{k^2_\perp+M_{u}^2}{x}+\frac{k^2_\perp+M_{s}^2}{\bar{x}}\right]\phi_{K^+}(x,k_\perp)\nonumber\\
       &&-\frac{2g_K\alpha_{K+}(P^+)}{\sqrt{x\bar{x}}}\sqrt{\mathcal{F}_u(k)\mathcal{F}_s(P-k)}\int_0^1 \frac{dy}{\sqrt{y\bar{y}}}\int\frac{d^2q_\perp}{(2\pi)^3}\phi_{K^+}(y,q_\perp)\nonumber\\        
        &&\times\left(\frac{q_\perp^2+y^2M_{s}^2+2y\bar{y}M_{u}M_{s}+\bar{y}^2M_{u}^2}{y\bar{y}}\right)\sqrt{\mathcal{F}_u(q)\mathcal{F}_s(P-q)}
\eea
The bound state equations for $K^0$ and $\bar{K}^0$,  follow by interchanging $u\leftrightarrow d$.

\subsection{Kaon spectrum}
The bound state masses are solution to the gap-like equation
\begin{equation}
\label{bseq_sol}
    1=\int_0^1dy\int d^2q_\perp\frac{V_X(y,q_\perp)}{y\bar{y}m_X^2-(q_\perp^2+M^2)}\mathcal{F}_u(q)\mathcal{F}_s(P-q)
\end{equation}
The contribution of different quark flavors will change as the momentum fraction change. When $y\sim0$,  most of the hadronic momentum is taken by the $s$ quark. The dynamics of the constituent quark mass will be dominated by the $u$ quark in that region. The mass difference between the two constituents,
 slightly shifts the center of the longitudinal momentum distribution to the lighter constituent, which  tends to carry more momentum fraction. The potential for $K$ and $K_5$ will be
\begin{equation}
    V_X=\begin{cases}
    +\frac{2g_K}{(2\pi)^3}\alpha_{K-}(P^+)\left(\frac{q_\perp^2+y^2M^2_s-2y\bar{y}M_uM_s+\bar{y}^2M_u^2}{y\bar{y}}\right)\ ,~\ \text{scalars}~\ K_5\\
    -\frac{2g_K}{(2\pi)^3}\alpha_{K+}(P^+)\left(\frac{q_\perp^2+y^2M^2_s+2y\bar{y}M_uM_s+\bar{y}^2M_u^2}{y\bar{y}}\right)\ ,~\ \text{pseudo scalars}~\ K
    \end{cases}
\end{equation}

To solve the gap equation (\ref{bseq_sol}) for the kaon spectra, we proceed as in the pion case. More specifically, 
we split the  $k^+$-integral in $\alpha_{K\pm}(P^+)$ to isolate the part carrying the longitudinal  momentum fraction $x$ in the bound state with $P^+$,
from the part  which is UV dominated by  the one-body integral in the gap equation
\begin{equation}
\label{ALPHAPM}
\begin{aligned}
\alpha_{K\pm}(P^+)^{-1}=&1\pm g_K\int\frac{dk^+d^2k_\perp}{(2\pi)^3}\frac{\epsilon(k^+)}{P^+-k^+}\left[\mathcal{F}_u(k)\mathcal{F}_s(P-k)+\mathcal{F}_s(k)\mathcal{F}_u(P-k)\right]\\
=&1\pm\frac{g_K}{(2\pi)^3}\int d^2k_\perp\left[\int_0^1 dx\frac{2}{x}-\int_0^\infty dx\frac{1}{x}+\int_{-\infty}^0 dx\frac{1}{x}\right]\left[\mathcal{F}_u(k)\mathcal{F}_s(P-k)+\mathcal{F}_s(k)\mathcal{F}_u(P-k)\right]\\
=&\begin{cases}
    \frac{1}{2}\left(\frac{m_s}{M_u}+\frac{m_u}{M_s}\right)-\frac{(M_u-M_s)^2}{2MM_s}+\frac{2g_K}{(2\pi)^3}\int d^2k_\perp\int_0^1 dx\frac{2}{x}\left[\mathcal{F}_u(k)\mathcal{F}_s(P-k)+\mathcal{F}_s(k)\mathcal{F}_u(P-k)\right] \\[10pt]
    2-\frac{1}{2}\left(\frac{m_s}{M_u}+\frac{m_u}{M_s}\right)+\frac{(M_u-M_s)^2}{2MM_s}-\frac{2g_K}{(2\pi)^3}\int d^2k_\perp\int_0^1 dx\frac{2}{x}\left[\mathcal{F}_u(k)\mathcal{F}_s(P-k)+\mathcal{F}_s(k)\mathcal{F}_u(P-k)\right]
    \end{cases}
\end{aligned}
\end{equation}
In the last split identity, we made use of the solution of the gap equation in (\ref{gap_eq_u}), namely
\begin{equation}
    1-g_K\int_{u}\frac{dk^+d^2k_\perp}{(2\pi)^3}\frac{\epsilon(k^+)}{k^+}\mathcal{F}^2_u(k)-g_K\int_{s}\frac{dk^+d^2k_\perp}{(2\pi)^3}\frac{\epsilon(k^+)}{k^+}\mathcal{F}^2_s(k)=\frac{1}{2}\left(\frac{m_s}{M_u}+\frac{m_u}{M_s}\right)-\frac{(M_u-M_s)^2}{2MM_s}
\end{equation}

Now we insert (\ref{ALPHAPM}) for  $\alpha_{K\pm}(P^+)$,  back in (\ref{bseq_sol}), and obtain the mass eigenvalue equation for the kaons
\begin{equation}
\begin{aligned}
           0=&
          2-\frac{1}{2}\left(\frac{m_s}{M_u}+\frac{m_u}{M_s}\right)+\frac{(M_u-M_s)^2}{2MM_s}\\
          -&\frac{2g_K}{(2\pi)^3}\int_0^1 dy\int d^2q_\perp\left[\frac{2}{y}-\frac{1}{y\bar{y}}+\frac{m_{K_5}^2-(M_u-M_s)^2}{y\bar{y}m_{K_5}^2-(q_\perp^2+yM_s^2+\bar{y}M_u^2)}\right]\left[\mathcal{F}_u(q)\mathcal{F}_s(P-q)+\mathcal{F}_s(q)\mathcal{F}_u(P-q)\right]\\[10pt]
          =&\frac{(M_u+M_s)^2-M_sm_s-M_um_u}{2M_uM_s}\\
          &-\frac{2g_K}{(2\pi)^3}\int_0^1dy\int d^2q_\perp\left[\frac{m_{K_5}^2-(M_s+M_u)^2}{y\bar{y}m_{K_5}^2-(q_\perp^2+yM_s^2+\bar{y}M_u^2)}\right]\left[\mathcal{F}_u(q)\mathcal{F}_s(P-q)+\mathcal{F}_s(q)\mathcal{F}_u(P-q)\right]
\end{aligned}
\end{equation}
and
\begin{equation}
\begin{aligned}
0=&\frac{1}{2}\left(\frac{m_s}{M_u}+\frac{m_u}{M_s}\right)-\frac{(M_u+M_s)^2}{2MM_s}\\
+&\frac{2g_K}{(2\pi)^3}\int dy\int d^2q_\perp\left[\frac{2}{y}-\frac{1}{y\bar{y}}+\frac{m_{K}^2-(M_s-M_u)^2}{y\bar{y}m_{K}^2-(q_\perp^2+yM_s^2+\bar{y}M_u^2)}\right]\left[\mathcal{F}_u(q)\mathcal{F}_s(P-q)+\mathcal{F}_s(q)\mathcal{F}_u(P-q)\right]\\[10pt]
=&\frac{1}{2}\left(\frac{m_s}{M_u}+\frac{m_u}{M_s}\right)-\frac{(M_u-M_s)^2}{2M_uM_s}\\
&+\frac{2g_K}{(2\pi)^3}\int_0^1dy\int d^2q_\perp\left[\frac{m_{K}^2-(M_u-M_s)^2}{y\bar{y}m_K^2-(q_\perp^2+yM_s^2+\bar{y}M_u^2)}\right]\left[\mathcal{F}_u(q)\mathcal{F}_s(P-q)+\mathcal{F}_s(q)\mathcal{F}_u(P-q)\right]
\end{aligned}
\end{equation}
The boost-invariant cut-off guarantees the cancellation between the integral $\int_0^1dy\frac{2}{y}$ and $\int_0^1dy \frac{1}{y\bar{y}}$. 
The integral
\bea
\label{INTX}
\int_0^1 dx\int d^2k_\perp\frac{1}{x}\left[\mathcal{F}^2_u(k)+\mathcal{F}^2_s(k)-\mathcal{F}_u(k)\mathcal{F}_s(P-k)-\mathcal{F}_s(k)\mathcal{F}_u(P-k))\right]\nonumber\\
\eea
\end{widetext}
in the last line will be dropped, since the difference between these two integrals only depends on the boost-invariant UV cut-off. 
Indeed, as the cutoff scale $1/\rho$ increases, (\ref{INTX}) becomes vanishingly small.

In  leading order of the chiral expansion around the small quark mass, we recover to the GOR rmass relation as a solution
\begin{equation}
    m^2_K=\frac{m_u+m_s}{f^2_K}|\langle \bar{u}u\rangle+\langle \bar{s}s\rangle| +\mathcal{O}(m_u^2,m_s^2,m_um_s)
\end{equation}
with the quark condensate
$$|\langle \bar{u}u\rangle+\langle \bar{s}s\rangle|=\frac{N_c}{2g_K}(M_u+M_s-m_u-m_s)$$ 
following from the gap equation.
The kaon weak decay constant $f_K$  in  leading order of the chiral expansion,  is  described by the quark form factor
\begin{widetext}
\bea
\label{eq:fK}
    f_K=\frac{\sqrt{N_c}M}{\sqrt{2}\pi}\left[
    \int_0^1dy\int_0^\infty dq_\perp^2\left(\frac{1}{q_\perp^2+yM_s^2+\bar{y}M_u^2}\right)
    \left[\mathcal{F}_u(k)\mathcal{F}_s(P-k)+\mathcal{F}_s(k)\mathcal{F}_u(P-k)\right]\right]^{1/2}
\eea
\end{widetext}
where $M=(M_u+M_s)/2$.

In the zero instanton size limit, we can use the transverse cutoff described in Appendix~\ref{APP_SHARP0} to analyze the kaon spectra,
and their dependence on the UV cutoff explicitly. For instance,  the eigenvalue equation for the scalar counterpart of the kaon or $K_5$, is
\begin{widetext}
\bea
\begin{aligned}
    \frac{(M_u+M_s)^2-M_sm_s-M_um_u}{2M_uM_s}
    =
    \frac{g_K}{4\pi^2}\int_0^1dy\int_0^{\Lambda^2} dq^2_\perp\left[\frac{m_{K_5}^2-(M_s+M_u)^2}{y\bar{y}m_{K_5}^2-(q_\perp^2+yM_s^2+\bar{y}M_u^2)}\right]\\
\end{aligned}
\eea
The mass of  $K_5$ is above the  cut-off scale, $m_{K_5}>\Lambda$, and clearly unbound. In contrast, the pseudoscalar kaons bind. Indeed, 
the mass eigenvalue $m_K$  is solution to
\begin{equation}
\label{Kaon_Mass}
\begin{aligned}
    &\frac{1}{2}\left(\frac{m_s}{M_u}+\frac{m_u}{M_s}\right)=\frac{(M_u-M_s)^2}{2M_uM_s}-\frac{g_K}{4\pi^2}\int_{0}^{1}dy\int_0^{\Lambda^2} dq^2_\perp\left[\frac{m_{K}^2-(M_u-M_s)^2}{y\bar{y}m_K^2-(q_\perp^2+yM_s^2+\bar{y}M_u^2)}\right]\\
    =&\frac{(M_u-M_s)^2}{2M_uM_s}+\frac{g_K}{4\pi^2}[m_{K}^2-(M_u-M_s)^2]\int_{0}^{1}dy\ln\left(1+\frac{\Lambda^2}{yM_s^2+\bar{y}M_u^2-y\bar{y}m_K^2}\right)
    \end{aligned}
\end{equation}
 with the condition that $(M_s-M_u)^2\leq m_K^2\leq (M_u+M_s)^2$. 
 The last integration can be carried out explicitly, with the final result for the zero instanton size
 \bea
&&\frac{1}{2}\left(\frac{m_s}{M_u}+\frac{m_u}{M_s}\right)=\frac{(M_u-M_s)^2}{2M_uM_s}\nonumber\\
   &&+\frac{g_K}{4\pi^2}[m_{K}^2-(M_u-M_s)^2]\bigg[\left(\frac{1}{2}+\frac{M_s^2-M_u^2}{2m_K^2}\right)\ln\left(1+\frac{\Lambda^2}{M_s^2}\right)+\left(\frac{1}{2}-\frac{M_s^2-M_u^2}{2m_K^2}\right)\ln\left(1+\frac{\Lambda^2}{M_u^2}\right)\nonumber\\
  &&-\sqrt{\frac{2(M_s^2+M_u^2)-m_K^2}{m_K^2}-\frac{(M_s^2-M_u^2)^2}{m_K^4}}\nonumber\\
  &&\times\left(\tan^{-1}\frac{1+\frac{M_s^2-M_u^2}{m_K^2}}{\sqrt{\frac{2(M_s^2+M_u^2)-m_K^2}{m_K^2}-\frac{(M_s^2-M_u^2)^2}{m_K^4}}}+\tan^{-1}\frac{1-\frac{M_s^2-M_u^2}{m_K^2}}{\sqrt{\frac{2(M_s^2+M_u^2)-m_K^2}{m_K^2}-\frac{(M_s^2-M_u^2)^2}{m_K^4}}}\right)\nonumber\\
    &&+\sqrt{\frac{4\Lambda^2+2(M_s^2+M_u^2)-m_K^2}{m_K^2}-\frac{(M_s^2-M_u^2)^2}{m_K^4}}\nonumber\\
    &&\times\left(\tan^{-1}\frac{1+\frac{M_s^2-M_u^2}{m_K^2}}{\sqrt{\frac{4\Lambda^2+2(M_s^2+M_u^2)-m_K^2}{m_K^2}-\frac{(M_s^2-M_u^2)^2}{m_K^4}}}+\tan^{-1}\frac{1-\frac{M_s^2-M_u^2}{m_K^2}}{\sqrt{\frac{4\Lambda^2+2(M_s^2+M_u^2)-m_K^2}{m_K^2}-\frac{(M_s^2-M_u^2)^2}{m_K^4}}}\right)\bigg]
\eea
\end{widetext}
In the second line of (\ref{Kaon_Mass}), the finiteness of the $q_\perp$-integration requires the inequality $y\bar{y}m_K^2<yM_u^2+\bar{y}M_s^2$ to hold. This inequality restricts the range of the $y$-integration. To ensure the $y$-integration can run over $0$ to $1$ smoothly, the kaon mass is constrained as noted.

\begin{figure*}
\subfloat[\label{fig_3pt1}]{%
  \includegraphics[height=5.5cm,width=.46\linewidth]{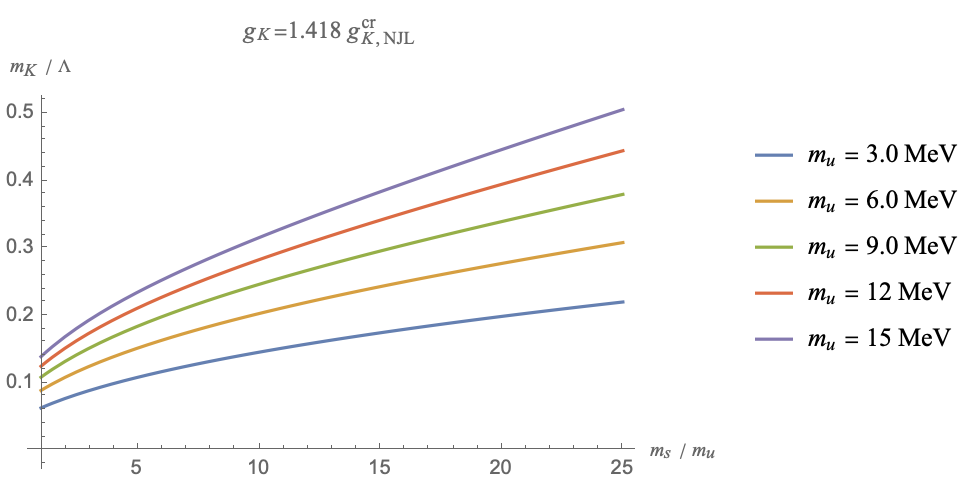}%
}\hfill
\subfloat[\label{fig_3pt6}]{%
  \includegraphics[height=5.5cm,width=.46\linewidth]{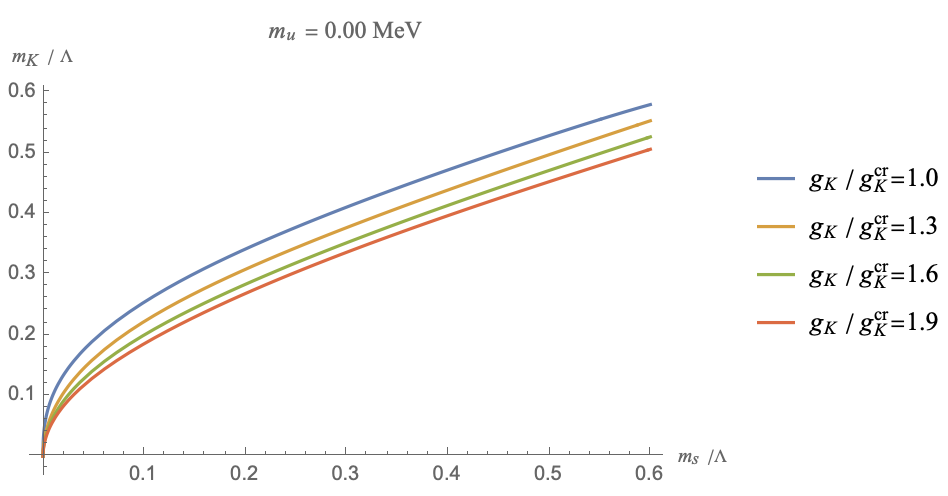}%
}\hfill
\subfloat[\label{fig_3pt1}]{%
  \includegraphics[height=5.5cm,width=.46\linewidth]{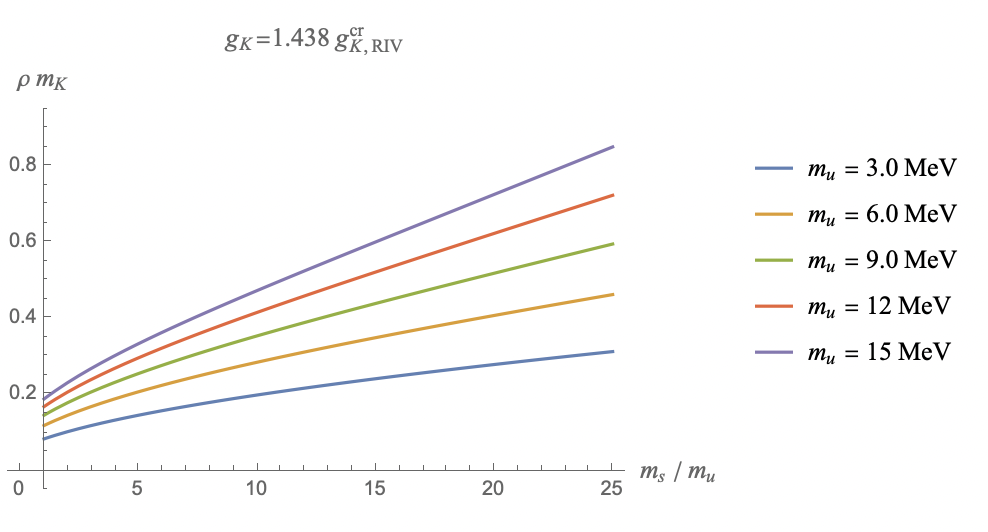}%
}\hfill
\subfloat[\label{fig_3pt6}]{%
  \includegraphics[height=5.5cm,width=.46\linewidth]{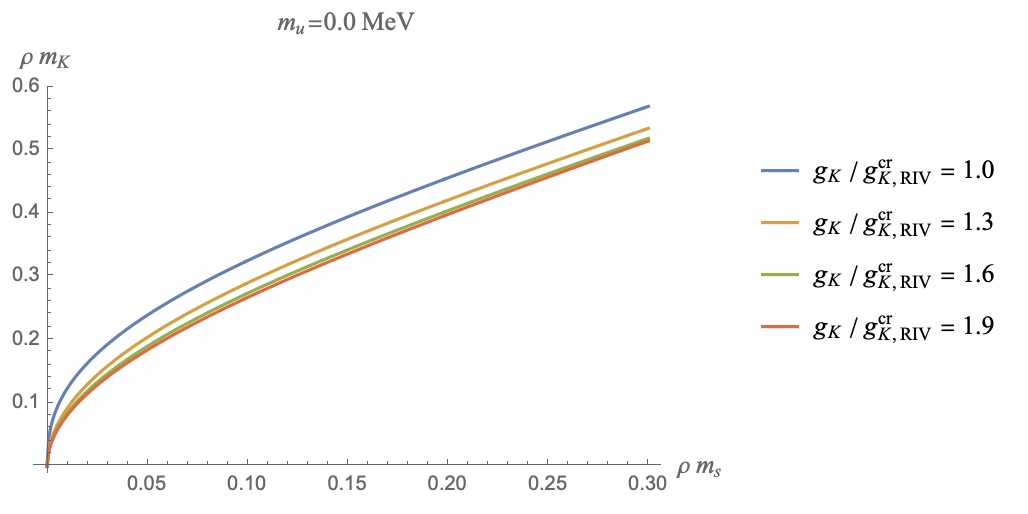}%
}
\caption{
a: The change in the kaon mass versus $m_s/m_u$, for different $m_u$ and fixed $g_K/g_K^{cr}=1.418$, in the zero instanton size;
b: Same as in a for fixed $m_u=0$ MeV and different $g_K/g_K^{cr}$, in the zero instanton size;
c,d: Same as in a,b in the ILM with a finite instanton size.}
\label{fig:mK}
\end{figure*}

We show in~Fig.\ref{fig:mK}, the dependence of $m_K$ on the current quark masses $m_u,m_s=$, as well as the multi-fermion coupling $g_K$.
Although the current quark mass of the strange quark $s$ is substantially larger than that of the $u,d$ quarks, the difference between the two constituent masses $M_u$ and $M_s$ remain small. If we expand Eq.\eqref{Kaon_Mass} in terms of the current mass difference $\Delta M/M$, where $\Delta M=(M_s-M_u)/2$, the squared kaon mass $m_K^2$ depends linearly  on the quark current masses $m_u$ and $m_s$ at leading order, leading the expected GOR relation
\begin{equation}
    m^2_K=\frac{m_u+m_s}{f^2_K}|\langle \bar{u}u\rangle+\langle \bar{s}s\rangle| +\mathcal{O}\left(\Delta M^4\right)
\end{equation}
where $M=(M_u+M_s)/2$, $\Delta M=(M_s-M_u)/2$ and 
\bea
|\langle \bar{u}u\rangle+\langle \bar{s}s\rangle|=\frac{N_c}{2g_K}(2M-m_u-m_s)
\eea
With the first order correction of the constituent mass difference around the chiral limit, the Kaon decay constant  is
\begin{widetext}
\bea
  f_K=\frac{\sqrt{N_c}M}{\sqrt{2}\pi}\ln\left(1+\frac{\Lambda^2}{M^2}\right)
\times \left[1-\frac{\Delta M^2}{M^2}\frac{1+3\Lambda^2/M^2}{6(1+\Lambda^2/M^2)^2\ln\left(1+\frac{\Lambda^2}{M^2}\right)}
+\mathcal{O}\left(\Delta M^4/M^4\right)\right]
\eea
\end{widetext}

In the zero size approximation, the ILM is NJL-like. We can fix its parameters by fixing  the kaon mass, kaon decay constant, and the ratio between the $s$ and $u$ current mass $m_s/m_u=22.7$. As a result, the transverse cut-off is fixed to $\Lambda=720.1$ MeV, slightly higher than  $\rho=0.313$ fm in the ILM. The current quark masses are then $m_u=12.7$ MeV and $m_s=287.2$ MeV,  and the fermionic coupling is  fixed to $g_K=1.418~g_{K,\mathrm{NJL}}^{cr}$. The constituent masses are found to be $M_u=155.8$ MeV and $M_s=466.1$ MeV. The quark condensate will thus be $|\langle \bar{u}u\rangle|^{1/3}=170.7$ MeV and $|\langle \bar{s}s\rangle|^{1/3}=158.4$ MeV. Conversely, by fixing $\Lambda, g_K, m_u/m_s$, the kaon 
 mass and the kaon decay constant are $m^{\mathrm{NJL}}_K=435$ MeV and $f^{\mathrm{NJL}}_K=155$ MeV, respectively.

In Fig.~\ref{fig:mK}a we show the behavior of the kaon mass $m_K/\lambda$ versus the ratio of strange to non-strange current masses
$m_s/m_u$, for fixed coupling $g_K/g_{K,NJL}^{cr}$ and for increasing light  quark mass $m_u$ (bottom to top). In Fig.~\ref{fig:mK}a we show the
same ratio for now fixed $m_u=0$ but increasing fermionic coupling $g_K$ (top to bttom). Figs.~\ref{fig:mK}a,b are for zero size
instantons, but a fixed cutoff $\Lambda$. In Fig.~\ref{fig:mK}c,d we show the same ratios for the ILM with fixed instanton size $\rho$.

In the ILM, the boost invariant cutoff for the kaon channels is
\begin{widetext}
\bea
   \sqrt{\mathcal{F}_u(k)\mathcal{F}_s(P-k)}\rightarrow\ {\cal F}\bigg(\frac{2P^+P^-}{\lambda_K^2}\bigg)
   =\left[(zF'(z))^2\right]\bigg|_{z=\frac{\rho}{2\lambda_K} \sqrt{\frac{k_\perp^2+M_u^2}{x}+\frac{k_\perp^2+M_s^2}{\bar{x}}}}
\eea
\end{widetext}
with  $\lambda_K$ a parameter of order 1.
The mass eigenvalue for $K_5$ is fixed by
\bea
&&\frac{(M_u+M_s)^2-M_sm_s-M_um_u}{2M_uM_s}=\nonumber\\
&&\frac{g_K}{2\pi^2}\left[(M_u+M_s)^2-m_{K}^2\right]\nonumber\\
&&\times\int_0^1dy\int_{\frac{\rho\sqrt{\bar{y}M^2_u+yM^2_s}}{2\lambda_K\sqrt{y\bar{y}}}}^\infty dz\frac{z}{z^2-\frac{\rho^2m_K^2}{4}}(zF'(z))^4\nonumber\\
\eea
Regardless of the chiral limit, the scalar kaon state  $K_5$ cannot be formed, due to a repulsive (positive) potential in this channel.
The mass eigenvalue for the pseudoscalar kaon is 
\bea
\label{kaon_mass_RIV}
&&\frac{1}{2}\left(\frac{m_s}{M_u}+\frac{m_u}{M_s}\right)=\frac{(M_u-M_s)^2}{2M_uM_s}\nonumber\\
&&+\frac{g_K}{2\pi^2}\left[m_{K}^2-(M_u-M_s)^2\right]\nonumber\\
&&\times \int_0^1dy\int_{\frac{\rho\sqrt{\bar{y}M^2_u+yM^2_s}}{2\lambda_K\sqrt{y\bar{y}}}}^\infty dz\frac{z}{z^2-\frac{\rho^2m_K^2}{4}}(zF'(z))^4\nonumber\\
\eea
The solution to (\ref{kaon_mass_RIV}) satisfies the GOR relation in  (\ref{eq:GOR}). The kaon weak decay  constant
$f_K$ is fixed by (\ref{eq:fK}) in the chiral limit, 
\bea
\label{kaon_decay}
    f_K&=&\frac{\sqrt{N_c}M}{\pi}\left[\int_0^1 dx\int_\frac{\rho \sqrt{\bar{x}M^2_u+xM_s^2}}{2\lambda_K\sqrt{x\bar{x}}}^\infty dz z^3(F'(z))^4\right]^{1/2}\nonumber\\
    &\approx&\frac{\sqrt{N_c}M}{\sqrt{2}\pi}\sqrt{\ln\frac{C}{\rho^2 M^2}}+\mathcal{O}(\rho)
\eea
Expanding in small  $\rho$, we have $$f^2_\pi=\frac{1}{2\pi^2}M^2N_c\ln\left(\frac{C}{\rho^2M^2}\right)$$ where $C\approx0.68471$ ia numerical constant.

With the empirical inputs $f_\pi=130.3$ MeV, $m_\pi=135$ MeV and $\rho=1/630~\mathrm{MeV}^{-1}$,  we  fix the ILM parameters in the kaon sector.
 $\lambda_K$ is fixed to be $4.5255$ in the chiral limit. The light current quark mass is $m=30.5$ MeV. The fermionic coupling is $g_S=2.191~g_{S,\mathrm{RIV}}^{cr}$. The mean constituent mass with fixed current mass and fixed coupling, is  $M=266.9$ MeV. The quark condensate is $|\langle \bar{\psi}\psi\rangle|^{1/3}=166.1$ MeV. 

\begin{widetext}
\subsection{Kaon DAs and PDFs}
The positive kaon LFWFs are generically of the form
\bea
    \phi_{K^+}(x,k_\perp)=\frac{1}{\sqrt{2x\bar{x}}}\frac{C_K}{m^2_K-\frac{k_\perp^2+xM_s^2+\bar{x}M_u^2}{x\bar{x}}}
   \sqrt{\mathcal{F}_u(k)\mathcal{F}_s(P-k)}
\eea
where $C_K$ is the normalization constant determined by (\ref{normal}). They enter in the definition of the kaon PDFs in the
leading twist-2 approximation, as defined by
\begin{equation}
\begin{aligned}
       q_{K}(x)=&\int\frac{d\xi^-}{4\pi}e^{-ixP^+\xi^-}\langle K(P)|\bar{\psi}(\xi^-)\gamma^+W(\xi^-,0)\psi(0)|K(P)\rangle\\
       =&\int \frac{ d^2k_\perp}{(2\pi)^3} \sum_{s_1,s_2}\left|\Phi_K(x,k_\perp,s_1,s_2)\right|^2
\end{aligned}
\end{equation}
where the gauge link $W(\xi^-,0)=\mathrm{exp}\left[-ig\int_0^{\xi^-}d\eta^- A^+(\eta^-)\right]$ is to 1. In the case of meson PDFs, quark and antiquark distributions are related by the charge symmetry $$\bar{q}_X(x)=q_{\bar{X}}(x)=q_X(1-x)=\bar{q}_{\bar{X}}(1-x)$$
For charged positive kaons, the $u$ quark PDF is
\bea
    u_{K^+}(x)=&&\frac{1}{4\pi^2}\int_0^{\infty} dk^2_\perp\frac{1}{x\bar{x}}\left|\frac{C_K}{m^2_K-\frac{k_\perp^2+xM_s^2+\bar{x}M_u^2}{x\bar{x}}}\right|^2\nonumber\\
    &&\times \frac{k_\perp^2+x^2M_s^2+2x\bar{x}M_sM_u+\bar{x}^2M_u^2}{x\bar{x}} \mathcal{F}_u(k)\mathcal{F}_s(P-k)\nonumber\\
    =&&\frac{C_K^2}{4\pi^2}\int_0^{\infty} dk^2_\perp\frac{k_\perp^2+x^2M_s^2+2x\bar{x}M_sM_u+\bar{x}^2M_u^2}{[x\bar{x}m^2_K-(k_\perp^2+xM_s^2+\bar{x}M_u^2)]^2}\mathcal{F}_u(k)\mathcal{F}_s(P-k)
\eea
In the ILM, the result is
\begin{equation}
\label{KPILM}
\begin{aligned}
    u_{K^+}(x)
    =&\frac{C_K^2}{2\pi^2}\int_{ \frac{\rho\sqrt{\bar{x}M_u^2+xM_s^2}}{2\lambda_K\sqrt{x\bar{x}}}}^{\infty} dz \frac{z^2-\frac{\rho^2}{4\lambda^2_K}(M_s-M_u)^2}{\left(\frac{\rho^2m_K^2}{4\lambda^2_K}-z^2\right)^2}z^5(F'(z))^4\\
\end{aligned}
\end{equation}
Note that charge symmetry implies 
$$u_{K^+}(x)=\bar{u}_{K^-}(x)=\bar{s}_{K^+}(1-x)=s_{K^-}(1-x)$$
Due to the unequal masses of $u$ and $s$ quark, the heavier $s$ quark will carry more momentum fraction inside the kaon. Thus, the  positive kaon PDF $u_{K^+}(x)$ will tilt to small $x$ region. 

For comparison, (\ref{KPILM}) for zero size instantons, and using the transverse cutoff detailed in Appendix~\ref{APP_SHARP2}, reduces to
\begin{equation}
    u_{K^+}(x)=\frac{C_K^2}{4\pi^2}\theta(x\bar{x})\left[\frac{(m^2_K-(M_s-M_u)^2)x\bar{x}\Lambda^2}{(xM_s^2+\bar{x}M_u^2-x\bar{x}m^2_K)(xM_s^2+\bar{x}M_u^2+\Lambda^2-x\bar{x}m^2_K)}+\ln\left(1+\frac{\Lambda^2}{xM_s^2+\bar{x}M_u^2-x\bar{x}m^2_K}\right)\right]\\
\end{equation}
The normalization constant $C_K$ is related to the effective kaon-quark coupling. In  leading order in an expansion using  $\Delta M/M$ expansion, we have
\bea
    C_K=-\frac{\sqrt{2N_c}M}{f_K}\left(1-\frac{m_K^2\Lambda^2}{M^4(1+\frac{\Lambda^2}{M^2})\ln(1+\frac{\Lambda^2}{M^2})}+\mathcal{O}(\Delta M^2/M^2,m_K^2/M^2)\right)
\eea
In the chiral limit, $C_K$ satisfies  the Goldberger-Treiman relation.
\end{widetext}

\begin{figure*}
\subfloat[\label{fig_3pt1}]{%
  \includegraphics[height=5.5cm,width=.46\linewidth]{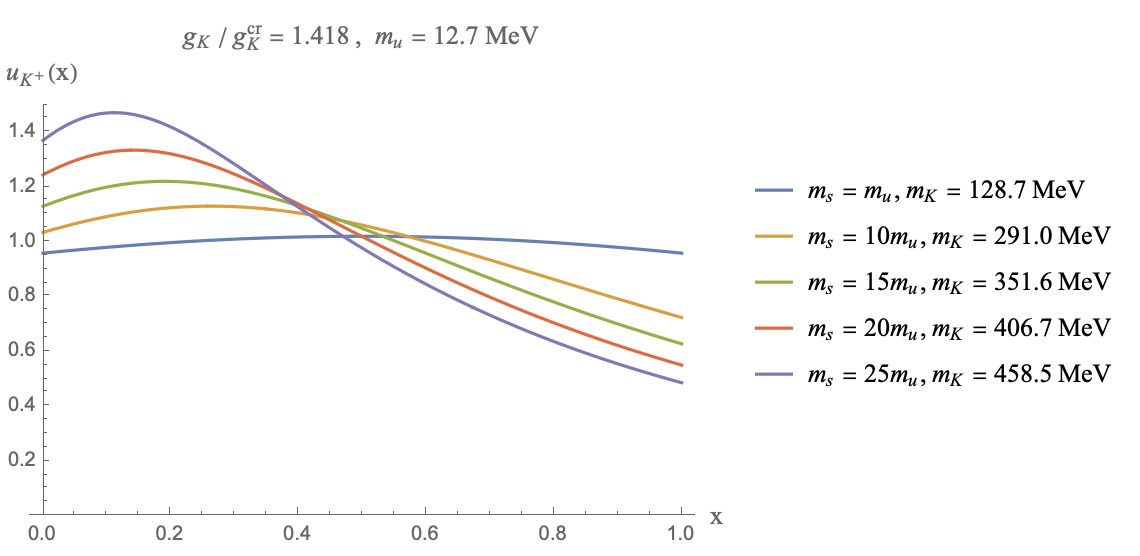}%
}\hfill
\subfloat[\label{fig_3pt1}]{%
  \includegraphics[height=5.5cm,width=.46\linewidth]{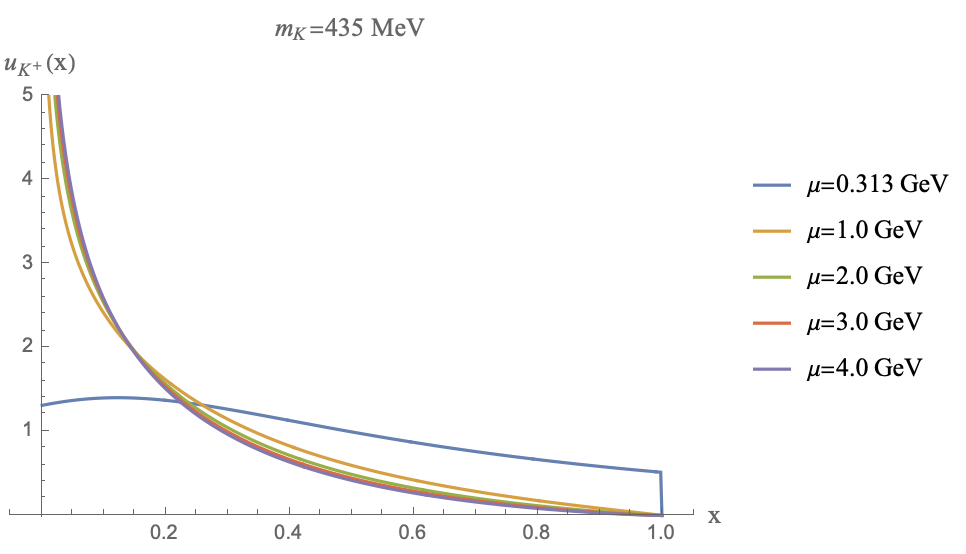}%
  }\hfill
\subfloat[\label{fig_3pt6}]{%
  \includegraphics[height=5.5cm,width=.46\linewidth]{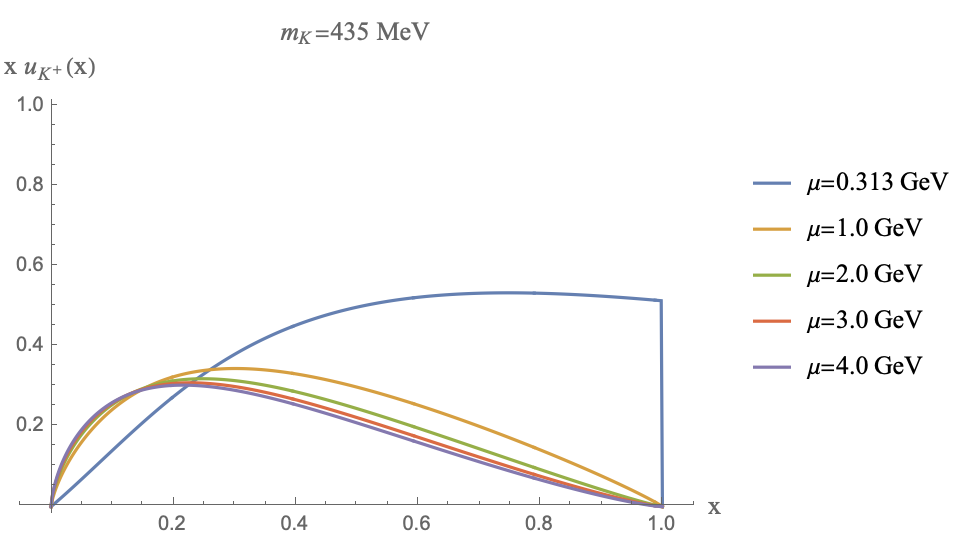}%
}\hfill
\subfloat[\label{fig_3pt6}]{%
  \includegraphics[height=5.5cm,width=.46\linewidth]{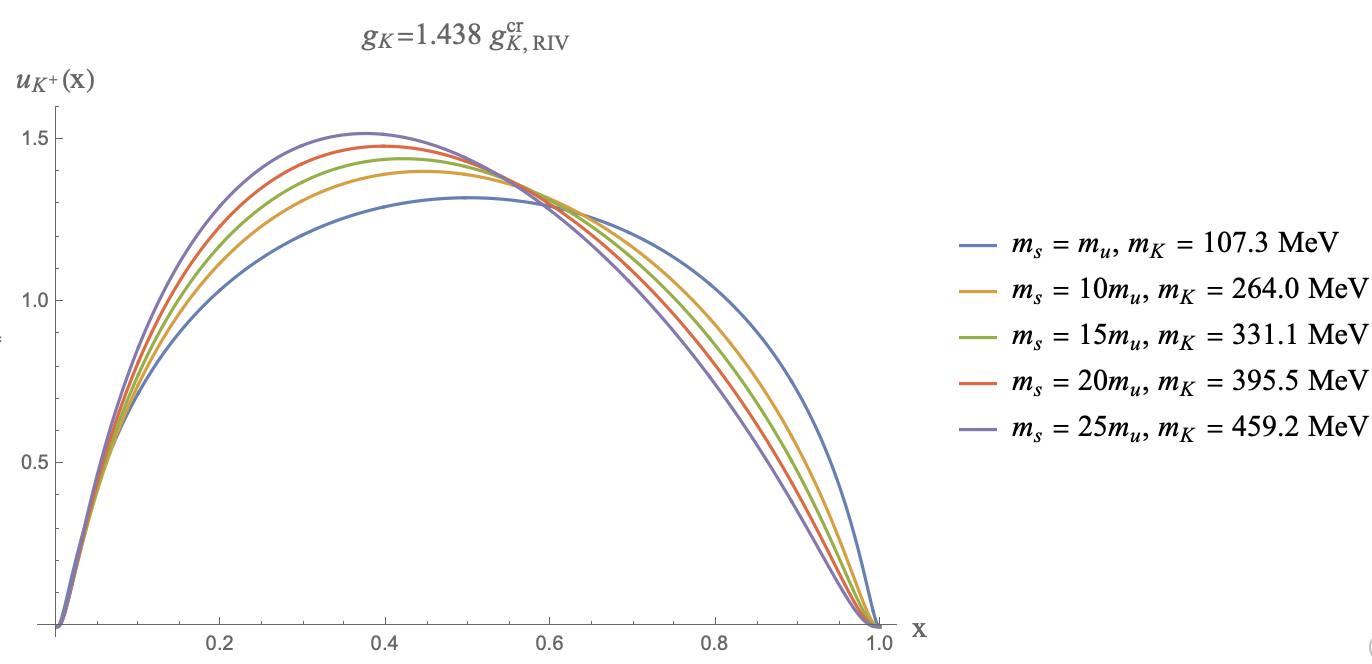}%
}\hfill
\subfloat[\label{fig_3pt6}]{%
  \includegraphics[height=5.5cm,width=.46\linewidth]{KpdfRIV.png}%
}\hfill
\subfloat[\label{fig_3pt6}]{%
  \includegraphics[height=5.5cm,width=.46\linewidth]{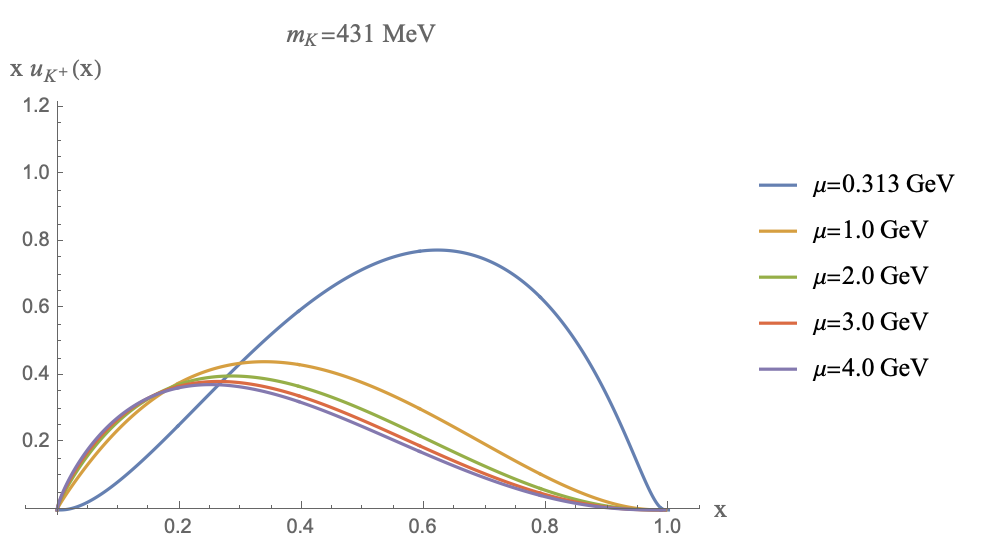}%
}
\caption{
a: The kaon charged PDF versus $x$ in the zero instanton size limit, for fixed coupling and $m_u$ but varying $m_s/m_u$;
b: DGLAP evolution from $\mu_0=0.313$ GeV to $\mu=4$ GeV, of the kaon charged PDF versus $x$ in the zero instanton size limit, 
for fixed kaon mass $m_K=435$ MeV; 
c: Same as b, for the charged kaon momentum distribution;
d,e,f: Same as a,b,c for the charged kaon in the ILM, with a finite instanton size $\rho=0.313$ fm.}
\label{fig_DGLAPKZ}
\end{figure*}

\begin{figure*}
\subfloat[\label{fig_3pt1}]{%
  \includegraphics[height=5.5cm,width=.46\linewidth]{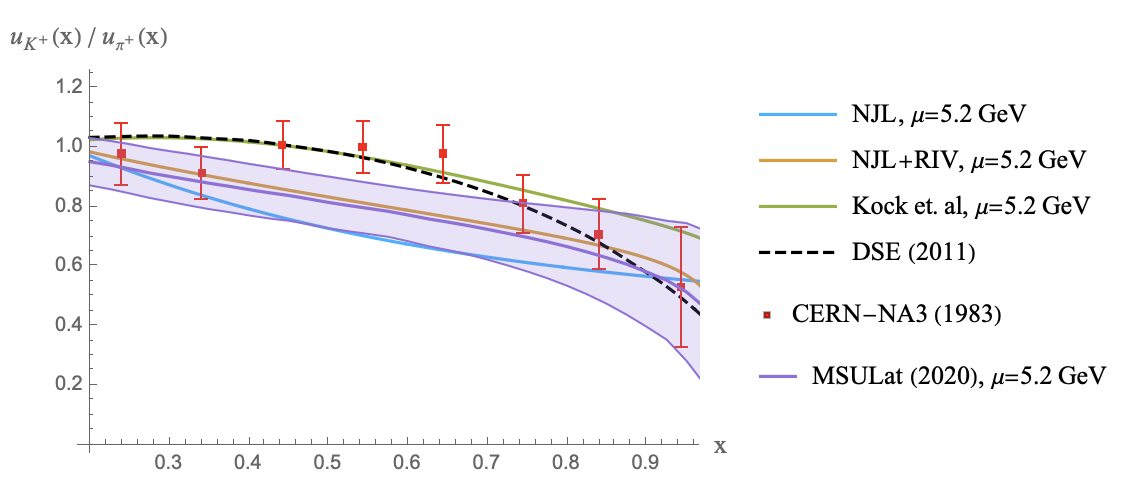}%
}\hfill
\subfloat[\label{fig_3pt6}]{%
  \includegraphics[height=5.5cm,width=.46\linewidth]{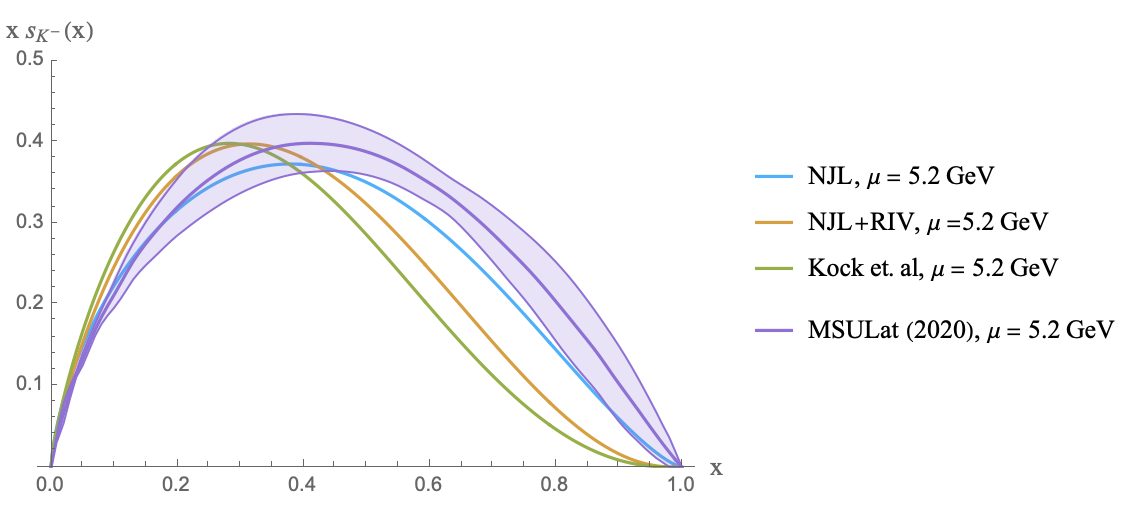}%
}
\caption{
a: The ratio of positive 
charge kaon to pion PDF versus parton $x$, in the zero instanton size limit, with $m_K=435$ MeV and evolved to $\mu=5.2$ GeV (solid-blue),
in the ILM with finite instanton size $\rho=0.318$ fm with $m_K=431$ MeV and evolved to $\mu=5.2$ GeV (solid-orange). The comparison is to the results of
the ILM using the LaMET extraction in~\cite{Kock:2020frx,Kock:2021spt} and evolved to $\mu=5.2$ GeV (solid-green), and the 
results from  the Dyson-Schwinger equation with full Bethe-Salpeter wavefunction~\cite{Nguyen_2011} (dashed-black). The data for the measured ratio (red)
are from~\cite{BADIER1980354} using  muon pair production, using the invariant mass cuts between $4.1$ GeV and $8.5$ GeV to eliminate the meson production on resonance. The recent lattice data MSULat (band-purple) are from~\cite{Huey-Wen_2021}, using the LaMET construction, at $\mu=5.2$ GeV.
b: The same as in b, but for the strange quark momentum distribution for the negative kaon $x s_{K^-}(x)$ versus $x$.
}
\label{fig_DGLAPK}
\end{figure*}

\subsection{Kaon PDF evolution}

As we noted earlier for the pion, the PDFs in the ILM are defined at a low renormalization scale $\mu_0$, below the instanton size resolution $1/\rho\sim 631$ MeV. To compare our result with the experiments in~\cite{Conway:1989fs,Lan:2019vui,Aicher:2010cb}, we evolve the PDFs in (\ref{pionPDF}) and (\ref{pionPDF_RIV}) starting from $\mu_0=313$ MeV.  The evolution will be carried to  $2$ GeV to compare with the available lattice data \cite{Sufian:2020vzb} as well. Since our analysis, reduced the PDFs to their valence quark content, only quark splitting will be considered in the DGLAP evolution. A more realistic evolution, starting from this low renormalization may require the iteration of the instanton interaction as recently suggested in~\cite{Shuryak:2022wtk}. This point will be considered elsewhere.

In Fig.~\ref{fig_DGLAPKZ}a we show the kaon PDF  at a low renormalization poit
$\mu_0=0.313$ GeV. The displayed
results are for fixed fermion coupling $g_K$ and light $u$ quark mass $m_u=12.7$ MeV, with increasing strange quark mass 
(bottom to top starting from the left).  In Figs.~\ref{fig_DGLAPKZ}b,c  we show the DGLAP evolution of the result with $m_K=435$ MeV,
from $\mu=\mu_0=0.313$ GeV to $\mu=$ 4 GeV (bottom to top starting from the left). All Figs.~\ref{fig_DGLAPKZ}a,b,c are carried
with zero size instantons but a finite cutoff $\Lambda$. The analogous results for the ILM with a finite size instanton $\rho=0.313$ fm are shown in 
Figs.~\ref{fig_DGLAPKZ}d,e,f. With increasing strange quark mass, the PDF is skewed in favor of $\bar s$.

In Fig.~\ref{fig_DGLAPK}a we show  our results for 
the  ratio of the $u$ quark PDF in $K^+$ to that in $\pi^+$ versus parton-x,
in comparison to experimental data, and lattice measurements.  The experimental data are from the CERN-NA3 (1983)~\cite{BADIER1980354}
(red dots), which measures the muon pair production within the invariant mass cut between $4.1$ GeV and $8.5$ GeV,  to eliminate the meson production on resonance. 
The lattice measurements are those reported by the MSULat(2020) in~\cite{Huey-Wen_2021} (purple-band), using LaMET at $\mu=5.2$ GeV. Our DGLAP evolved result from $\mu=\mu_0=0.313$ GeV to $\mu=5.2$ GeV, correspond to the zero instanton size limit (solid-blue), the ILM with finite instanton size (solid-orange) and the LaMET in the ILM~\cite{Kock:2020frx,Kock:2021spt} (solid-green).
For additional comparison, we show the results from the Dyson-Schwinger equation with full Bethe-Salpeter wavefunction from~\cite{Nguyen_2011} (black-dashed).

\begin{figure*}
\subfloat[\label{fig_3pt1}]{%
  \includegraphics[height=5.5cm,width=.46\linewidth]{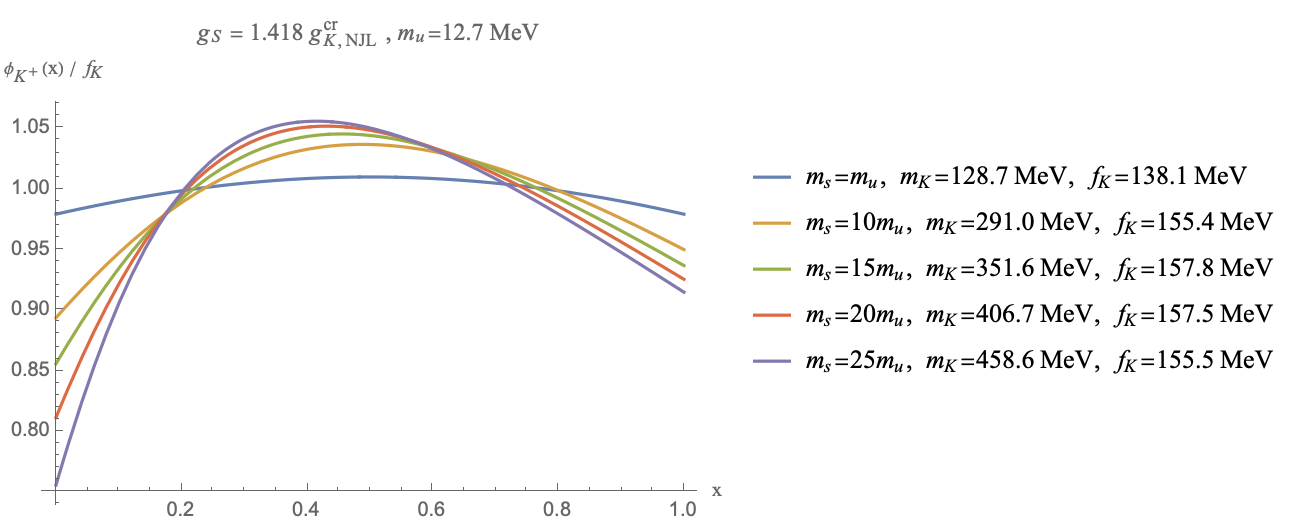}%
}\hfill
\subfloat[\label{fig_3pt6}]{%
  \includegraphics[height=5.5cm,width=.46\linewidth]{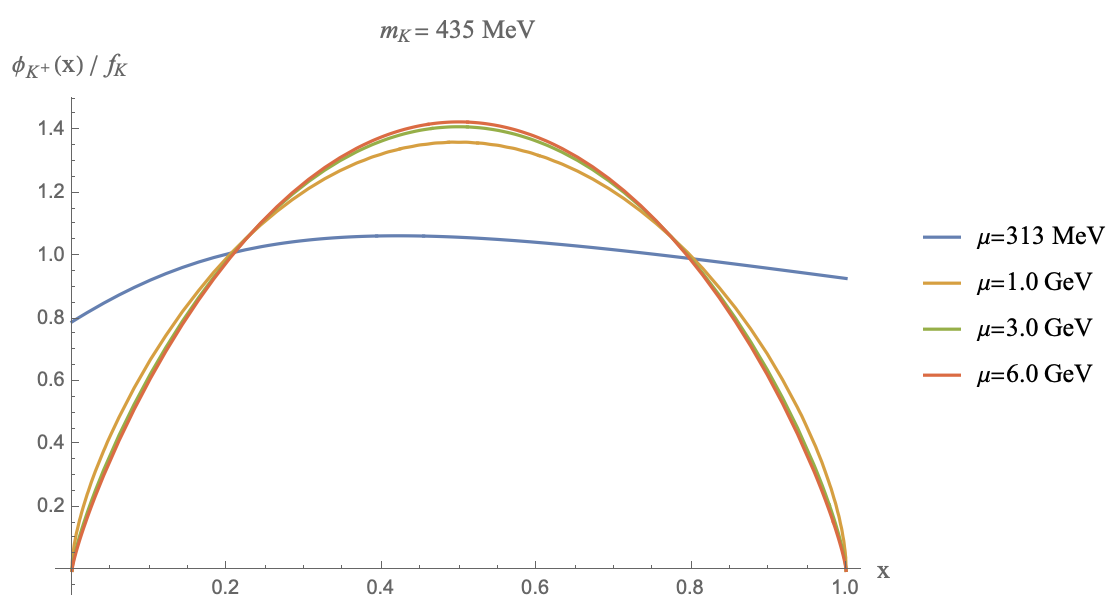}%
}\hfill
\subfloat[\label{fig_3pt6}]{%
  \includegraphics[height=5.5cm,width=.46\linewidth]{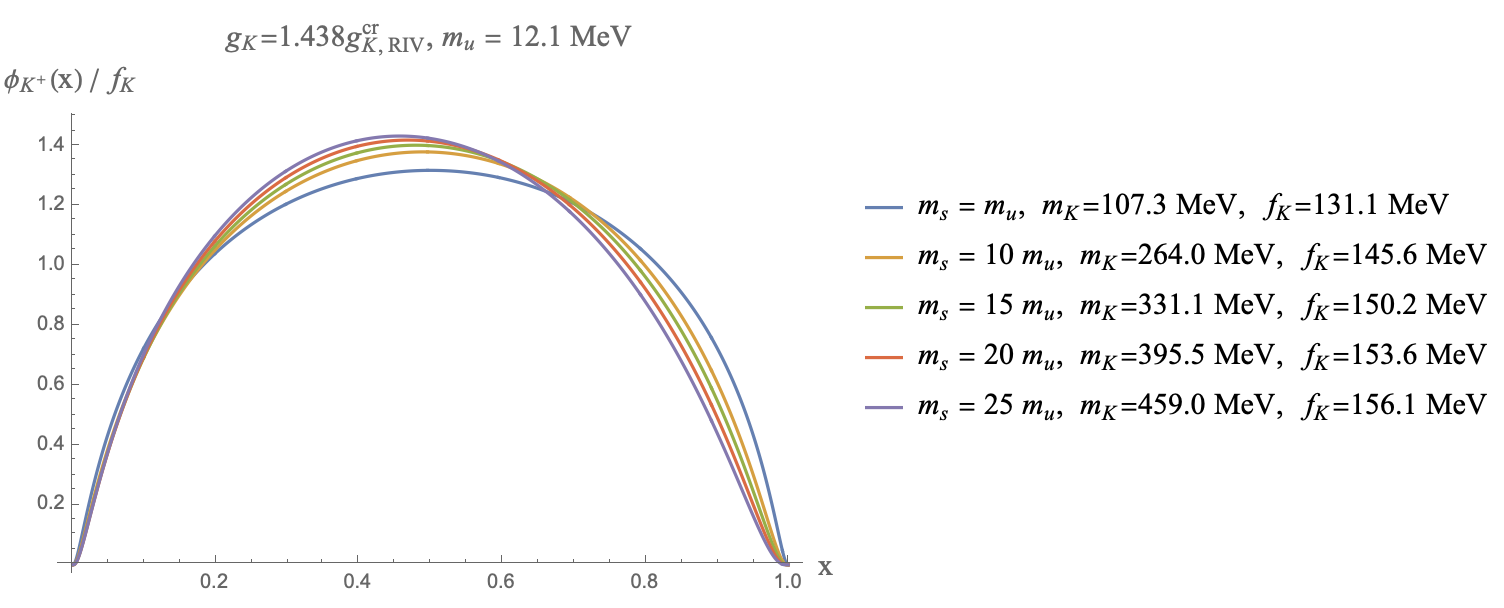}%
}\hfill
\subfloat[\label{fig_3pt6}]{%
  \includegraphics[height=5.5cm,width=.46\linewidth]{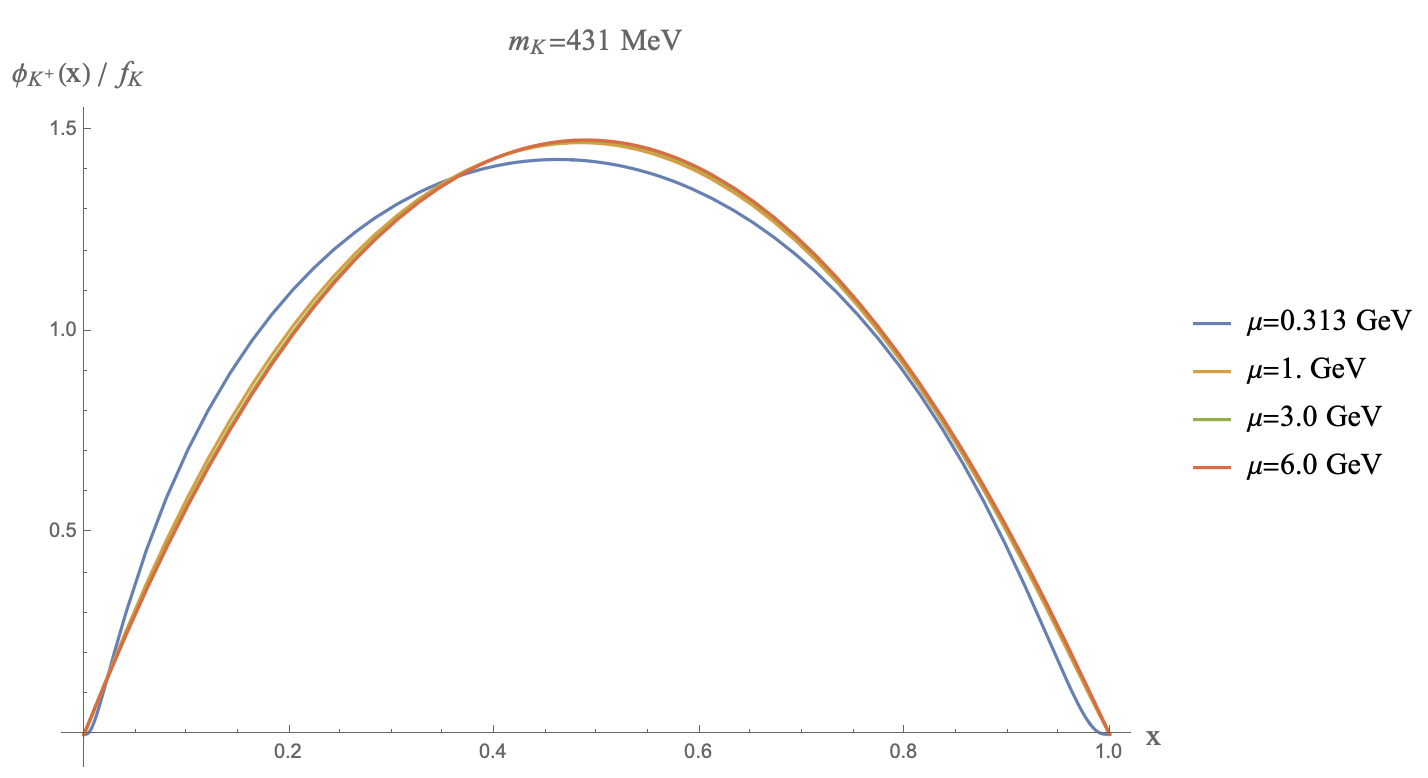}%
}
\caption{
a: The positive charged kaon DA versus parton $x$, for fixed coupling $g_S/g_{K}^{cr}=1.418$ and fixed $m_u=12.7$ MeV,
but different ratios $m_s/m_u$, in the zero instanton size limit;
b: The positive charged kaon DA versus parton $x$, for fixed $m_K=435$ MeV using the ERBL evolution from the initial 
$\mu=0.313$ GeV to $\mu=6$ GeV;
c,d: Same as a,b but in the ILM, for a fixed instanton size $\rho=0.313$ fm.
}
\label{fig_DGLAPKDAX}
\end{figure*}

\subsection{ERBL evolution of kaon DA}
The kaon DAs are also tied to the kaon LFWFs
\begin{widetext}
\begin{equation}
\begin{aligned}
        \phi_{K}(x)=&-i\int\frac{d\xi^-}{2\pi}e^{ixP^+\xi^-}\langle0|\bar{\psi}(0)\gamma^+\gamma^5\frac{\tau^a}{\sqrt{2}}W(0,\xi^-)\psi(\xi^-)|K^a(P)\rangle\\
        =&\frac{\sqrt{N_c}}{\sqrt{2}\pi^2}\int dk^2_\perp \frac{\phi_K(x,k_\perp)}{\sqrt{2x\bar{x}}}[xM_s\mathcal{F}_s(P-k)+\bar{x}M_u\mathcal{F}_u(k)]
\end{aligned}
\end{equation}
with the standard normalization
\begin{equation}
    \langle0|\bar{\psi}\gamma^+\gamma^5\frac{\tau^a}{\sqrt{2}}\psi|K^a(P)\rangle=if_K P^+
\end{equation}
In the ILM,  the kaon DA is
\begin{equation}
\begin{aligned}
        \phi_{K}(x)&=\frac{\sqrt{N_c}(xM_s+\bar{x}M_u)}{2\sqrt{2}\pi^2}\int_0^{\infty} dk^2_\perp \frac{C_{K}}{x\bar{x}m^2_K-(k_\perp^2+xM_s^2+\bar{x}M_u^2)}\mathcal{F}_u(k)\mathcal{F}_s(P-k)\\
\end{aligned}
\end{equation}
From the normalization condition of the kaon DA, we can extract the kaon mass dependence of the kaon weak decay constant.
\bea
   \frac{ f_K(m_K)}{f_K}=&&\Bigg[
\frac{\int_0^1dx\int_0^\infty dk_\perp^2\left(\frac{1}{k_\perp^2+\bar{x}M_u^2+xM_s^2-x\bar{x}m_K^2}\right)\mathcal{F}_u(k)\mathcal{F}_s(P-k)}{\left(\int_0^1dx\int_0^\infty dk_\perp^2\frac{(k_\perp^2+\bar{x}M_u^2+xM_s^2)}{(k_\perp^2+\bar{x}M_u^2+xM_s^2-x\bar{x}m_K^2)^2}\mathcal{F}_u(k)\mathcal{F}_s(P-k)\right)^{1/2}}\Bigg]\nonumber\\
&&\times \left[\int_0^1dx\int_0^\infty dk_\perp^2\left(\frac{1}{k_\perp^2+\bar{x}M_u^2+xM_s^2}\right)\mathcal{F}_u(k)\mathcal{F}_s(P-k)\right]^{-\frac 12}
\eea
with $f_K=f_\pi$ given  in (\ref{eq:fpi}).
In the ILM, the kaon DA is
\bea
\label{kaonDARIV}
        \phi_{K^+}(x)=&&\frac{\sqrt{N_c}(xM_s+\bar{x}M_u)}{\sqrt{2}\pi^2}
      C_{K}\int_{ \frac{\rho\sqrt{\bar{x}M_u^2+xM_s^2}}{2\lambda_K\sqrt{x\bar{x}}}}^{\infty} dz  \frac{z^5}{\frac{\rho^2m^2_K}{4\lambda_K^2}-z^2} (F'(z))^4
\eea
Its evolution starting from $\mu_0=313$ MeV, will also follows from (\ref{eq:ERBL}) in the ERBL regime.

For completeness, we note that in the zero instanton size limit, the form factor will reduce to the transverse cut-off discussed in Appendix~\ref{AP_TRANS}, with
the result for the kaon DA as
\bea
\label{kaonDA}
        \phi_{K^+}(x)&=&\frac{\sqrt{N_c}(xM_s+\bar{x}M_u)}{2\sqrt{2}\pi^2}\theta(x\bar{x})\int_0^{\Lambda^2} dk^2_\perp \frac{C_{K}}{x\bar{x}m^2_{K}-(k_\perp^2+xM_s^2+\bar{x}M_u^2)}\nonumber\\
        &=&-\frac{\sqrt{N_c}(xM_s+\bar{x}M_u)C_{K}}{2\sqrt{2}\pi^2}\theta(x\bar{x})\ln\left(1+\frac{\Lambda^2}{xM_s^2+\bar{x}M_u^2-x\bar{x}m_K^2}\right)
\eea
with $\phi_{K^-}(x)=\phi_{K^+}(1-x)$. The Kaon decay constant $f_K$ simplifies to
\bea
    f_K=C_{K}\frac{\sqrt{N_c}M}{2\sqrt{2}\pi^2}\int_0^1dx\frac{xM_s+\bar{x}M_u}{M}
\ln\left(1+\frac{\Lambda^2}{xM_s^2+\bar{x}M_u^2-x\bar{x}m_K^2}\right)
\eea
\end{widetext}

\begin{figure*}
    \centering
   \includegraphics[height=5.5cm,width=.56\linewidth]{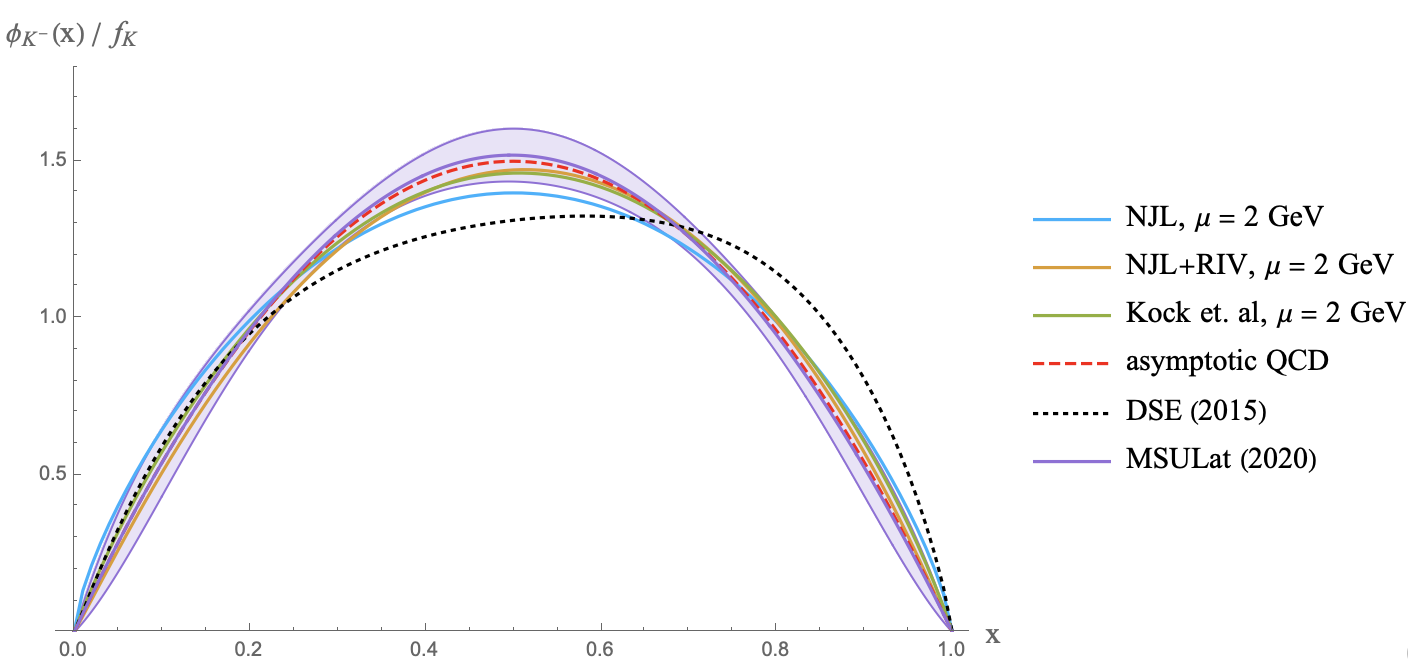}
    \caption{Evolution of the negative kaon DA  in~\eqref{kaonDA} with $m_K=435$ MeV,  from $\mu=0.313$ GeV to $\mu=2$ GeV  in the zero instanton size limit (solid-blue), and in the ILM with a finite instanton size in~\eqref{kaonDARIV} with $m_K=431$ MeV (solid-orange). 
    The results are compared to those obtained also from the ILM (solid-green), using the
    LaMET construction in~\cite{Kock:2020frx,Kock:2021spt}, also evolved to $\mu=2$ GeV. The results using the 
    Dyson-Schwinger equation with Bethe-Salpeter wavefunctions (dashed-black) are  from~\cite{Shi2015}.
    The QCD asymptotic result of $6x\bar x$ (dashed-red) is from\cite{EFREMOV1980245}. The recent lattice data MSULat(2020) (purple-band) are from 
    \cite{Zhang2020} using the LaMET construction.}
\label{KDA_EXPTXX}
\end{figure*}

In Fig.~\ref{fig_DGLAPKDAX}a we show the kaon DA at the low renormalization point $\mu_0=0.313$ MeV, 
with increasing quark mass (top to bottom from the left). The results are for instantons of zero size, a fixed
cutoff $\Lambda$, a fixed fermion coupling $g_S$, and a light quark mass $m_u=12.7$ MeV. In Fig.~\ref{fig_DGLAPKDAX}b
the same results for $m_K=435$ MeV are evolved from $\mu=\mu_0=0.313$ GeV to $\mu=6$ GeV ((top to bottom from the left).
Figs.~\ref{fig_DGLAPKDAX}c,d are the results for the ILM with a finite instanton size, with a slight difference in the input parameters. 

In Fig.~\ref{KDA_EXPTXX} our results for the kaon DA are compared to the recent lattice results MSULat (2020)~\cite{Zhang2020}
using LaMET (purple-band). Our results  are for instantons of zero size evolved to $\mu=2$ GeV (blue-solid), the ILM with finite size instantons
(orange-solid) and the ILM using the LaMET construction~\cite{Kock:2020frx,Kock:2021spt} (solid-green).  For further comparison, we show the asymptotic QCD 
result of $6x\bar x$~\cite{EFREMOV1980245}  (dashed-red), and the results from the  Dyson-Schwinger equation with Bethe-Salpeter wavefunctions~\cite{Shi2015}
(dashed-black).

\section{Conclusions}
\label{SEC_CONC}
We presented a comprehensive analysis of the spontaneous breaking of chiral symmetry on the light front,
in the context of the ILM with induced non-local multi-fermion interactions. The new element in the light
front analysis, is the splitting of the effective fermion fields into good plus bad components, with the latter
non-propagating in the light front direction. It is a constraint, that once eliminated generates additional 
multi-fermion interactions between solely the good fermionic components, in the emergent multi-flavor
and non-local $^\prime$t Hooft interaction.  This construction generalizes the original arguments 
presented in~\cite{Bentz:1999gx,Itakura:2000te,Naito_2004} using local NJL interactions. The non-locality
is important in the characterization of the partonic distributions.

In the mean field approximation (leading contribution in $1/N_c$), the spontaneous 
breaking of chiral symmetry parallels that in the rest frame. A running constituent mass and a chiral condensates
are generated, that are identical to the ones derived in the rest frame, thanks to boost invariance. More importantly,
we have shown that the pion and kaon DAs and PDFs derived on the light front in the mean field approximation, 
are all in agreement with those established in the ILM using the LaMET construction. 

The results we presented for the pion and kaon partonic distributions are all evaluated at 
a low factorization scale $\mu_0<1/\rho\sim $ 630 MeV.  A comparison with experimental data 
and lattice results at larger scales, requires evolution. For simplicity, we have assumed that 
factorization holds at this relatively low scale, and used perturbative QCD evolution. Good agreements with the existing data
for the pion and kaon were found.

The present analysis of the pion and kaon parton distributions relies on the mean-field approximation in the ILM.
It is the leading contribution in a $1/N_c$ book-keeping analysis,  that can be improved a priori. Clearly, the present analysis
can be extended to study the formation of light diquarks on the light front, as well the nucleon and delta baryons. 

A major goal of the upcoming physics at the electron ion collider (EIC) is to understand the partonic composition
of nucleons and nuclei, as they enter in their composition of mass and spin. The present analysis shows that for
pions and kaons, most of their composition follows from the QCD vacuum. At low resolution, it is mostly due to the
emerging multi-fermion $^\prime$t Hooft interactions induced by the light quark zero modes as captured by the ILM. 
The pion and kaon longitudinal parton distributions are sensitive to the nature of the quark zero modes in the vacuum.

contrary to common lore, on the light front the vacuum is anythingh but trivial. Its main effect is to induce the spontaneous
breaking of chiral symmetry with a running constituent quark mass for the valence partons. It also, induces nonperturbative
multi-fermion interactions among the flying leading partons, and a scalar chiral condensate much like in the rest frame.
How this analysis in the context of the ILM, squares with the recently suggested superfluid at zero x-parton~\cite{Ji:2020baz},
would be of future interest.

\vskip 1cm
{\bf Acknowledgements}
This work is supported by the Office of Science, U.S. Department of Energy under Contract No. DE-FG-88ER40388.

\appendix 

\section{Conventions used on the LF}
\label{Appx:LFspinor}
Throughout this paper, our conventions of the light front frame follows Kogut-Soper convention based on the Weyl chiral basis of the gamma matrices
\begin{eqnarray}
&\gamma^0=\begin{pmatrix}
0 & \mathds{1} \\
\mathds{1} & 0 \\
\end{pmatrix} ~\
&\gamma^{i}=\begin{pmatrix}
0 & \sigma^i \\
-\sigma^i  & 0 \\
\end{pmatrix}
\end{eqnarray}
The light front components are normalized to be
\begin{equation}
\gamma^\pm=\frac{\gamma^0\pm \gamma^3}{\sqrt{2}}
\end{equation}
with light front projection defined 
\begin{equation}
\mathcal{P}_+=\frac{1}{2}\gamma^-\gamma^+=\begin{pmatrix}
1 & 0 & 0 & 0 \\
0 & 0 & 0 & 0 \\
0 & 0 & 0 & 0 \\
0 & 0 & 0 & 1 \\
\end{pmatrix}
\end{equation}
and $\mathcal{P}_-=1-\mathcal{P}_+$
The free LF spinors for the quarks and anti-quarks are 
\begin{equation}
u_s(p)=\frac{1}{\sqrt{\sqrt{2}p^+}}\left(\slashed{p}+M\right)\begin{pmatrix}\chi_s \\ \chi_s \end{pmatrix}
\end{equation}
and
\begin{equation}
v_s(p)=\frac{1}{\sqrt{\sqrt{2}p^+}}\left(\slashed{p}-M\right)\begin{pmatrix}2s\chi_s \\ -2s\chi_s \end{pmatrix}
\end{equation}
respectively, 
with $\chi_s$  a 2-spinor with a spin pointing in the $z$-direction, and $M$  the constituent quark mass.

To denote the spin states in the creation of a quark-anti-quark pair, the matrix elements with different spin states can be written as a matrix with $[s_1s_2]$ entries,
\begin{equation}
   \bar{u}_{s_1}(k)v_{s_2}(P-k)= \frac{1}{\sqrt{x\bar{x}}}\begin{pmatrix}
    k_L  & M(\bar{x}-x) \\
     M(\bar{x}-x)  & -k_R
    \end{pmatrix}
\end{equation}
for the scalar, and
\begin{equation}
   \bar{u}_{s_1}(k)i\gamma^5v_{s_2}(P-k)=
        \frac{i}{\sqrt{x\bar{x}}}\begin{pmatrix}
     -k_L & M \\
     -M & -k_R
    \end{pmatrix}
\end{equation}
for the pseudoscalar,
with $k_{L/R}=k^1\mp ik^2$ and $x=k^+/P^+$ the quark momentum fraction on the LF.  Similarly, for the quark-quark pair, 
the spin matrix elements with entries  $[s' s]$ are
\begin{widetext}
\begin{equation}
   \bar{u}_{s'}(p)u_{s}(k)= \sqrt{p^+k^+}\begin{pmatrix}
    M(\frac{1}{p^+}+\frac{1}{k^+}) & \frac{p_L}{p^+}-\frac{k_L}{k^+}\\
    \frac{k_R}{k^+}-\frac{p_R}{p^+} & M(\frac{1}{p^+}+\frac{1}{k^+}) 
    \end{pmatrix}
\end{equation}

\begin{equation}
   \bar{u}_{s'}(p)\gamma^5u_{s}(k)= \sqrt{p^+k^+}\begin{pmatrix}
    M(\frac{1}{p^+}-\frac{1}{k^+}) & \frac{k_L}{k^+}-\frac{p_L}{p^+}\\
    \frac{k_R}{k^+}-\frac{p_R}{p^+} & -M(\frac{1}{p^+}-\frac{1}{k^+}) 
    \end{pmatrix}
\end{equation}

\begin{equation}
   \bar{v}_{s}(k)v_{s'}(p)= \sqrt{p^+k^+}\begin{pmatrix}
    -M(\frac{1}{p^+}+\frac{1}{k^+}) & \frac{k_L}{k^+}-\frac{p_L}{p^+}\\
    \frac{p_R}{p^+}-\frac{k_R}{k^+} & -M(\frac{1}{p^+}+\frac{1}{k^+}) 
    \end{pmatrix}
\end{equation}

\begin{equation}
   \bar{v}_{s}(k)\gamma^5v_{s'}(p)= \sqrt{p^+k^+}\begin{pmatrix}
    -M(\frac{1}{p^+}-\frac{1}{k^+}) & \frac{p_L}{p^+}-\frac{k_L}{k^+}\\
    \frac{p_R}{p^+}-\frac{k_R}{k^+} & M(\frac{1}{p^+}-\frac{1}{k^+}) 
    \end{pmatrix}
\end{equation}

The lowest Fock states for the scalar $\sigma$ and pseudoscalar $\pi$ are  explicitly
\begin{equation}
\begin{aligned}
   |\sigma,P\rangle=&\frac{1}{\sqrt{N_c}}\int_0^1 \frac{dx}{\sqrt{2x\bar{x}}}\int\frac{d^2k_\perp}{(2\pi)^3}\bigg\{\phi^0_\sigma(x,k_\perp)\left[b^\dagger_{\uparrow}(k) c^\dagger_{\downarrow}(P-k)+b^\dagger_{\downarrow}(k) c^\dagger_{\uparrow}(P-k)\right]\\
    &+\phi^{+1}_\sigma(x,k_\perp)b^\dagger_{\uparrow}(k) c^\dagger_{\uparrow}(P-k)+\phi^{-1}_\sigma(x,k_\perp)b^\dagger_{\downarrow}(k) c^\dagger_{\downarrow}(P-k)\bigg\}|0\rangle
\end{aligned}
\end{equation}
\begin{equation}
\begin{aligned}
    |\pi,P\rangle=&\frac{1}{\sqrt{N_c}}\int_0^1 \frac{dx}{\sqrt{2x\bar{x}}}\int\frac{d^2k_\perp}{(2\pi)^3}\bigg\{\phi^0_\pi(x,k_\perp)\left[b^\dagger_{\uparrow}(k) c^\dagger_{\downarrow}(P-k)-b^\dagger_{\downarrow}(k)c^\dagger_{\uparrow}(P-k)\right]\\
    &+\phi^{+1}_\pi(x,k_\perp)b^\dagger_{\uparrow}(k) c^\dagger_{\uparrow}(P-k)+\phi^{-1}_\pi(x,k_\perp)b^\dagger_{\downarrow}(k)c^\dagger_{\downarrow}(P-k)\bigg\}|0\rangle
\end{aligned}
\end{equation}
with the light front normalizations subsumed, provided that
\begin{equation}
    \int_0^1 dx\int\frac{d^2k_\perp}{(2\pi)^3}~|\phi^0_\sigma|^2+|\phi^{+1}_\sigma|^2+|\phi^{-1}_\sigma|^2=1
\end{equation}
and
\begin{equation}
    \int_0^1 dx\int\frac{d^2k_\perp}{(2\pi)^3}~|\phi^0_\pi|^2+|\phi^{+1}_\pi|^2+|\phi^{-1}_\pi|^2=1
\end{equation}
Each spin-wavefunction can also be redefined, to extract out the corresponding normalization of their spin state, with
 \begin{eqnarray}
\phi^0_\sigma=\frac{M(\bar{x}-x)}{\sqrt{x\bar{x}}}\phi_\sigma; &\ ~\ \phi^{+1}_\sigma=\frac{k_L}{\sqrt{x\bar{x}}}\phi_\sigma; &\ ~\ \phi^{-1}_\sigma=\frac{-k_R}{\sqrt{x\bar{x}}}\phi_\sigma
 \end{eqnarray}
 and
  \begin{eqnarray}
 \phi^0_\pi=\frac{iM}{\sqrt{x\bar{x}}}\phi_\pi; &\ ~\ \phi^{+1}_\pi=\frac{-ik_L}{\sqrt{x\bar{x}}}\phi_\pi; &\ ~\ \phi^{-1}_\pi=\frac{-ik_R}{\sqrt{x\bar{x}}}\phi_\pi
 \end{eqnarray}

\section{Two-body t$^\prime$Hooft Hamiltonian on the light front}
\label{APP_HLF}
The general form of the two-body bound state equation can be written as
\begin{equation}
\int\frac{dq^{+}d^2q_\perp}{(2\pi)^3}\hat{H}_{s_1s_2s_1's_2'}(k,P-k,q,P-q)\Phi_X(q,P-q,s'_1,s'_2)=m^2_X\Phi_X(k,P-k,s_1,s_2)
\end{equation}
where the light front Hamiltonian in the momentum space is
\bea
&&\hat{H}_{s_1s_2s_1^\prime s_2^\prime }(k,P-k,k^\prime ,P-k^\prime )=\nonumber\\
&&\left[\frac{k^2_\perp+M^2}{x}+\frac{k^2_\perp+M^2}{\bar{x}}\right](2\pi)^3\delta^3_+(k-k^\prime )\
+\frac{1}{\sqrt{2x\bar{x}}}\mathcal{V}_{s_1,s_2,s^\prime _1,s^\prime _2}(k,P-k,k^\prime ,P-k^\prime )\frac{1}{\sqrt{2y\bar{y}}P^+}
\eea
for a pair of indentical constituents. The labels $1,2$ refer to the two constituents. 
 To address the s-channel and t-channel on equal footing, we use the Fierzed re-arranged instanton induced t$^\prime$Hooft interaction
\bea
\label{eqn:effective_kernel}
\mathcal{L}_{\rm tHooft}=\frac{G}{8(N_c^2-1)}\bigg(&&\frac{N_c^2-1}{N_c^2}\left[(\bar{\psi}\psi)^2-(\bar{\psi}i\gamma^5\psi)^2-(\bar{\psi}\tau^a\psi)^2+(\bar{\psi}i\gamma^5\tau^a\psi)^2\right]\nonumber\\
      &&+\frac{N_c-2}{4N_c}\left[(\bar{\psi}\lambda^\alpha\psi)^2-(\bar{\psi}i\gamma^5\lambda^\alpha\psi)^2-(\bar{\psi}\lambda^\alpha\tau^a\psi)^2+(\bar{\psi}i\gamma^5\tau^a\lambda^\alpha\psi)^2\right]\nonumber\\
      &&+\frac{1}{8}(\bar{\psi}\sigma^{\mu\nu}\lambda^\alpha\psi)^2-\frac{1}{8}(\bar{\psi}\sigma^{\mu\nu}\lambda^\alpha\tau^a\psi)^2\bigg) 
\eea
\\
\\
{\bf s-channel:}
\\
The interaction in the s-channel  follows solely from the first contribution in (\ref{eqn:effective_kernel}),
through the substitution $(\bar{\psi}\psi)^2\rightarrow\bar{u}_{1^\prime} v_{2^\prime}\bar{v}_2u_1$, 
\begin{equation}
\begin{aligned}
&\mathcal{V}^S_{s_1,s_2,s'_1,s'_2}(k_1,k_2,k'_1,k'_2)\\
=&-\frac{G}{4(N_c^2-1)}\bigg[\alpha_+(P^+)\bar{u}_{s'_1}(k'_1)v_{s'_2}(k_2')\bar{v}_{s_2}(k_2)u_{s_1}(k_1)-\alpha_-(P^+)\bar{u}_{s'_1}(k'_1)\tau^av_{s'_2}(k_2')\bar{v}_{s_2}(k_2)\tau^au_{s_1}(k_1)\\
&-\alpha_-(P^+)\bar{u}_{s_1}(k_1)i\gamma^5v_{s_2}(k_2)\bar{v}_{s'_2}(k'_2)i\gamma^5u_{s_1'}(k_1')+\alpha_+(P^+)\bar{u}_{s_1}(k_1)i\gamma^5\tau^av_{s_2}(k_2)\bar{v}_{s'_2}(k'_2)i\gamma^5u_{s'_1}(k'_1)\bigg]
\end{aligned}
\end{equation}
in leading order in $1/N_c$.   The colored vertices in (\ref{eqn:effective_kernel}) do not contribute
in the s-channel to this order, since the bound two-body eigenstate is colorless. The extra vertex factors of $\alpha_\pm(P^+)$ resum the tadpoles.
\\
\\
{\bf t-channel:}
\\
The interaction in the t-channel  follows from the second and thirs contributions in (\ref{eqn:effective_kernel}),
through the substitution $(\bar{\psi}\psi)^2\rightarrow-\bar{u}_{1^\prime}u_{1}\bar{v}_{2}v_{2^\prime}$,
\begin{equation}
\begin{aligned}
\label{VTT}
&\mathcal{V}^T_{s_1,s_2,s'_1,s'_2}(k_1,k_2,k_1',k_2')\\
=&\frac{G}{4(N_c^2-1)}(1-\tau^a_1\tau^a_2)\bigg[\frac{N_c^2-1}{2N_c}\bigg(\bar{u}_{s_1}(k_1)u_{s'_1}(k'_1)\bar{v}_{s'_2}(k'_2)v_{s_2}(k_2)-\bar{u}_{s'_1}(k_1')i\gamma^5u_{s_1}(k_1)\bar{v}_{s_2}(k_2)i\gamma^5v_{s'_2}(k'_2)\bigg)\\
&+\frac{N_c^2-1}{4N_c}\bar{u}_{s'_1}(k_1')\sigma^{\mu\nu}u_{s_1}(k_1)\bar{v}_{s_2}(k_2)\sigma_{\mu\nu}v_{s'_2}(k'_2)\bigg]
\end{aligned}
\end{equation}
also in leading order in $1/N_c$. The minus sign arises  from Fermi statistics. We made use of the color averaging 
$\langle\lambda^\alpha_1\cdot\lambda^\alpha_2\rangle=2\frac{N_c^2-1}{N_c}$ in the t-channel,  since the interaction kernel acts on a color-singlet 
two-body eigenstate.
\\
\\
{\bf Eikonal approximation:}
\\
In the eikonal or large momentum limit,  $k_1^+\approx k_1'^+$ and  $k_2^+\approx k_2'^+$ as the in-out particles fly on straight trajectories.
The bound state equation simplifies  by replacing the $y$-dependence in the interaction kernel by $x$, with the result
\begin{equation}
\begin{aligned}
        &m_X^2\Phi_X(x,k_\perp,s_1,s_2)=\frac{k_\perp^2+M^2}{x\bar{x}}\Phi_X(x,k_\perp,s_1,s_2)\\
        &-\frac{1}{2x\bar{x}}\sqrt{\mathcal{F}(k)\mathcal{F}(P-k)}\int_0^1 dy\int\frac{d^2q_\perp}{(2\pi)^3}\sum_{s,s'}\left[\mathcal{V}^{S/T}_{s_1,s_2,s,s'}(k,P-k,q,P-q)\sqrt{\mathcal{F}(q)\mathcal{F}(P-q)}\right]\bigg|_{x=y}\Phi_X(y,q_\perp,s,s')
\end{aligned}
\end{equation}

In coordinate space,  the LFWF is defined as
\begin{equation}
\Phi(\xi^-,b_\perp)=\int\frac{dxd^2k_\perp}{(2\pi)^3}\Phi(x,k_\perp)e^{-ixP^+\xi^-+ik_\perp\cdot b_\perp}
\end{equation}
and the bound state equation is now a standard Schrödinger equation 
\begin{equation}
       \left[\frac{-\nabla_\perp^2+M^2}{x\bar{x}}\delta_{s_1s_1'}\delta_{s_2s_2'}+\hat{V}_{s_1s_2,s_1's_2'}(\xi^-,b_\perp)\right]\Phi_X(\xi^-,b_\perp,s_1',s_2')=m_X^2\Phi_X(\xi^-,b_\perp,s_1,s_2)
\end{equation}
with the spin-flavor dependent potential     
\begin{equation}
\label{KV12}
        \hat{V}(\xi^-,b_\perp)= -2g_S(1-\tau^a_1\tau^a_2)\frac{1}{2x\bar{x}}\mathcal{F}^2\left(\frac{-\nabla^2_\perp+M^2}{x\bar{x}}\right)V(1,2)
\end{equation}
The momentum fraction in coordinate space is  $\frac{1}{x}=\frac{-i}{\partial_-}$ with $\partial_-=\frac{1}{P^+}\frac{\partial}{\partial \xi^-}$. 
In the large $N_c$ limit,  the interaction kernel in (\ref{KV12}) simplifies
\bea
\label{V12XX}
&&V(1,2)=\nonumber\\
&&\bigg[M\mathds{1}_1 M\mathds{1}_2-\frac{1}{2}(\sigma_{1\perp}\times\nabla_\perp)(M\mathds{1}_2)+\frac{1}{2}(M\mathds{1}_1)(\sigma_{2\perp}\times\nabla_\perp)\nonumber\\
&&-\frac{1}{4}(\sigma_{1\perp}\times\nabla_\perp)(\sigma_{2\perp}\times\nabla_\perp)
-\frac{1}{4}(i\sigma_{1\perp}\cdot\nabla_\perp)(i\sigma_{2\perp}\cdot\nabla_\perp)\bigg]\delta^2(b_\perp)\nonumber\\
&&-\bigg[M\sigma_{1z}M\sigma_{2z}+\frac{1}{2}(i\sigma_{1\perp}\cdot\nabla_\perp) (M\sigma_{2z})-\frac{1}{2} (M\sigma_{1z})(i\sigma_{2\perp}\cdot\nabla_\perp)\nonumber\\
&&-\frac{1}{4}(i\sigma_{1\perp}\cdot\nabla_\perp)(i\sigma_{2\perp}\cdot\nabla_\perp)
-\frac{1}{4}(\sigma_{1\perp}\times\nabla_\perp)(\sigma_{2\perp}\times\nabla_\perp)\bigg]\delta^2(b_\perp)\nonumber\\
&&+[(M\sigma_{1\perp}\sigma_{2z}-M\sigma_{1z}\sigma_{2\perp})\cdot i\nabla_{\perp}+(\sigma_{1\perp}\cdot\nabla_{\perp})(\sigma_{2\perp}\cdot\nabla_{\perp})-(\sigma_{1\perp}\times\nabla_{\perp})(\sigma_{2\perp}\times\nabla_{\perp})]\delta^2(b_\perp)\nonumber\\
&&+\frac{1}{2}[(\sigma_{1\perp}\cdot\nabla_{\perp}\delta^2(b_\perp))(\sigma_{2\perp}\cdot\nabla_{\perp})+(\sigma_{1\perp}\cdot\nabla_{\perp})(\sigma_{2\perp}\cdot\nabla_{\perp}\delta^2(b_\perp))]\nonumber\\
&&-\frac{1}{2}[(\sigma_{1\perp}\times\nabla_{\perp}\delta^2(b_\perp))(\sigma_{2\perp}\times\nabla_{\perp})+(\sigma_{1\perp}\times\nabla_{\perp})(\sigma_{2\perp}\times\nabla_{\perp}\delta^2(b_\perp))]
\eea
The first two lines in  (\ref{V12XX})  stem from the non-tensor parts in (\ref{eqn:effective_kernel}), in agreement with~\cite{Shuryak:2021mlh}.
The remaining contributions in (\ref{eqn:effective_kernel}) are from the tensor part in (\ref{eqn:effective_kernel}).
Note that the minus sign from the Fermi statistics is compensated by the minus sign from the anti-spinor contraction $\bar{v}v$, with net attraction 
in the t-channel for the pion and sigma.

\section{DGLAP evolution and ERBL evolution}

 Dokshitzer–Gribov–Lipatov–Altarelli–Parisi (DGLAP) evolution is governed by
\begin{equation}
    \mu\frac{d}{d\mu}\begin{pmatrix}
    q(x,\mu) \\
    g(x,\mu) 
    \end{pmatrix}
    =\sum_{q'}\frac{\alpha_s(\mu)}{\pi}\int_x^1\frac{dy}{y}\begin{pmatrix}
    P_{qq'}(x/y) & P_{qg}(x/y)\\
    P_{gq'}(x/y) & P_{gg}(x/y)
    \end{pmatrix}
    \begin{pmatrix}
    q'(y,\mu) \\
    g(y,\mu) 
    \end{pmatrix}
\end{equation}
with the splitting functions defined as follows
$$P_{qq}(x)=C_F\left[\frac{1+x^2}{(1-x)_+}+\frac{3}{2}\delta(1-x)\right]$$

$$P_{qg}(x)=\frac{1}{2}\left[x^2+\left(1-x\right)^2\right]$$

$$P_{gq}(x)=C_F\left[\frac{1+(1-x)^2}{x}\right]$$

$$P_{gg}(x)=2C_A\left[\frac{x}{(1-x)_+}+\frac{1-x}{x}+x(1-x)\right]+\frac{\beta_0}{2}\delta(1-x)$$
where $C_F=\frac{N_c^2-1}{2N_c}$, $C_A=N_c$, and the one loop running coupling $\beta_0=\frac{11}{3}N_c-\frac{2}{3}n_f$.
\end{widetext}
The full evolution of the quark parton distribution inside a hadron,  is described by the DGLAP equations. At low energy,  where the valence  degrees of freedom dominate, the valence quark distribution is sufficient to describe the hadronic structure. Only the quark-to-quark splitting process denoted by $P_{qq}(x)$, has to be taken into account. As the energy scale increases, the sea quark and gluon production spreads the momentum distribution  on more partons, effectively shifting the momentum valence quark distribution to small $x$, and eventually soften the $x=1$ tail behavior.

On the other hand, the evolution of the light front wave function (LFWF) and its associated distribution amplitudes (DA), is governed by the Efremov-Radyushkin-Brodsky-Lepage (ERBL) equation, which expands the LFWF and DA from the asymptotic form $6x(1-x)$,  in a complete basis of Gegenbauer polynomials.
\begin{equation}
\label{eq:ERBL}
    \phi(x,Q)=6x\bar{x}\sum_{n=0}^\infty a_n(Q_0)\left(\frac{\alpha_s(Q^2)}{\alpha_s(Q^2_0)}\right)^{\gamma_n/\beta_0}C_n^{3/2}(x-\bar{x})
\end{equation}
The anomalous dimension associated with the evolution is defined as
\begin{equation}
    \gamma_n=C_F\left[-3+4\sum_{j=1}^{n+1}\frac{1}{j}-\frac{2}{(n+1)(n+2)}\right]
\end{equation}
where $$\alpha_s(Q)=\frac{4\pi}{\beta_0\ln\left(\frac{Q^2}{\Lambda_{QCD}^2}\right)}$$ and $\Lambda_{QCD}=226$ MeV. $C^m_n(z)$ are  Gegenbauer polynomials. Due to the orthogonality of the Gegenbauer polynomials, the initial coefficients can be evaluated by 
\begin{equation}
    a_n(Q_0)=\frac{2(2n+3)}{3(n+1)(n+2)}\int_0^1dyC^{3/2}_n(y-\bar{y})\phi(y,Q_0)
\end{equation}

\section{Random Instanton Vacuum}
\label{Appx:RIVappx}
Here we briefly review the emergence of the running constituent quark mass $M(k)$ in the ILM, by resummation of the leading $1/N_c$ 
contributions~\cite{Diakonov:1985eg,Nowak:1989jd}. The analysis is in the rest frame in Euclidean signature. 
For a sufficiently  dilute vacuum composed of instantons and anti-instantons, the gauge fields are 
\begin{equation}
\label{GAUGE}
    A(x)=\sum^{N_-}_{I=1}A_I(x)+\sum_{\bar{I}}^{N_+}A_{\bar{I}}(x)
\end{equation}
where $I, \bar I$ refer to the instanton and anti-instanton moduli.
More specifically, each instanton is localized at $z_I$, with size $\rho_I$ and color orientation $U_I$
\begin{equation}
    A_{I}(x)=U_IA(x-z_I,\rho_I)U_I^\dagger
\end{equation}
For a topologically neutral vacuum, the mean number of instantons match that of the 
 anti-instantons $N_+=N_-$. 
 
 Light quarks in the ILM scatter randomly at each instanton and anti-instanton site,
 with the light quark propagator 
\begin{equation}
    S(x,y)=\left\langle\langle x|\frac{1}{-i\slashed{\partial}-\slashed{A}-im}|y\rangle\right\rangle_A
\end{equation}
averaged over (\ref{GAUGE}) with  $N=N_++N_-$ fixed in an Euclidean 4-volume  $V$.
The average runs the entire instanton ensemble with equal sampling. The planar 
rescattering contributions  to the light quark propagator in leading order 
 in $1/N_c$  counting, gives~\cite{Diakonov:1985eg,Pobylitsa:1989uq}
\begin{widetext}
\begin{equation}
\label{SSGAP}
    S^{-1}-S_0^{-1}=\frac{N}{2N_cV}\langle x|\left[\sum_I\int d^4z_IdU_I\frac{1}{S-\slashed{A}_I^{-1}}+\sum_{\bar{I}}\int d^4z_{\bar{I}}dU_{\bar{I}}\frac{1}{S-\slashed{A}_{\bar{I}}^{-1}}\right]|y\rangle
\end{equation}
\end{widetext}
with the apparent diluteness factor $\rho^4 N/N_c V\sim \kappa_{I+\bar I}/N_c$.   The dominant contribution to
(\ref{SSGAP}) stems from the quark zero modes, with a running constituent quark mass~\cite{Kock:2020frx}
\bea
   && M(k,m)=m+\frac{M(0,0)}{(1+\xi^2)^{1/2}+\xi}\nonumber\\
    &&\times \left[z(I_0(z)K_0(z)-I_1(z)K_1(z))'\right]^2\bigg|_{z=\frac{k\rho}{2}}
\eea
with $$\xi=\frac{N_cmM(0,0)}{4\pi^2\rho^2N/V}$$
If we define $M=M(0,m)$ and  $M(k^2)=M(k,m)$, then
\begin{equation}
\label{RIV_mass}
    M(k^2)=M(zF'(z))^2+m\left[1-(zF'(z))^2\right]\bigg|_{z=\frac{k\rho}{2}}
\end{equation}
with $$F(z)=I_0(z)K_0(z)-I_1(z)K_1(z)$$
the profile of the quark zero mode in singular gauge.
The runnung quark mass asymptotes the current mass $m$ for $k\rho \gg 1$,
and reduces to the constant mass $M$ for $k\rho\ll 1$.


\begin{figure*}
\subfloat[\label{fig_3pt1}]{%
  \includegraphics[height=5.5cm,width=.46\linewidth]{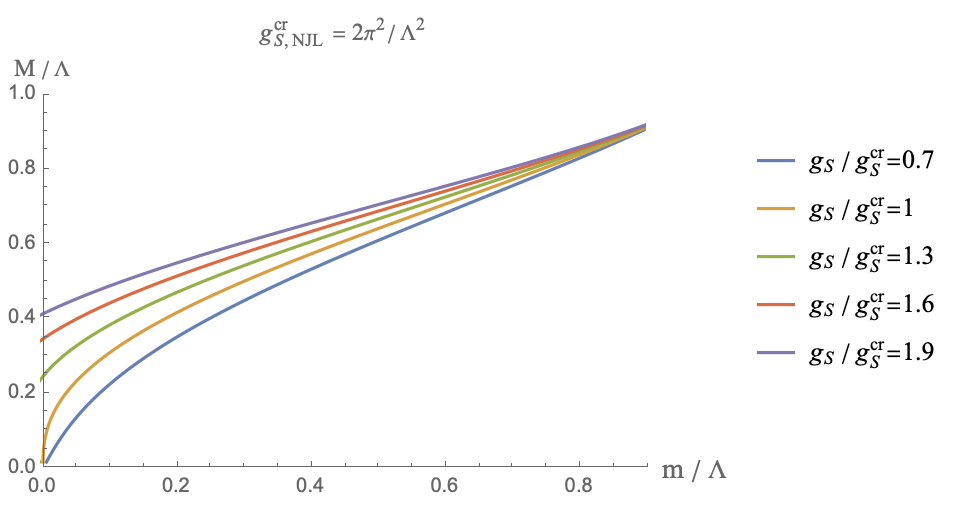}%
}\hfill
\subfloat[\label{fig_3pt6}]{%
  \includegraphics[height=5.5cm,width=.46\linewidth]{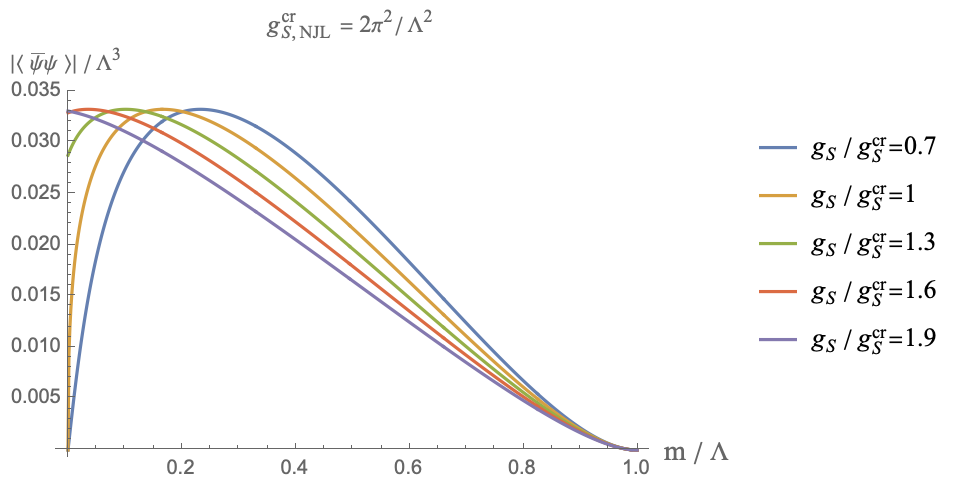}%
}
\caption{
a: The change of the constituent mass $M$ versus the current mass $m$, for different couplings $g_S/g_S^{cr}$ with $g_S^{cr}=2\pi^2/\Lambda^2$,
in the zero instanton size limit; 
b: Same as in a for the chiral condensate.}
\label{fig_DGLAPKYY}
\end{figure*}

\section{Gap equation for zero instanton size}
\label{APP_SHARP0}
In the zero instanton size limit, the induced form factor from the quark zero modes ${\mathcal F}(k^2)\rightarrow 1$,
and the induced $^\prime$t Hooft effective interactions become local in Euclidean space. While their continuation to
Minkowski space is straightforward, they lead to diverging results both in the IR and UV in the mean field treatment.
This situation is familiar for local fermionic or NJL type interactions in the rest frame, and therefore carry to the light front.

For virtual loops with running momenta $k$ in Euclidean space,
 the cutoff choice should preserve $O(4)$ symmetry, which naturally means $\sqrt{k^2}<\Lambda_E$. 
 For virtual loops in Minkowski space, the cutoff should preserve boost invariance on the light front, which 
 translates to $\sqrt{2}k^\pm <\Lambda$ or equivalently~\cite{Itakura:2000te,Naito_2004} 
\begin{equation}
    \frac{k^2_\perp+M^2}{\sqrt{2}\Lambda}<k^+<\frac{\Lambda}{\sqrt{2}}
\end{equation}
The lower bound follows from the on-shell condition $k^2=2k^+k^--k_\perp^2=M^2$.

With this in mind, the diverging integrals in (\ref{gap_eq}) and (\ref{cond_eq}) after the boost invariant regularization, yield
\begin{widetext}
\bea
\label{mass_integral}
   \int\frac{dk^+d^2k_\perp}{(2\pi)^3}\frac{\epsilon(k^+)}{k^+}\theta(\Lambda/\sqrt{2}-k^+)\theta(\Lambda/\sqrt{2}-k^-)\bigg|_{k^-
   =\frac{k_\perp^2+M^2}{2k^+}}=\frac{\Lambda^2}{4\pi^2}\left[1-\frac{M^2}{\Lambda^2}+\frac{M^2}{\Lambda^2}\ln\frac{M^2}{\Lambda^2}\right]
\eea
\end{widetext}
As a result, the gap equation in the zero instanton size limit is
\begin{equation}
    \frac{m}{M}=1-\frac{g_S\Lambda^2}{2\pi^2}\left[1-\frac{M^2}{\Lambda^2}+\frac{M^2}{\Lambda^2}\ln\frac{M^2}{\Lambda^2}\right]
\end{equation}
and the quark condensate $\langle\bar{\psi}\psi\rangle$ is
\begin{equation}
    \frac{\langle\bar{\psi}\psi\rangle}{\Lambda^3}=-\frac{N_c}{2\pi^2} \left(\frac{M}{\Lambda}\right)\left[1-\frac{M^2}{\Lambda^2}+\frac{M^2}{\Lambda^2}\ln\frac{M^2}{\Lambda^2}\right]
\end{equation}
A nonzero constituent quark mass $M$ develops, whenever the fermionic interaction is sufficiently
attrative with  $g_S>g_{S,\mathrm{NJL}}^\mathrm{cr}=\frac{2\pi^2}{\Lambda^2}$.

In Fig.~\ref{fig_DGLAPKYY}a we show the dependence of the constituent quark mass $M/\Lambda$ versus the current quark mass $m/\Lambda$,
with increasing fermionic coupling $g_S$ (bottom to top from left), for a fixed cutoff $\Lambda$ and a critical coupling $g_{S,NJL}^{cr}=2\pi^2/\Lambda^2$.
The same display for the chiral condensate is shown in In Fig.~\ref{fig_DGLAPKYY}b.

\section{Bound states for zero instanton size}
\label{APP_SHARP2}
\begin{figure}
    \centering
     \includegraphics[scale=0.5]{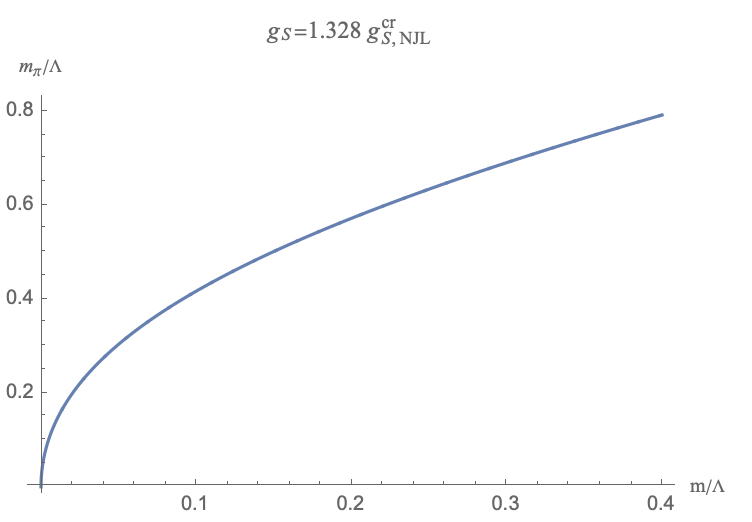}
     \caption{
Pion mass versus the current quark mass, with fixed cutoff $\Lambda$ and a fermion coupling $g_S$, in the zero instanton size limit.
}
 \label{fig_MPINJLX}
\end{figure}

In the zero instanton size limit, the emergent form factor from the quark zero modes ${\cal F}\rightarrow 1$. 
To regulate the vacuum or 0-body integrals to one loop, we use the parity even and boost invariant sharp cutoff 
discussed in Appendix~\ref{APP_SHARP0}. In an n-body state on the light front, the natural cutoff from time
ordered perturbation theory, relates to the light front energy $K^-=\sum_ik_i^-$ of free valence quarks~\cite{Lepage:1980fj},
\bea
K^-=\sum_{i=1}^n\frac{k_{\perp i}^2+M_i^2}{x_i K^+}<2\Lambda^2
\eea
Here $k_{\perp i}$ and $x_i$, are the transverse momentum,  and the longitudinal momentum fraction of the valence parton-i, respectively.
This cutoff is boost invariant and parity even. For 2-body bound states with unequal masses,  it translates to
\bea
\frac{k_\perp^2+M_u^2}{x}+\frac{k_\perp^2+M_s^2}{\bar x}<2\Lambda^2\sim \frac 1{\rho^2}
\eea
A similar cutoff was used in~\cite{Itakura:2000te} for equal quark masses, and independently argued in the context of the LaMET analysis of the ILM in~\cite{Kock:2020frx}.
\\
\\
{\bf sigma meson:}
\\
\begin{widetext}
\begin{equation}
\begin{aligned}
    \frac{m}{M}=&-\frac{g_s}{4\pi^2}\int_{0}^{1}dy\int_0^{\Lambda^2} dq^2_\perp\left[\frac{m_\sigma^2-4M^2}{y\bar{y}m_\sigma^2-(q_\perp^2+M^2)}\right]\\
    =&-\frac{g_s}{4\pi^2}(4M^2-m_\sigma^2)\int_0^1dy\ln\left(1+\frac{\Lambda^2}{M^2-y\bar{y}m_\sigma^2}\right)~,\ m_\sigma^2<4M^2\\
    =&-\frac{g_s}{2\pi^2}(4M^2-m_\sigma^2)\\
    &\times\left[\frac{1}{2}\ln\left(1+\frac{\Lambda^2}{M^2}\right)-\sqrt{\frac{4M^2-m_\sigma^2}{m_\sigma^2}}\left(\tan^{-1}\frac{1}{\sqrt{\frac{4M^2-m_\sigma^2}{m_\sigma^2}}}-\sqrt{\frac{4(M^2+\Lambda^2)-m_\sigma^2}{4M^2-m_\sigma^2}}\tan^{-1}\frac{1}{\sqrt{\frac{4(M^2+\Lambda^2)-m_\sigma^2}{m_\sigma^2}}}\right)\right]\\
\end{aligned}
\end{equation}
The sigma is a treshold state with a mass $m_\sigma=2M$ in the chiral limit, but otherwise unbound in ILM with a sharp or a smooth cutoff
on the light front. 
\\
\\
{\bf pi$_5$ meson:}
\\
\begin{equation}
\begin{aligned}
    2-\frac{m}{M}=&\frac{g_s}{4\pi^2}\int_{0}^{1}dy\int_0^{\Lambda^2} dq^2_\perp\left[\frac{m_{\pi_5}^2-4M^2}{y\bar{y}m_{\pi_5}^2-(q_\perp^2+M^2)}\right]\\
    =&\frac{g_s}{4\pi^2}(4M^2-m_{\pi_5}^2)\int_0^1dy\ln\left(1+\frac{\Lambda^2}{M^2-y\bar{y}m_{\pi_5}^2}\right)~,\ m_{\pi_5}^2<4M^2\\
    =&\frac{g_s}{2\pi^2}(4M^2-m_{\pi_5}^2)\\
    &\times\left[\frac{1}{2}\ln\left(1+\frac{\Lambda^2}{M^2}\right)-\sqrt{\frac{4M^2-m_{\pi_5}^2}{m_{\pi_5}^2}}\tan^{-1}\frac{1}{\sqrt{\frac{4M^2-m_{\pi_5}^2}{m_{\pi_5}^2}}}+\sqrt{\frac{4(M^2+\Lambda^2)-m_{\pi_5}^2}{m_{\pi_5}^2}}\tan^{-1}\frac{1}{\sqrt{\frac{4(M^2+\Lambda^2)-m_{\pi_5}^2}{m_{\pi_5}^2}}}\right]\\
\end{aligned}
\end{equation}
\\
\\
{\bf sigma$_5$ meson:}
\\
\begin{equation}
\begin{aligned}
    2-\frac{m}{M}=&\frac{g_s}{4\pi^2}\int_0^1dy\int_0^{\Lambda^2} dq^2_\perp\left[\frac{m_{\sigma_5}^2}{y\bar{y}m_{\sigma_5}^2-(q_\perp^2+M^2)}\right]\\
    =&-\frac{g_s}{2\pi^2}m_{\sigma_5}^2\\
    &\times\left[\frac{1}{2}\ln\left(1+\frac{\Lambda^2}{M^2}\right)-\sqrt{\frac{4M^2-m_{\sigma_5}^2}{m_{\sigma_5}^2}}\left(\tan^{-1}\frac{1}{\sqrt{\frac{4M^2-m_{\sigma_5}^2}{m_{\sigma_5}^2}}}-\sqrt{\frac{4(M^2+\Lambda^2)-m_{\sigma_5}^2}{4M^2-m_{\sigma_5}^2}}\tan^{-1}\frac{1}{\sqrt{\frac{4(M^2+\Lambda^2)-m_{\sigma_5}^2}{m_{\sigma_5}^2}}}\right)\right]
\end{aligned}
\end{equation}
The instanton induced interactions are repulsive in both of the $\sigma_5, \pi_5$ channels on the light front. 
There are no bound states in this channel in the chiral limit or otherwise.
\\
\\
{\bf pi meson:}
\\
\begin{equation}
\begin{aligned}
    \frac{m}{M}=&-\frac{g_s}{4\pi^2}\int_{0}^{1}dy\int_0^{\Lambda^2} dq^2_\perp\left[\frac{m_\pi^2}{y\bar{y}m_\pi^2-(q_\perp^2+M^2)}\right]\\
    =&\frac{g_s}{4\pi^2}m_\pi^2\int_{0}^{1}dy\ln\left(1+\frac{\Lambda^2}{M^2-y\bar{y}m_\pi^2}\right)~,\ m_\pi^2<4M^2\\
    =&\frac{g_s}{2\pi^2}m_\pi^2\left[\frac{1}{2}\ln\left(1+\frac{\Lambda^2}{M^2}\right)-\sqrt{\frac{4M^2-m_\pi^2}{m_\pi^2}}\tan^{-1}\frac{1}{\sqrt{\frac{4M^2-m_\pi^2}{m_\pi^2}}}+\sqrt{\frac{4(M^2+\Lambda^2)-m_\pi^2}{m_\pi^2}}\tan^{-1}\frac{1}{\sqrt{\frac{4(M^2+\Lambda^2)-m_\pi^2}{m_\pi^2}}}\right]
\end{aligned}
\end{equation}
\end{widetext}
This is by far the strongest channel in the ILM, with a strongly bound $\pi$ state. 
The dependence of the pion $m_\pi$ on the current mass $m$ for fixed $\Lambda$ is shown in Fig.~\ref{fig_MPINJLX}.
The pion mass $m_\pi$ increases as $m$ increases,  to eventually reach the 2-constituent quark  threshold $2M$. 
It is a true Goldstone mode, with the mass vanishing in the chiral limit
\begin{equation}
    m^2_\pi=\frac{m/M}{\frac{g_s}{4\pi^2}\ln\left(1+\frac{\Lambda^2}{M^2}\right)}=\frac{2m}{f^2_\pi}|\langle \bar{\psi}\psi\rangle| +\mathcal{O}(m^2)
\end{equation}
which is the expected Gell-Mann-Oakes-Renner relation.
In chiral limit, the pion decay constant is 
\begin{equation}
    f_\pi=\frac{\sqrt{N_c}M}{\sqrt{2}\pi}\left[\ln\left(1+\frac{\Lambda^2}{M^2}\right)\right]^{1/2}
\end{equation}

With the physical input of pion decay constant  and pion mass, we can fix the best parameters for our NJL-type model. The transverse cut-off is fixed to be $\Lambda=729$ MeV, roughly higher than inverse instanton size $\rho=0.313$ fm. The current quark mass we obtained is $m=12.4$ MeV and the coupling will be fixed to be $g_S=1.314~g_{S,\mathrm{NJL}}^{cr}$. The constituent mass with the given current mass and fixed coupling will be $M=207.7$ MeV. The quark condensate will thus be $|\langle \bar{\psi}\psi\rangle|^{1/3}=228.9$ MeV. The pion mass and the pion decay constant with the given parameters will be $m^{\mathrm{NJL}}_\pi=135.0$ MeV and $f^{\mathrm{NJL}}_\pi=130.3$ MeV respectively, which are very consistent with the physical values.

\section{Fermionic constraint in the operator form}
One of the key feature of the light front approach, is the elimination of the fermionic constraint on the 
bad component of the fermionic field. In the main text, we have used an auxillary field approach and the
mean-field approximation, to eliminate the bad component in leading order in $1/N_c$. Here, we briefly
show that the elimination can be enforced solely at the operator level, without recourse to auxillary fields,
using power counting in $1/N_c$. 

The multi-fermion interactions generated in the ILM are essentially determinantal, owing to the induced
$U_A(1)$ flavor symmetry breaking by instantons. By Fierzing, 
they consist of  combinations of only twist-3 operators, 
$\sigma=\bar{\psi}\psi$, $\pi=\bar{\psi}i\gamma^5\psi$, $\sigma^a=\bar{\psi}\tau^a\psi$, and $\pi^a=\bar{\psi}i\tau^a\gamma^5\psi$,
which form a closed set. The elimination of the bad component $\psi_-$ in favor of the good component
$\psi_+$ amounts to the closed hierarchy in $1/N_c$
\begin{widetext}
\begin{equation}
\begin{aligned}
\label{xscalarx}
    &\sigma(x)=\sigma^{(0)}(x)-\frac{g_S}{2N_c}\left[\hat{V}(x)\sigma(x)-\hat{A}(x)\pi(x)-\hat{V}^a(x)\sigma^a(x)+\hat{A}^a(x)\pi^a(x)\right]\\
\end{aligned}
\end{equation}
\begin{equation}
\begin{aligned}
\label{pseudoscalar}
    &\pi(x)=\pi^{(0)}(x)-\frac{g_S}{2N_c}\left[-\hat{A}(x)\sigma(x)-\hat{V}(x)\pi(x)+\hat{A}^a(x)\sigma^a(x)+\hat{V}^a(x)\pi^a(x)\right]\\
\end{aligned}
\end{equation}
\begin{equation}
\begin{aligned}
    \sigma^a(x)=&\sigma^{a(0)}(x) -\frac{g_S}{2N_c}
    \left[\hat{V}^a(x)\sigma(x)-\hat{A}^a(x)\pi(x)-\hat{V}^{ab}(x)\sigma^b(x)+\hat{A}^{ab}(x)\pi^b(x)\right]
\end{aligned}
\end{equation}
\begin{equation}
\begin{aligned}
\label{xpix}
    \pi^a(x)=&\pi^{a(0)}(x)-\frac{g_S}{2N_c}
    \left[-\hat{A}^a(x)\sigma(x)-\hat{V}^a(x)\pi(x)+\hat{A}^{ab}(x)\sigma^b(x)+\hat{V}^{ab}(x)\pi^b(x)\right]
\end{aligned}
\end{equation}
with $g_S=G_SN_c$ assumed of fixed order in $N_c$, and
\begin{equation}
\begin{aligned}
\label{s}
        \sigma^{(0)}(x)=\bar{\psi}_+(x)\frac{1}{2}\left(i\gamma^i_\perp\overrightarrow{\partial_i}+m\right)\gamma^+\left(\frac{-i}{\partial_-}\psi_+(x)\right)+\left(\frac{i}{\partial_-}\bar{\psi}_+(x)\right)\frac{1}{2}\left(i\gamma^i_\perp\overleftarrow{\partial}_i+m\right)\gamma^+\psi_+(x)
\end{aligned}
\end{equation}
\begin{equation}
\begin{aligned}
\label{s5}
        \pi^{(0)}(x)=\bar{\psi}_+(x)\frac{1}{2}\left(-i\gamma^i_\perp\overrightarrow{\partial_i}+m\right)i\gamma^5\gamma^+\left(\frac{-i}{\partial_-}\psi_+(x)\right)-\left(\frac{i}{\partial_-}\bar{\psi}_+(x)\right)\frac{1}{2}\left(i\gamma^i_\perp\overleftarrow{\partial}_i+m\right)i\gamma^5\gamma^+\psi_+(x)
\end{aligned}
\end{equation}
\begin{equation}
\begin{aligned}
\label{p5}
        \sigma^{a(0)}(x)=\bar{\psi}_+(x)\frac{1}{2}\left(-i\gamma^i_\perp\overrightarrow{\partial_i}+m\right)\gamma^+\tau^a\left(\frac{-i}{\partial_-}\psi_+(x)\right)+\left(\frac{i}{\partial_-}\bar{\psi}_+(x)\right)\frac{1}{2}\left(i\gamma^i_\perp\overleftarrow{\partial}_i+m\right)\gamma^+\tau^a\psi_+(x)
\end{aligned}
\end{equation}
\begin{equation}
\begin{aligned}
\label{p}
        \pi^{a(0)}(x)=\bar{\psi}_+(x)\frac{1}{2}\left(-i\gamma^i_\perp\overrightarrow{\partial_i}+m\right)i\gamma^5\gamma^+\tau^a\left(\frac{-i}{\partial_-}\psi_+(x)\right)-\left(\frac{i}{\partial_-}\bar{\psi}_+(x)\right)\frac{1}{2}\left(i\gamma^i_\perp\overleftarrow{\partial}_i+m\right)i\gamma^5\gamma^+\tau^a\psi_+(x)
\end{aligned}
\end{equation}
In this hierarchy, the active operators generate
\begin{align}
        &\hat{V}(x)[~\cdot~]=\bar{\psi}_+(x)\gamma^+\frac{-i}{\partial_-}(\psi_+(x)[~\cdot~])+\frac{i}{\partial_-}(\bar{\psi}_+(x)[~\cdot~])\gamma^+\psi_+(x) \\
        &\hat{A}(x)[~\cdot~]=\bar{\psi}_+(x)i\gamma^+\gamma^5\frac{-i}{\partial_-}(\psi_+(x)[~\cdot~])-\frac{i}{\partial_-}(\bar{\psi}_+(x)[~\cdot~])i\gamma^+\gamma^5\psi_+(x) \\
        &\hat{V}^a(x)[~\cdot~]=\bar{\psi}_+(x)\gamma^+\tau^a\frac{-i}{\partial_-}(\psi_+(x)[~\cdot~])+\frac{i}{\partial_-}(\bar{\psi}_+(x)[~\cdot~])\gamma^+\tau^a\psi_+(x) \\
        &\hat{A}^a(x)[~\cdot~]=\bar{\psi}_+(x)i\gamma^+\gamma^5\tau^a\frac{-i}{\partial_-}(\psi_+(x)[~\cdot~])-\frac{i}{\partial_-}(\bar{\psi}_+(x)[~\cdot~])i\gamma^+\gamma^5\tau^a\psi_+(x) \\
        &\hat{V}^{ab}(x)[~\cdot~]=\bar{\psi}_+(x)\gamma^+\tau^a\tau^b\frac{-i}{\partial_-}(\psi_+(x)[~\cdot~])+\frac{i}{\partial_-}(\bar{\psi}_+(x)[~\cdot~])\gamma^+\tau^b\tau^a\psi_+(x) \\
        &\hat{A}^{ab}(x)[~\cdot~]=\bar{\psi}_+(x)i\gamma^+\gamma^5\tau^a\tau^b\frac{-i}{\partial_-}(\psi_+(x)[~\cdot~])-\frac{i}{\partial_-}(\bar{\psi}_+(x)[~\cdot~])i\gamma^+\gamma^5\tau^b\tau^a\psi_+(x) \\
\end{align}
In the light front vacuum $|0\rangle$ empty of valence particles and anti-particles, these operators can develop a vevs.
Parity and isospin symmetry (chiral limit) restric these vev$^\prime$s
\begin{equation}
\label{XSCA}
    \langle \sigma^{(0)}(x)\rangle=-2N_cm\int\frac{d^4k}{(2\pi)^4}\frac{4ik^+}{(k^2-M^2)k^+}=-2N_cm\int\frac{dk^+d^2k_\perp}{(2\pi)^3}\frac{\epsilon(k^+)}{k^+}
\end{equation}
\begin{equation}
    \langle \hat{V}(x)\rangle =-4N_c\int\frac{dk^+d^2k_\perp}{(2\pi)^3}\frac{\epsilon(k^+)}{k^+}
\end{equation}
\begin{equation}
    \langle \hat{V}^{ab}(x)\rangle =\langle \hat{V}(x)\rangle\delta^{ab}
\end{equation}
\begin{equation}
    \langle \hat{A}(x)\rangle=\langle \hat{V}^a(x)\rangle=\langle \hat{A}^a(x)\rangle=\langle \hat{A}^{ab}(x)\rangle =0
\end{equation}
Here $\epsilon(k^+)={\rm sgn}(k^+)$ is the signum function. A non-vanishing vev for the scalar operator in (\ref{XSCA}) reflects
on the spontaneous breaking of chiral symmetry, and is mainly driven by the accumulation of the $k^+=0$ zero modes on the light front.
With this in mind, the vev of the scalar $\sigma=\bar\psi\psi$ operator can be unwound to all orders in $1/N_c$
\begin{equation}
\label{XTAD}
    \langle \sigma(x)\rangle=\langle \sigma^{(0)}\rangle\left[1+2g_s\int\frac{dk^+d^2k_\perp}{(2\pi)^3}\frac{\epsilon(k^+)}{k^+}+4g_s^2\left(\int\frac{dk^+d^2k_\perp}{(2\pi)^3}\frac{\epsilon(k^+)}{k^+}\right)^2+\dots\right]
\end{equation}
which is a resummation of all the tadpole diagrams in leading order in $1/N_c$. The elimination of the bad component of the fermionic
field on the light front, amounts to a mass renormalization.

To characterize the operator hierarchy (\ref{xscalarx}-\ref{xpix}) in different Fock sectors, it is best to shift to momentum space
\begin{equation}
    \sigma(P)=\int dx^-d^2x_\perp\sigma(x)e^{iP^+x^--P_\perp\cdot x_\perp}
\end{equation}
\begin{equation}
    \pi(P)=\int dx^-d^2x_\perp\sigma_5(x)e^{iP^+x^--P_\perp\cdot x_\perp}
\end{equation}
\begin{equation}
    \sigma^a(P)=\int dx^-d^2x_\perp\pi_5^a(x)e^{iP^+x^--P_\perp\cdot x_\perp}
\end{equation}
\begin{equation}
    \pi^a(P)=\int dx^-d^2x_\perp\pi^a(x)e^{iP^+x^--P_\perp\cdot x_\perp}
\end{equation}
with the expectation values 
$$\langle\hat{V}(x)[e^{-iPx}]\rangle=4N_c\int\frac{dk^+d^2k_\perp}{(2\pi)^3}\frac{\epsilon(k^+)}{P^+-k^+}e^{-iPx}$$
$$\langle\hat{V}^{ab}(x)[e^{-iPx}]\rangle=\delta^{ab}\langle\hat{V}(x)[e^{-iPx}]\rangle$$
In the 2-body Fock space, with $P$ the total momentum of the pair, we have
\begin{equation}
\begin{aligned}
    &\sigma(P)=\sigma^{(0)}(P)-2g_s\int\frac{dk^+d^2k_\perp}{(2\pi)^3}\frac{\epsilon(k^+)}{P^+-k^+}\sigma(P)\\
\end{aligned}
\end{equation}
\begin{equation}
\begin{aligned}
    &\pi(P)=\pi^{(0)}(P)+2g_s\int\frac{dk^+d^2k_\perp}{(2\pi)^3}\frac{\epsilon(k^+)}{P^+-k^+}\pi(x)\\
\end{aligned}
\end{equation}
\begin{equation}
\begin{aligned}
    &\sigma^a(P)=\sigma^{a(0)}(P)+2g_s\int\frac{dk^+d^2k_\perp}{(2\pi)^3}\frac{\epsilon(k^+)}{P^+-k^+}\sigma^a(P)\\
\end{aligned}
\end{equation}
\begin{equation}
\begin{aligned}
    \pi^a(P)=&\pi^{a(0)}(P)-2g_s\int\frac{dk^+d^2k_\perp}{(2\pi)^3}\frac{\epsilon(k^+)}{P^+-k^+}\pi^a(P)
\end{aligned}
\end{equation}
Again, the elimination of the bad component of the fermionic field, amounts to a resummation of the tadpole contributions,
which is equivalent to a vertex renormalization of the strating multi-fermion interaction by the factor
$$\alpha^\pm(P^+)=\left[1\pm 2g_s\int\frac{dk^+d^2k_\perp}{(2\pi)^3}\frac{\epsilon(k^+)}{P^+-k^+}\right]^{-1}$$
More specifically, we have
\begin{equation}
\begin{aligned}
    \sigma(P)=&\sigma^{(0)}(P)\left[1-2g_s\int\frac{dk^+d^2k_\perp}{(2\pi)^3}\frac{\epsilon(k^+)}{P^+-k^+}+4g_s^2\left(\int\frac{dk^+d^2k_\perp}{(2\pi)^3}\frac{\epsilon(k^+)}{P^+-k^+}\right)^2+\dots\right]\\
\end{aligned}
\end{equation}
\begin{equation}
\begin{aligned}
    \pi(P)=&\pi^{(0)}(P)\left[1+2g_s\int\frac{dk^+d^2k_\perp}{(2\pi)^3}\frac{\epsilon(k^+)}{P^+-k^+}+4g_s^2\left(\int\frac{dk^+d^2k_\perp}{(2\pi)^3}\frac{\epsilon(k^+)}{P^+-k^+}\right)^2+\dots\right]\\
\end{aligned}
\end{equation}
\begin{equation}
\begin{aligned}
    \sigma^a(P)=&\sigma^{a(0)}(P)\left[1+2g_s\int\frac{dk^+d^2k_\perp}{(2\pi)^3}\frac{\epsilon(k^+)}{P^+-k^+}+4g_s^2\left(\int\frac{dk^+d^2k_\perp}{(2\pi)^3}\frac{\epsilon(k^+)}{P^+-k^+}\right)^2+\dots\right]\\
\end{aligned}
\end{equation}
\begin{equation}
\begin{aligned}
    \pi^a(P)=&\pi^{a(0)}(P)\left[1-2g_s\int\frac{dk^+d^2k_\perp}{(2\pi)^3}\frac{\epsilon(k^+)}{P^+-k^+}+4g_s^2\left(\int\frac{dk^+d^2k_\perp}{(2\pi)^3}\frac{\epsilon(k^+)}{P^+-k^+}\right)^2+\dots\right]\\
\end{aligned}
\end{equation}
The zero-th order operators in $1/N_c$ in momentum space are 
\begin{equation}
    \sigma^{(0)}(P)=\int[d^3k]_+\int[d^3q]_+\bar{\psi}_+(k)\left[\frac{\gamma^+(\vec{\gamma}_\perp\cdot\vec{q}+m)}{2q^+}+\frac{(\vec{\gamma}_\perp\cdot\vec{k}+m)\gamma^+}{2k^+}\right]\psi_+(q)(2\pi)^3\delta_+^3(P+k-q)
\end{equation}
\begin{equation}
    \pi^{(0)}(P)=\int[d^3k]_+\int[d^3q]_+\bar{\psi}_+(k)\left[\frac{i\gamma^5\gamma^+(\vec{\gamma}_\perp\cdot\vec{q}+m)}{2q^+}-\frac{i\gamma^5\gamma^+(\vec{\gamma}_\perp\cdot\vec{k}+m)}{2k^+}\right]\psi_+(q)(2\pi)^3\delta_+^3(P+k-q)
\end{equation}
\begin{equation}
    \sigma^{a(0)}(P)=\int[d^3k]_+\int[d^3q]_+\bar{\psi}_+(k)\left[\frac{\gamma^+\tau^a(\vec{\gamma}_\perp\cdot\vec{q}+m)}{2q^+}+\frac{(\vec{\gamma}_\perp\cdot\vec{k}+m)\gamma^+\tau^a}{2k^+}\right]\psi_+(q)(2\pi)^3\delta_+^3(P+k-q)
\end{equation}
\begin{equation}
    \pi^{a(0)}(P)=\int[d^3k]_+\int[d^3q]_+\bar{\psi}_+(k)\left[\frac{i\gamma^5\gamma^+\tau^a(\vec{\gamma}_\perp\cdot\vec{q}+m)}{2q^+}-\frac{i\gamma^5\gamma^+\tau^a(\vec{\gamma}_\perp\cdot\vec{k}+m)}{2k^+}\right]\psi_+(q)(2\pi)^3\delta_+^3(P+k-q)
\end{equation}
with $[d^3k]_+=\frac{dk^+d^2k_\perp}{2k^+(2\pi)^3}\epsilon(k^+)$

\end{widetext}

\section{Light front zero modes}
\label{Appx:Gap}
In the ILM, the spontaneous breaking of chiral symmetry and the emergence of a scalar quark condensate,
 are due to the instanton and anti-instanton fermionic zero modes as we noted above. In the rest frame and in
 Euclidean signature, this is manifest in (\ref{MFK}) with the $k^2=0$ pole for $M\rightarrow m$. In Minkowski signature
 on the light front, this is manifest  in (\ref{gap_eq}) with the $k^+=0$ pole. Technically, there is a subtlety in trying
 to relate  (\ref{MFK}) to (\ref{gap_eq}) by analytical continuation, since 
 \begin{widetext}
\bea
\label{IM2X}
I(M^2)=\int\frac{d^2k_\perp dk^+dk^-}{(2\pi)^4}\frac{4i}{2k^+k^--k_\perp^2-M^2+i\epsilon}\mathcal{F}^2(k)
\rightarrow
   \int \frac{dk^+d^2k_\perp}{(2\pi)^3}\frac{\epsilon(k^+)}{k^+}\mathcal{F}^2(k)
\eea
\end{widetext}
assumes that the pole $k^-=\frac{k_\perp^2+M^2}{2k^+}$ remains within the closing contour. However, as 
$k^+\rightarrow 0$, the pole pinches the contour, hence the subtlety~\cite{Yan1973}. Another way to see this is to rewrite
(\ref{IM2X}) using the Schwinger trick. 

In the zero instanton size limit,
\bea
   && I(M^2)\rightarrow \nonumber\\
    &&\int\frac{d^2k_\perp dk^+dk^-}{(2\pi)^4}\int_0^\infty d\alpha \ 4e^{i\alpha(2k^+k^--k_\perp^2-M^2+i\epsilon)}\nonumber\\
\eea
the $k^-$ integration  yields a singular delta-function
\bea
    \int_{-\infty}^\infty\frac{dk^-}{2\pi} e^{i\alpha 2k^+k^-}\rightarrow \frac{1}{2\alpha}\delta(k^+)
\eea
The spontaneous breaking of chiral symmetry on the light front stems from the accumulation of
the $k^+=0$ fermionic modes in the gap equation. The singularity can be regulated by the covariant BPHZ  subtraction scheme
\bea
    \frac{\partial I(M^2)}{\partial M^2}\rightarrow 
    -\int\frac{d^2k_\perp }{(2\pi)^3}\frac{2}{k_\perp^2+M^2}
\eea
hence
\bea
    \int \frac{dk^+d^2k_\perp}{(2\pi)^3}\frac{\epsilon(k^+)}{k^+}=2\int\frac{d^2k_\perp }{(2\pi)^3}\ln\left(\frac{\Lambda^2}{k_\perp^2+M^2}\right)\nonumber\\
\eea
with $\Lambda$ a dimensionfull scale left over by the substraction.

\bibliography{reference}
\end{document}